\documentclass[%
  twoside,
  floatfix, 
  reprint,
  amsmath,amssymb,
  aps,
  prx,
  nofootinbib,
  superscriptaddress,
  a4paper
]{revtex4-2}

\usepackage{graphicx}%
\usepackage[usenames,dvipsnames]{xcolor}
\usepackage{siunitx}
\usepackage{physics}  
\usepackage{orcidlink}

\usepackage[english]{babel} 
\usepackage{subeqnarray}    
\usepackage{multirow}       
\usepackage{upgreek}        
\usepackage{mathrsfs}       
\usepackage{enumitem}       
\usepackage{titlesec}       

\usepackage{fourier}
\usepackage{framed}

\DeclareSIUnit{\au}{{a.u.}}
\usepackage{gensymb} 

  \newcommand{\fig}[1]{Fig.\,\ref{#1}}
  
  \newcommand{\sect}[1]{Sec.~\ref{#1}}
  \newcommand{\eq}[1]{Eq.~\eqref{#1}}

  \definecolor{kclred}{RGB}{226,35,26}

  \definecolor{cornellred}{rgb}{0.7, 0.11, 0.11}
  \definecolor{carnelian}{rgb}{0.7, 0.11, 0.11}
  \definecolor{cerulean}{rgb}{0.0, 0.48, 0.65}
  \definecolor{darkpastelred}{rgb}{0.76, 0.23, 0.13}
  \definecolor{alizarin}{rgb}{0.82, 0.1, 0.26} 
  \definecolor{cobalt}{rgb}{0.0, 0.28, 0.67} 
  \definecolor{glaucous}{rgb}{0.38, 0.51, 0.71} 
  \definecolor{blue-violet}{rgb}{0.54, 0.17, 0.89}
  \definecolor{tealgreen}{rgb}{0.0, 0.51, 0.5}
  \definecolor{amber}{rgb}{1.0, 0.49, 0.0}
\definecolor{cadet}{rgb}{0.33, 0.41, 0.47}    

\usepackage[utf8]{inputenc}
\usepackage[T1]{fontenc}

\usepackage{lipsum}

\usepackage[
  centering, includefoot,
  text={7.1in,10.2in},
  total={6.3in,8.75in},
  top=0.8in, left=0.62in,
  ]{geometry}

\usepackage{hyperref}

\hypersetup{
  colorlinks,
  linkcolor=cobalt,
  urlcolor=cobalt,
  citecolor=cadet,
  pdftitle={Picard-Lefschetz theory for strong-field physics}, 
  pdfauthor={Anne Weber}
}

\usepackage{natbib}

\newcommand{\citenistsec}[1]{\cite[\href{http://dlmf.nist.gov/#1}{\S{}#1}]{NIST_handbook}}

\usepackage{enumitem}
\setlist[itemize]{noitemsep}
\usepackage{subcaption}

  \usepackage{ragged2e} 

  \makeatletter
  \long\def\@makecaption#1#2{%
    \vskip\abovecaptionskip
    \sbox\@tempboxa{\small\justifying \textbf{#1.} #2}%
    \ifdim \wd\@tempboxa >\hsize
      \small\justifying \textbf{#1.} #2\par
    \else
      \global \@minipagefalse
      \hb@xt@\hsize{\hfil\box\@tempboxa\hfil}%
    \fi
    \vskip\belowcaptionskip}
  \makeatother

  \newcommand{\Sati}{S_\mathrm{ATI}}

  \newcommand{\Shhg}{S_\mathrm{HHG}}

  \newcommand{\ti}{t_\mathrm{i}}
  \renewcommand{\tr}{t_\mathrm{r}}
  \newcommand{\tis}{t_{\mathrm{i},s}}
  \newcommand{\trs}{t_{\mathrm{r},s}}

  \newcommand{\Ip}{\mathcal{I}_\mathrm{p}}
  
  \newcommand{\Upond}{U_\mathrm{p}}

\renewcommand{\real}{\mathrm{Re}}
\newcommand{\imag}{\mathrm{Im}}

\renewcommand{\d}{\,\mathrm{d}}

\newcommand{\e}{\mathrm{e}}

\newcommand{\im}{\mathrm{i}}
\renewcommand{\vec}[1]{\mathbf{#1}}

  \usepackage[colorinlistoftodos]{todonotes}

  \makeatletter \def\@captype{figure} \makeatother

\usepackage{mathtools}

\begin{document}
\title{A universal approach to saddle-point methods in attosecond science}

  \author{Anne Weber
  \orcidlink{0000-0002-1651-5063}}
  \email{anne.weber@kcl.ac.uk}
  \affiliation{Attosecond Quantum Physics Laboratory, Department of Physics, King’s College London, Strand Campus, London WC2R 2LS, UK}

  \author{Job Feldbrugge
  \orcidlink{0000-0003-2414-8707}}
  \affiliation{Higgs Centre for Theoretical Physics, University of Edinburgh, Edinburgh EH8 9YL, UK}

  \author{Emilio Pisanty
  \orcidlink{0000-0003-0598-8524}}
  \affiliation{Attosecond Quantum Physics Laboratory, Department of Physics, King’s College London, Strand Campus, London WC2R 2LS, UK}

\date{October 14, 2025}

\begin{abstract}
Light-matter interactions within the strong-field regime, where intense laser fields can ionise a target via tunnelling, give rise to fascinating phenomena such as the generation of high-order harmonic radiation (HHG) and, correspondingly, light pulses of attosecond duration. 
On the atomic scale, these strong-field processes are naturally described in terms of highly-oscillatory time integrals which are often approximated using saddle-point methods.
These methods simultaneously simplify the calculations and let us understand the physical processes in terms of semi-classical electron trajectories, or quantum orbits. 
However, applying saddle-point methods for HHG driven by polychromatic laser fields without clear dynamical symmetries has remained challenging. 
Here we introduce Picard-Lefschetz theory as a universal and robust link between the time integrals and the semi-classical trajectories, for arbitrary driving laser fields.
The continuous deformation of the integration contour towards so-called Lefschetz thimbles allows an exact evaluation of the integral, as well as the identification of relevant quantum orbits, independently of dynamical laser field symmetries or quantum orbit classification heuristics.
The latter is realised via the ``necklace algorithm'', a novel solution to the open problem of determining the relevance of saddle points for a two-dimensional integral, which we introduce here.
We demonstrate the versatility and rigour of Picard-Lefschetz methods by studying Stokes transitions and spectral caustics arising in HHG driven by two-colour laser fields.
For example, we showcase a quantum-orbit analysis of the colour switchover, which links the regime of perturbative two-colour fields with that of fully bichromatic driving fields.
With this work, we set the foundation for a rigorous application of quantum-orbit based approaches in attosecond science that enables the interpretation of state-of-the-art experimental setups, and guides the design of future ones.

\end{abstract}

\maketitle


\section{Introduction}

  In quantum mechanics, transitions between states can be described using Feynman's path integral formalism, which considers all possible trajectories connecting initial and final states. 
  Saddle-point methods (SPMs) allow this highly oscillatory integral to be approximated as a discrete sum over dominant, classical-like paths, providing both computational efficiency and physical insight.
  In attosecond science, a field with the goal of measuring atomic, electronic and molecular dynamics on their natural timescale, SPMs have played a central role from the outset \cite{lewenstein1994theory,keldysh1964ionization,perelomov1967IonizationII}.  
  To probe the ultrafast dynamics, attosecond science relies on highly nonlinear light-matter effects such as above-threshold ionisation (ATI) and high-order harmonic generation (HHG).
  The theoretical description of those optical processes within the strong-field approximation has been linked to a semi-classical picture of discrete electron trajectories, known as the \emph{quantum orbit formalism} \cite{becker2002abovethreshold,kopold2000quantum,salieres2001feynmans}.
  However, the technological development of the past decades has enabled the use of more complex laser fields to drive the nonlinear optical processes.
  With that, the framework of SPMs has begun to show its limitations. 
  Specifically, in the absence of temporal symmetries in the laser waveforms it becomes unclear which trajectories (i.e., which saddle points) contribute meaningfully to the dynamics.

  Here, we introduce the methods of Picard-Lefschetz theory \cite{lefschetz1924lanalysis,picard1897theorie,pham1983vanishing,witten2010analytic}~--~the formalised generalisation of all SPMs~--~as a rigorous and universal approach to the most common integrals arising in attosecond science. 
  Picard-Lefschetz theory suggests a rigorous mathematical definition of the Feynman path integral in terms of steepest descent manifolds connected to saddle points of the phase function \cite{feldbrugge2023existence}.
  This allows the analysis of phenomena that were previously inaccessible to semi-classical methods like spectral caustics \cite{raz2012spectral,birulia2019spectral,raab2025xuv,dong2024caustic,facciala2018highorder,facciala2016probe}, and with that a deeper understanding of the underlying quantum dynamics. 

  Within attosecond science, the strong-field approximation (SFA) is the main theoretical framework to describe the microscopic response of a gas atom subjected to a strong driving laser field (peak intensity comparable to the Coulomb force).
  The SFA offers an intuitive description of the process of HHG in terms of the three steps (1) tunnel ionisation, (2) propagation in the continuum and acceleration by the driving field, and (3) recombination with the parent ion to emit a high-energetic photon
  \cite{lewenstein1994theory,corkum1993plasma,kulander1993superintense}.
  The spectrum of high-energy photons typically covers a long range of frequencies (the HHG plateau), temporally corresponding to a train of short flashes of light -- laser pulses of attosecond duration. 
  To understand and control the properties of this attosecond pulse it is necessary to understand the process of HHG for a range of different driving laser fields.

  Non-symmetric driving fields are of growing interest because they have demonstrated precise control over the temporal and polarisation characteristics of the created HHG spectra \cite{jin2014waveforms,cirmi2023optical,mitra2020suppression,mansten2008spectral,chipperfield2009ideal} and offer additional insights to the quantum dynamics at play \cite{kneller2022look,he2010interference}.
  These tailored light fields are often a combination of laser fields of different frequencies and polarisation.
  For example, a weak commensurate second colour field can be used to modify strong-field ionisation such that an additional phase delay scan allows to measure the intricate details of the tunnelling process \cite{shafir2012resolving,dudovich2006measuring,zhao2013determination,eicke2019Attoclock}.
  Going beyond the perturbative regime when adding a second field can extend the range of generated harmonic frequencies, increase the overall signal and ultimately 
  change the spectral and temporal properties of the created attosecond pulse
  \cite{mitra2020suppression, mauritsson2009subcycle, mansten2008spectral,he2010interference, ruiz2009control, roscamabbing2020divergence, haessler2014optimization}.
  Lastly, by creating three-dimensional polarisation states of light it becomes possible to distinguish between chiral enantiomers of molecules \cite{baykusheva2018chiral,ayuso2019synthetic}.
  The heuristics that were established to describe the atomic response in terms of discrete quantum orbits (corresponding to the different saddle points in the SPM), however, were developed for simple, monochromatic and one-dimensional, driving field shapes.
  They fail for those generic drivers with arbitrary waveform.

  To overcome these limitations we need to advance our understanding of SPMs.
  Their mathematical backbone and the generalised approach to highly-oscillatory and only conditionally convergent integrals is Picard-Lefschetz theory \cite{lefschetz1924lanalysis,picard1897theorie,pham1983vanishing}. 
  As such, it has been introduced to solve high-dimensional path integrals in quantum field theories and was applied to solve path integrals in other areas of physics \cite{witten2010analytic,feldbrugge2017lorentzian,tanizaki2014realtime,bharathkumar2020lefschetz,feldbrugge2023oscillatory,feldbrugge2025efficient}. 
  The fundamental idea is that there exists a continuous deformation (`the downwards flow') of the integration contour into the complex plane such that -- evaluated along this new contour -- the integrand oscillations minimise, while giving an exact formulation of the integral. 
  This new contour is called the \emph{Lefschetz thimble} and passes through the relevant saddle points of the phase function of the integrand. 
  Vice versa, this means that we can identify relevant saddle points by checking if the inverse deformation (i.e., the upwards flow) leads us back to the original integration contour.


  The aim of this paper is to introduce Picard-Lefschetz theory to attosecond science.
  We demonstrate its effectiveness by addressing two challenges that remain inaccessible to the established semi-classical quantum-orbit methods: 
  tackling configurations that exhibit caustics (where multiple saddle-point solutions coalesce), and, more generally, obtaining the strong-field response throughout any continuous parameter scan that changes the number of relevant trajectories.
  For that, this work is structured as follows.
  We begin in \sect{sec:quantum-orbits} with a brief overview of the existing quantum-orbit formalism for the ionisation amplitude for strong-field tunnelling and the dipole response for HHG, as examples of integrals over one and two dimensions, respectively.
  \sect{sec:picard_lefschetz_theory} introduces Picard-Lefschetz theory, and in particular the two procedures for evaluating highly-oscillatory integrals that can be derived from it. 
  The first is the numerical implementation of the continuous downwards flow that deforms the integration contour into Lefschetz thimbles.
  Secondly, we have developed the ``necklace algorithm'' to identify which of the critical points are relevant contributors in the saddle-point approximation of a two-dimensional integral.

  These methods are then applied to strong-field physics in \sect{sec:hhg_driven_by_two_colour_laser_fields}, where we consider HHG from two-colour driving fields as an example.
  There, we present the harmonic response for driving field configurations that produce swallowtail caustics over a two-colour intensity ratio and phase delay scan.
  Moreover, we use the saddle-point based approach for the analysis of relevant electron trajectories in a setup where we gradually replace a monochromatic driver with its second harmonic -- a technique termed colour switchover \cite{weber2025quantum} -- to demonstrate the versatility of our methods. 
  Our specific implementation has been deposited in Ref.~\cite{figuremaker-code}.

\section{Quantum-orbit approaches in strong-field theory}
  \label{sec:quantum-orbits}
    The response of an atom to illumination by a laser pulse whose peak intensity is comparable with the strength of the binding Coulomb forces between nucleus and electrons can be described using the so-called strong-field approximation (SFA) framework \cite{le2016strongfield,lewenstein1994theory,popruzhenko2014keldysh,smirnova2013multielectron}.
    The SFA consists of a set of approximations, most importantly assuming only one single active electron which is either in the ground state or in a Volkov-type continuum state where its motion is dictated entirely by the driving laser field and the drift momentum, neglecting the ion's Coulomb potential.
    Generally, these approximations can be extremely restrictive, but for laser peak intensities in the order of $10^{14} \SI{}{W/cm^2}$ and atomic gas targets, the SFA is indeed the preferred theoretical model, and in good agreement with experimental measurements \cite{smirnova2013multielectron}.
    Apart from numerical simplicity, the SFA also offers a quite intuitive understanding of the processes happening in these parameter ranges.
    Strong-field effects well-described within the SFA include above-threshold ionisation (ATI), non-sequential double-ionisation, high-harmonic generation (HHG) etc. \cite{amini2019symphony}

    In this paper, we focus on the two processes of direct ATI (via strong-field tunnelling) and HHG. 
    For both processes, the atomic response can be written as a Feynman path integral with the semi-classical action as an exponentiated phase function in the integrand \cite{salieres2001feynmans, kopold2000quantum, lewenstein1994theory}.
    That is, for ATI we consider the ionisation probability and for HHG we consider the radiation dipole associated with the emitted photons.
    These types of integrals are highly oscillatory and can be solved using methods of stationary phase, also known as saddle-point methods \cite{bleistein1975asymptotic,wong2020asymptotic}.%
    By identifying stationary points of the exponent, i.e., saddle points of the action, the atomic response can be expressed in terms of Gaussian contributions from distinct ionisation events (for ATI) and quantum orbits (for HHG), in analogy to the least-action principle in Feynman's path integral formalism.

    However, in rewriting the continuous path integral to a sum over discrete contributions there are (at least) two intricacies that are often overlooked: 
    Firstly, the summation only runs over a strict subset of stationary solutions of the action.
    For example, all saddle point solutions have their complex-conjugated counterpart which are typically neglected because the resulting contribution would be exponentially large and hence, unphysical \cite{popruzhenko2014keldysh,jasarevic2020application,milosevic2002role,figueirademorissonfaria2000phasedependent}.
    And secondly, in situations were saddle points are in close vicinity, their contribution is not actually of Gaussian shape.
    A prominent instance of this is the high-harmonic cutoff where the saddle-point solutions for the `short' and `long' trajectories perform a missed approach and their joint contributions is modelled in terms of an Airy function ~\cite{figueirademorissonfaria2002highorder,pisanty2020imaginary,milosevic2025application, milosevic2002role}. 
    Both of these issues become particularly relevant when we start driving the processes not with simple one-dimensional monochromatic fields with dynamical symmetry, but instead with more complicated fields composed of multiple components of different polarisations, frequencies etc.
    With every additional frequency component new saddle-point solutions arise.
    These potentially constitute new relevant quantum orbits, depending on the amplitude ratio and phase shifts between the constituent fields~\cite{habibovic2025complete}.

    In the following we briefly describe the established procedures for the two considered processes of ATI and HHG and allude to how our current understanding of saddle-point methods is insufficient to describe state-of-the-art experimental setups.

\subsection{Direct photoelectrons from above-threshold ionisation in the strong-field tunnelling regime} 
    \label{sec:ATI}

    Electrons that tunnel through the barrier formed by the combination of Coulomb potential and the strong laser field's vector potential $\vec{A}(t)$ can be observed at a detector as direct photoelectrons.
    The spectrum of drift momenta $\vec{p}$ of those electrons is typically expressed in terms of the photoelectron ionisation amplitude, given in atomic units as
    \begin{equation}
      \Psi(\vec{p}) = \int_{-\infty}^{\infty} P(\vec{p} + \vec{A}(t) ) \e^{-\im \Sati(t)}  \d t
      \label{eq:ATI-integral}
    \end{equation}
    where the phase function 
    \begin{equation}
      \Sati(\vec{p},t) = \int_{-\infty}^t \left[ \Ip + \frac{1}{2} \left(\vec{p} + \vec{A}(t') \right)^2\right] \d t' \,,
      \label{eq:Sati}
    \end{equation}
    is the semi-classical action of the electron with the ionisation potential $\Ip$ \cite{popruzhenko2014keldysh}.
    The integration in \eq{eq:ATI-integral} runs over the past time, and essentially includes the interaction time with the laser field that drives the process.
    The prefactor $P(\vec{k})$ incorporates any information about the ground state of the atom and is assumed to depend smoothly on the canonical momentum $\vec{k}$.
    Saddle points $t_s$ of this semi-classical action are defined by
    \begin{equation}
    \frac{\partial \Sati}{\partial t}\bigg|_{t=t_s} =
    \Ip + \frac{1}{2} \left(\vec{p} + \vec{A}(t_s) \right)^2 = 0 \,, 
    \label{eq:Satidrv}
    \end{equation}
    generally complex-valued for $\Ip > 0$, and are interpreted as the discrete ionisation times at which the electron escapes the Coulomb barrier \cite{popruzhenko2014keldysh}. 
    The ionisation amplitude \eq{eq:ATI-integral} can then be rewritten as
  \begin{equation}
    \Psi(p) \approx \sum_{t_s} \sqrt{\frac{2 \pi}{\im \Sati''}} P(\vec{p} + \vec{A}(t_s)) \e^{-\im \Sati(t_s)}
    \label{eq:ATI-sum}
  \end{equation}
    where the square-root term with the second derivative of the action comes in as we expand the exponential term around the saddle points $t_s$ into a Gaussian shape and analytically integrate those, which is known as the standard saddle-point method \cite{bleistein1975asymptotic,wong2020asymptotic}. 

    To determine the subset of relevant saddle points out of all solutions to \eq{eq:Satidrv} we need to identify a connected steepest-descent path that leads us in positive $\real(t)$ direction, consistent with the original integration domain from $t=-\infty$ to $\infty$ in \eq{eq:ATI-integral}.
    
    That is, we seek a steepest-descent path of values of 
    The magnitude of the integral is dictated by
    For that, we allow our integration variable to take complex values, and plot the value of $\imag(\Sati(t))$ in this complex plane of~$t$, as this dictates the magnitude of the integral $|\e^{-\im \Sati(t)}| = \e^{\imag(\Sati(t))}$.    
    Because of the Cauchy-Riemann relations for complex analytic functions, the paths of steepest descent of $\imag(\Sati(t))$ are then given by lines of constant $\real(\Sati(t))$%
    \footnote{See \sect{sec:pl-maths} for the derivation.}.
    A suitable integration path can therefore be found by plotting the contour level lines $\real(\Sati(t)) = \real(\Sati(t_s))$ for each saddle point and then identifying a connected path.

\begin{figure}[t]
      \includegraphics[width=\linewidth]{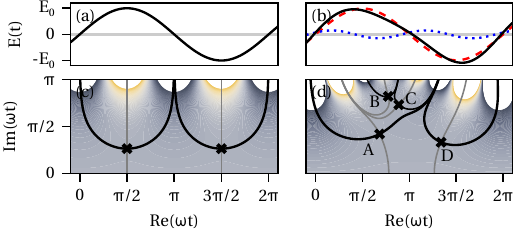}
      \caption{
      Bottom row: Contour plots of $\imag(\Sati(t))$ for the complex $t$ plane for the two driving fields shown on top (panels (a) and (b)) and drift momenta $p=0$ and $p=\SI{1.2}{\au}$ respectively.
      Steepest-descent and steepest-ascent contour lines (black and grey lines, respectively) are attached to saddle points \eq{eq:Satidrv} (black markers, labelled in (d) for convenience), with the resulting integration path drawn as a heavy black line.
      The electric field in (b) is composed of the two constituent fields of frequency $\omega$ (red dashed) $2 \omega$ (blue dotted) with phase shift $\varphi = 0.5$ and amplitude ratio $E_2/E_1 = 0.15$ acc. to \eq{eq:tc-field}.
      }
      \label{fig:ATI-landscapes}
    \end{figure}

    For the simple monochromatic driving field $\vec{E}(t) = E_0 \sin(\omega t)$ (shown in \fig{fig:ATI-landscapes}(a))
    with 
    $I_0 = E_0^2 = 0.92\times 10^{14} \SI{}{W/cm^2}$ 
    ($ E_0 = \SI{0.05}{\au})$, 
    $\lambda = \SI{1030}{nm}$ 
    ($\omega = \SI{0.044}{\au}$) 
    this is shown in \fig{fig:ATI-landscapes}(c), where we have used
    $\Ip = \SI{15.8}{eV}$ and $\vec{A}(t) = \-\int \vec{E}(t) \d t$.
    In these contour plots of $\imag(\Sati(t))$ (hills in yellow, valleys in dark grey) we find
    there are two saddle points in the upper complex-half plane, at $\real(\omega t) \approx \pi/2$ and $\approx 3\pi/2$ around the maxima of the field.
    Both are part of the connected steepest-descent integration path (heavy black line) and hence relevant contributors to the summation of ionisation events \eq{eq:ATI-sum}.
    This confirm the intuitive understanding of the process at these times the the distortion of the Coulomb potential is most significant and it is easiest for the electron to tunnel out.

    In the case of a more complicated driving field, as shown in \fig{fig:ATI-landscapes}(b), the situation requires a bit more attention.
    Technically, there are four saddle points, which -- for convenience -- we have labelled A,B,C,D in panel (d). 
    Drawing the respective contour level lines for each saddle point shows that a connected path in positive $\real(t)$ direction can only be formed with the lines passing through points A and D, but not B and C.
    This lets us conclude that only A and D are contributors to the summation \eq{eq:ATI-sum}, whereas B and C need to be neglected.
 
    While this approach of determining the relevance of specific saddle points seems intuitively promising, it is surprisingly non-trivial to develop an algorithm that finds connected steepest-descent contours in a robust and rigorous fashion \cite{gibbs2024numerical,shanin2022saddle}. 
    Numerical instabilities are expected as soon as singularities enter the region of interest (vis. hills and valleys that are far away from the real axis, yellow and dark grey regions in \fig{fig:ATI-landscapes} respectively) or when saddle points are in close vicinity.
    Furthermore, this approach still leaves us uninformed on how to solve the integral in the case of Stokes transitions or coalescing saddles, because in those cases the steepest-descent contours do not have a unique definition.

    Ultimately, even though we could technically decide over the relevance of saddle-points on a case-by-case basis by examining the action landscapes, so far there is no robust method to determine the steepest-descent integration path.
    While the one-dimensional integration is at least heuristically understood, for the two-dimensional time integration even the case-by-case procedure fails to provide a meaningful strategy.

\subsection{High-harmonic generation}

  The process of high-harmonic generation is typically understood in terms of the three-step model: 
  The electron escapes the atomic Coulomb potential via strong-field tunnel ionisation (step 1), then propagates in the continuum where it is accelerated by the driving laser field (step 2) until it finally recombines with its parent ion. 
  Upon recombination a high-energy photon is emitted (step 3) whose frequency is a integer multiple, a harmonic, of the fundamental driver's frequency $\omega$ \cite{keldysh1964ionization,faisal1973multiple,reiss1980effect,lewenstein1994theory}.
  The spectrum of the emitted radiation typically covers a long range of frequencies (the HHG plateau) followed by a sharp drop in intensity at the high-harmonic cutoff.
  The theoretical description that supports this understanding is the above-mentioned SFA~\cite{lewenstein1994theory}.
  Starting from the time-dependent Schrödinger equation and incorporating this set of assumptions and approximations (find detailed explanations in e.g.~\cite{nayak2019saddle,smirnova2013multielectron}) ultimately yields the so-called Lewenstein integral which describes the time-dependent dipole moment created at the final photoemission step.
  Therein, the measured quantity in an experiment is the spectral power, or rather, the spectral intensity for harmonic frequency $q \omega$,
  \begin{equation}
    I(q \omega) = (q \omega)^2 |\vec{D}(q \omega)|^2
    \label{eq:spectral-intensity}
  \end{equation}
  which uses the Fourier transform of the Lewenstein integral and $q \in \mathbb{N}$. 
  The dipole moment $\vec{D}(q \omega)$ is given as the two-dimensional integral over ionisation and recombination times $\ti$ and $\tr$:
  \begin{align}
    \vec{D}(q \omega) &= \im \int_{-\infty}^{+\infty} \d \tr \int_{-\infty}^{\tr} \d \ti \, 
    \vec{d}\left( \vec{p}_s(\ti, \tr) + \vec{A}(\tr)\right) \nonumber \\
    & \quad \Upsilon\left( \vec{p}_s(\ti, \tr) + \vec{A}(\ti)\right) \nonumber \\ 
    & \quad \left( \frac{2 \pi}{\im (\tr - \ti)} \right)^{3/2}
    \e^{-\im \Shhg(\ti, \tr)}
     \,. \label{eq:hhg-integral}
  \end{align}
  with the semi-classical action 
  \begin{align}
    \Shhg(\ti, \tr) &= \frac{1}{2} \int_{\ti}^{\tr} \left( \vec{p}_s(\ti, \tr) + \vec{A}(t)\right)^2 \d t + \nonumber \\ 
    & \quad + \left(\tr - \ti \right) \Ip - q \omega \tr  \label{eq:Shhg} 
  \end{align}
  The scalar factor $\Upsilon(\vec{k})$ denotes the transition dipole from the ground state into the excited state, while $\vec{d}(\vec{k})$ is the recombination matrix element \cite{smirnova2013multielectron} for the kinematic momentum $\vec{k}$.
  The stationary momentum associated with a given electron path between $\ti$ and $\tr$ is given by
  \begin{equation}
    \vec{p}_s(\ti, \tr) = -\frac{1}{\tr-\ti} \int_{\ti}^{\tr} \vec{A}(t) \d t
  \end{equation}
  as a result of applying the saddle-point method to the integration over possible intermediate drift momenta $\vec{p}$.

  \begin{figure}[t]
    \includegraphics[width=\linewidth]{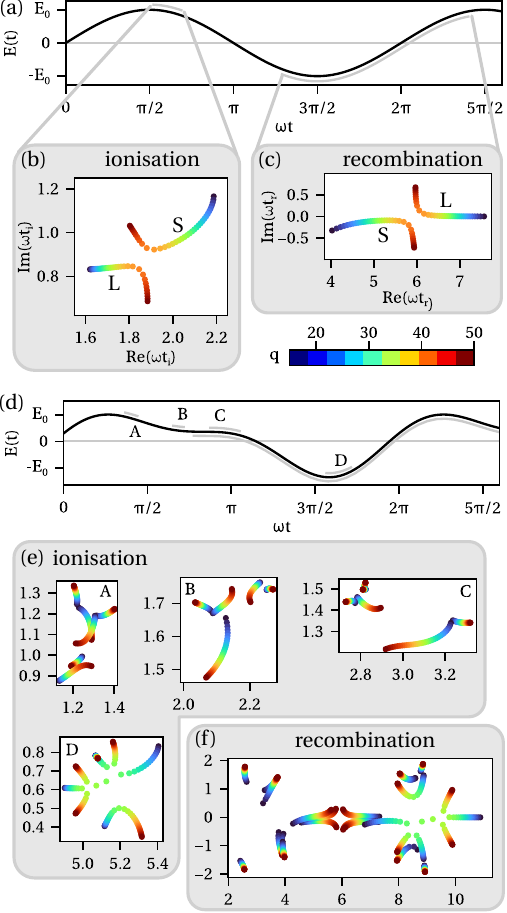}
    \caption{
    Typical structure of the complex saddle point solutions for HHG in the complex plane of ionisation and recombination times.
    For the monochromatic driving field shown in (a) the solutions across a range of harmonic orders (colour bar) follow lines in the complex planes (b) and (c) and can be classified as `short' (S) and 'long' (L) trajectories.
    For the electric field shown in (d), a two-colour field with $\varphi = 0.75$ and $E_2/E_1 = 0.44$, as per \eq{eq:tc-field},
    the solutions still trace lines in the complex plane and form several ionisation windows (labelled A,B,C,D)), but their structure is more intricate, hindering a classification.}
    \label{fig:HHG-saddles-in-cp}
    \end{figure}

  Similar to the aforementioned one-dimensional time-integral for ATI, the exponentiated action $\Shhg(\ti, \tr)$ makes the integral \eq{eq:hhg-integral} highly oscillatory.
  Analogously, the integral can be understood as a sum of contributions from several quantum orbits, each associated with a stationary point $(\tis, \trs)$ of the action $\Shhg$.
  The integral \ref{eq:hhg-integral} is therefore often approximated by
  \begin{align}
    \vec{D}(q \omega) &\approx \sum_s \frac{2 \pi}{\sqrt{-\mathrm{det}(\Shhg''(\tis,\trs))}} 
    \vec{d}\left( \vec{p}_s(\tis, \trs) + \vec{A}(\trs)\right) \nonumber \\
    & \quad \vec{\Upsilon}\left( \vec{p}_s(\tis, \trs) + \vec{A}(\tis)\right) \nonumber \\ 
    & \quad \left( \frac{2 \pi}{\im (\trs - \tis)} \right)^{3/2}
    \e^{-\im \Shhg(\tis, \trs)}
     \,. \label{eq:hhg-sum-spm}
  \end{align}
  The stationary points are time pairs $(\tis, \trs)$ 
  for which the first derivatives vanish,
  \begin{equation}
    \frac{\partial \Shhg}{\partial \ti}  = 0 \quad \text{and} \quad 
    \frac{\partial \Shhg}{\partial \tr} = 0 \,, \label{eq:speqs}
  \end{equation}
  and are saddle points in the complex plane.
  Each of those saddle points $(\tis, \trs)$ represents a semi-classical electron trajectory giving rise to the quantum orbit formalism  \cite{becker2002abovethreshold,milosevic2006quantumorbit}

  For a given harmonic order $q$ there will be multiple solutions to \eq{eq:speqs}, such that the total harmonic response typically consists of contributions from various interfering quantum orbits.
  Inspecting the ionisation and recombination times for the different values of $q \omega$, i.e.\,, throughout the harmonic spectrum, traces `lines' in the complex planes for $\ti$ and $\tr$ respectively; see \fig{fig:HHG-saddles-in-cp}(b) and (c) for the solutions around one ionisation window of a monochromatic driving field shown in panel (a).
  Those lines are often associated with different types of trajectories, mainly classified by their travel time $\real(\trs-\tis)$ into `short' and `long' orbits, or associated with various ionisation bursts within one cycle of the driving field \cite{pisanty2020imaginary,milosevic2002role,milosevic2019xray}. 
  
  As in the case of ATI, not all mathematical solutions to \eq{eq:speqs} are actually relevant quantum orbits. 
  For example, we typically only consider those solutions where the ionisation times have a positive imaginary part, $\imag(\tis) > 0$.
  Furthermore, for monochromatic driving fields, the short trajectories have to be discarded after the high-harmonic cutoff of the spectrum \cite{pisanty2020imaginary}.
  These heuristics, while often physically motivated, lack a consistent theoretical foundation and fail to generalise to arbitrary or time-varying laser fields.
  They have been established without mathematical rigour and are often based on the fact that including other solutions leads to diverging integral values \cite{jasarevic2020application,figueirademorissonfaria2000phasedependent,milosevic2002role}. 
  Moreover, the heuristics rely on the classification of the solutions.

  For a generic driving laser field the structure of ionisation and recombination times in the complex plane may be much more complicated, as shown in \fig{fig:HHG-saddles-in-cp} panel (e) and (f) for the electric driving field in panel (d).
  In this particular case, we may attribute solutions to separate ionisation windows (here labelled A - D) but a classification scheme for the individual solutions cannot easily be derived. 
  Furthermore, the classification scheme breaks down once we consider smooth transitions of driving fields, for example a scan through a phase delay between the two components of a two-colour field \cite{figueirademorissonfaria2000phasedependent}.
  In such cases, the saddle-point structures may change qualitatively, including coalescences and branch cuts, causing the existing heuristics to fail.
  As a result, we are unable to determine relevant saddle points and therewith the contributing quantum orbits.
  This underscores the need for a more robust and systematic approach to applying saddle-point methods to the integral \eq{eq:hhg-integral} for the case of arbitrary driving fields.
  Without such an approach, our ability to interpret strong-field phenomena in terms of quantum orbits remains fundamentally limited.
  The following section introduces Picard-Lefschetz theory as the rigorous framework to address these issues, in a generic form.


\section{Picard-Lefschetz theory}
  \label{sec:picard_lefschetz_theory}

\begin{figure}[t]
      \includegraphics[width=\linewidth]{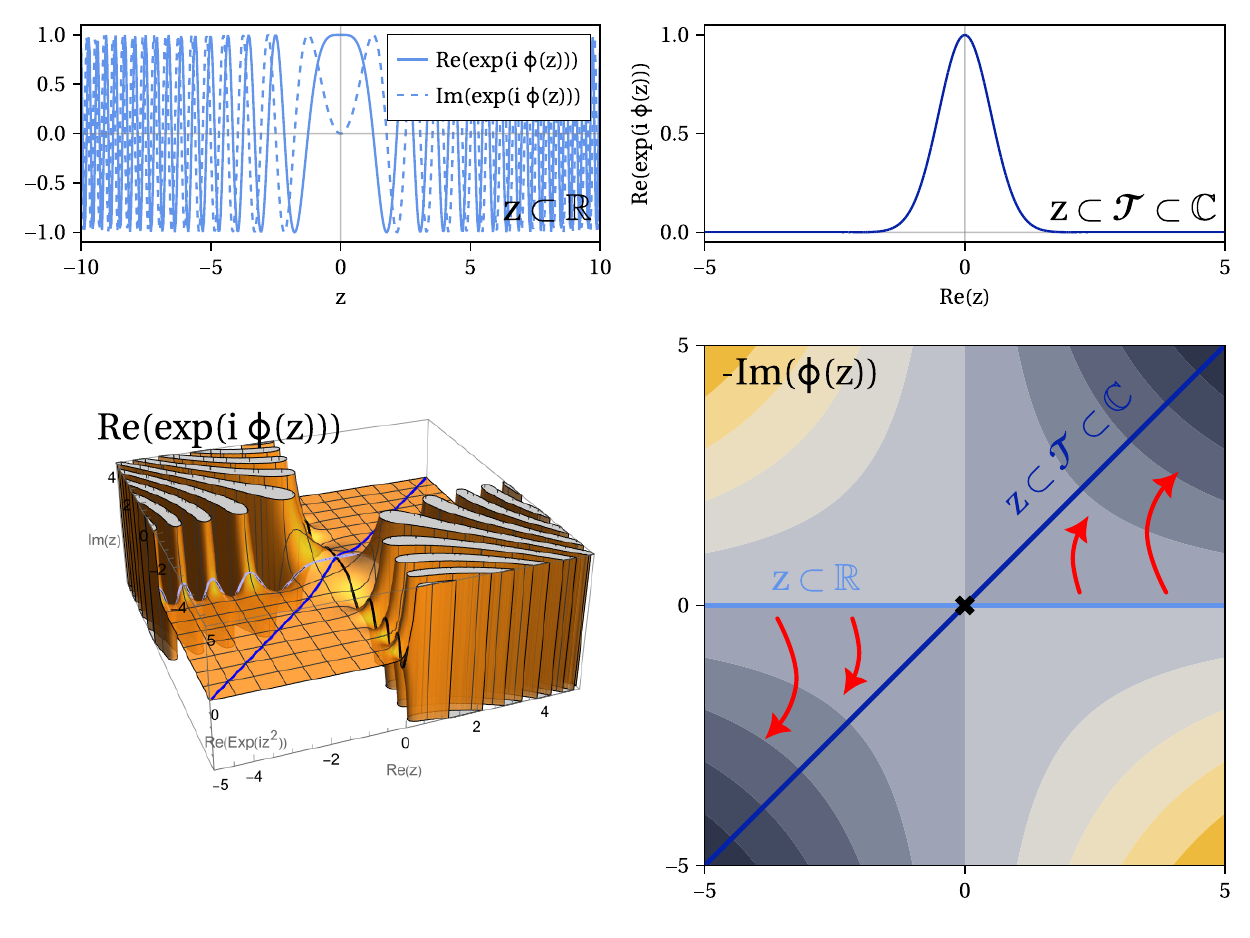}
      \caption{Fundamental idea of Picard-Lefschetz theory, shown on the toy model function $\phi(z) = z^2$: The integrand $\e^{\im z^2}$ is highly oscillatory when evaluated along the real axis (top left panel).
      The continuation of $z$ into the complex plane (bottom left panel) shows, that the oscillations (along the light blue line) vanish if we evaluate the integrand along a different contour (dark blue line).
      The contour that localises the integrand by minimising the oscillations follows steepest-descent paths of $\imag(z^2)$ (contour plot in the bottom right panel) and is identified by deforming the original integration domain according to the downwards flow (red arrows in the bottom right panel) and leads across the saddle point at $z=0+0\im$, where  $\phi'(z)=0$.
      Often, the integrand along the new contour has Gaussian shape (top right panel) and can be calculated analytically.
      }
      \label{fig:PL-schematics}
    \end{figure}

    Integrals of the form
    \begin{equation}
      I = \int\displaylimits_{\mathclap{\mathcal{C}_0 \subset \mathbb{R}^N}}
      \e^{\im \phi(\vec{x})/\hbar} 
      \d \vec{x} \,,
      \label{eq:PL-integral}
    \end{equation}
    such as \eq{eq:ATI-integral} and (\ref{eq:hhg-integral}),
    evaluated along a path $\mathcal{C}_0$ in real space with the real-valued phase function $\phi(\vec{x})$ are highly oscillatory and only conditionally convergent. 
    This makes them notoriously difficult to evaluate numerically, especially in the semi-classical limit $\hbar \rightarrow 0$ (see \fig{fig:PL-schematics}(a) for an example).
    These types of 
    path integrals appear across a vast range of research areas and each research area has developed different methods to solve them.
    As such, Picard-Lefschetz theory \cite{picard1897theorie,lefschetz1924lanalysis,arnold2012singularities,pham1983vanishing} was applied in physics in the context of Chern-Simons quantum field theory where it aids to solve the QCD sign problem \cite{witten2010analytic}. 
    After that, it was used in quantum cosmology to solve the conformal-factor problem \cite{feldbrugge2017lorentzian,feldbrugge2017no}, 
    developed into a numerical technique for lensing problems in radio astronomy \cite{feldbrugge2023oscillatory, feldbrugge2023multiplane, 2025PhRvD.111f3061B}
    and combined with Hamiltonian Monte-Carlo techniques in an attempt to solve the sign problem in lattice field theory \cite{fujii2013hybrid,collaboration2012high,cristoforetti2013monte}.
    Most recently, it was used to develop a rigorous definition of the real-time path integral \cite{feldbrugge2023existence} and efficiently evaluate real-time path integrals in quantum mechanics \cite{feldbrugge2025efficient, feldbrugge2025realtime, 2025PhRvD.111h5027F}.
    We here give a very brief overview of the main ideas, illustrated by a one-dimensional example, before we lay out further details on the mathematical background, the numerical implementations, and the application to caustics and catastrophe theory, in the subsequent sections.

    \begin{figure*}[t]
      \includegraphics[width=\textwidth]{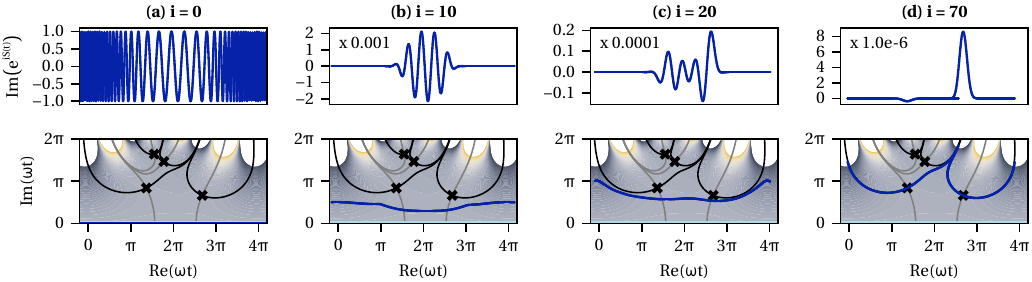}
      \caption{Flowing the integration contour (dark blue) acc.\ to the downwards flow \eq{eq:downwards-flow} for the same scenario as in \fig{fig:ATI-landscapes}(d) for discretised flow steps $i$ to minimise the oscillations of the integrand. 
      The integrand evaluated along the contour is shown in the top rows.
       }
      \label{fig:ATI-integrand-contours}
    \end{figure*}

    The fundamental insight of Picard-Lefschetz theory, shown on a toy model function in \fig{fig:PL-schematics}, 
    is to apply Cauchy's integral theorem and deform the integration contour \mbox{$\mathcal{C} \subset \mathbb{R}^N$} into the complex space $\mathbb{C}^N$ such that along this new contour the integrand no longer oscillates, the integral converges absolutely, and hence is easier (if not trivial) to evaluate. 
    Of course, the key question is: How do we find this optimal integration contour?

    Assuming that $\phi(\vec{z})$ is a locally analytic function%
    \footnote{That is, it is meromorphic, i.e., locally complex differentiable, such that it can be approximated by a Taylor series almost everywhere in the complex space.
    A remark on notation: We are using $\vec{z}$ instead of $\vec{x}$ to highlight the continuation into the complex space $\mathbb{C}^N$.}
    of $\vec{z} \in \mathbb{C}$ we continuously (!) deform the integration domain towards contours along which the amplitude of the integrand $|\e^{\im \phi}| = \e^{\real(\im \phi)}$ decreases as rapidly as possible.
    The direction of the deformation is therefore given by the \emph{downwards flow}
    \begin{equation}
      \frac{\mathrm{d} \vec{z}}{\mathrm{d} \lambda} =
       - \left(\frac{\partial \im \phi}{\partial \vec{z}} \right)^*
       \label{eq:downwards-flow}
    \end{equation}
    and the resulting integration domain -- as a function of the flow parameter $\lambda$ -- is the contour
    \begin{equation}
        \mathcal{T}(\lambda) = \{ \vec{z}(\lambda, \vec{z}_0)\,|\, \vec{z}(\lambda=0, \vec{z}_0) = \vec{z}_0 \in \mathcal{C}_0\} \,,
    \end{equation}
    where the flow is initialised along the original integration domain $\mathcal{C}_0$.
    In the limit $\lambda \rightarrow \infty$ the deformation of the contour $\mathcal{T}(\lambda)$ converges to yield a set%
    \footnote{We acknowledge that strictly speaking \eq{eq:thimbles} should be a union rather than a sum.
    However, for consistency with the literature and the \emph{summation} over integral contributions in \eq{eq:PL-integrals} we use a sum here as well.}
    of steepest-descent manifolds $\mathcal{T_\sigma}\subset \mathbb{C}^N$, so-called Lefschetz ``thimbles'':
    \begin{equation}
        \mathcal{T} = \lim_{\lambda \to \infty} \mathcal{T}(\lambda) = \sum_{\sigma} n_\sigma \mathcal{T}_\sigma\,.
      \label{eq:thimbles}
    \end{equation}
    Each thimble is attached to a critical point $\vec{z}_\sigma$ of the phase function where $\phi'(\vec{z}_\sigma) =0$ (i.e., saddle points in the complex space) as these are stationary solutions to the downwards flow \eq{eq:downwards-flow}~\cite{tanizaki2014realtime}.

    Importantly, the deformed integration contour $\mathcal{T}$ only includes a subset of all critical points.
    This subset is specified by the intersection number $n_\sigma \in \mathbb{Z}$ which counts whether the steepest-\emph{ascent} manifold attached to a critical point $\vec{z}_\sigma$ intersects the original integration domain $\mathcal{C}_0 \subset \mathbb{R}^N$.
    That is, when the thimble of the critical point $\vec{z}_\sigma$ is relevant to the integral, there exists a point on the original integration domain such that the flow eventually reaches it in the limit $\lambda \rightarrow \infty$. 
    Intuitively, because the downwards flow defines a continuous, and hence ``unambiguous'', contour transformation, we can reverse this procedure.
    Relevant critical points are therefore those which have the steepest-ascent manifold connecting back to the original integration domain.


    Let us briefly show how this contour deformation works in practice, using the example of the ionisation probability amplitude for strong-field tunnel ionisation, i.e., \eq{eq:ATI-integral}, which is a 1D integration over time, see \sect{sec:ATI}.
    The phase term of the integrand is $\im \phi(t)/\hbar = -\im \Sati(t)$, and the original integration contour $\mathcal{C}_0$ is the real $t$ axis.%
    \footnote{Note the sign change!}
    The continuous deformation of the integration contour according to the downwards flow \eq{eq:downwards-flow} (where the gradient on the right-hand side is now given by
    $-\tfrac{\partial \im \Sati(t)}{\partial t}$)
    is shown in \fig{fig:ATI-integrand-contours}, from left to right, for the parameter configuration as in \fig{fig:ATI-landscapes}(d).
    The bottom row shows contour plots of $\real(-\im \Sati)=\imag(\Sati)$
    with the integration contour drawn in dark blue and the  
    integrand $\exp(-\im \Sati(t))$ evaluated along this contour is shown in the panels above.
    While the integrand is highly oscillatory along the original, real-valued integration contour (panel (a), iteration step $i=0$), the very rapid oscillations disappear as soon as the contour is deformed even only slightly into the complex plane, i.e., after a few flow steps ((b), $i=10$).
    The flow ultimately converges ((d), $i=70$) to the steepest-descent contours attached to saddle points (black lines across the black markers), confirming the integration route shown in \fig{fig:ATI-landscapes}(d).
    Vice versa, the relevant saddle points to \eq{eq:PL-integrals} are those which have a steepest-ascent contour connecting to the real axis and hence $n_\sigma =1$.
    As $\imag(-\im \Sati)$ is constant along the steepest-ascent contours, in this one-dimensional case these are simply the level lines at $\imag(-\im \Sati(t_s))$. 
    For the two saddle points with $\imag(\omega t_s) \approx 1.25$ (A and D from \fig{fig:ATI-landscapes}(d)) we find level lines connect almost straight down to the real axis, whereas the steepest-ascent contours of the other two saddles (B and C, both around $2.5+2.5\im$) ultimately lead up into direction of higher imaginary parts, i.e., towards the light yellow-shaded regions -- `hills' -- of the contour plot, rendering $n_\sigma =0$.

    Upon the deformation of the contour according to the downwards flow, the integral (\ref{eq:PL-integral}) can ultimately be expressed as a sum over contributions from the separate thimbles%
    \footnote{In this paper, we focus on the constructive interference of the integrand at the stationary points of exponent $\phi$. 
    However, in general, the sum of thimbles includes both steepest-descent manifolds associated with the stationary points of the exponent and the stationary points of the exponent restricted to the boundary of the original integration domain, i.e.\, $\phi|_{\mathcal{C}_0}$. These boundary thimbles are always relevant. However, the integral along the first set of thimbles typically dominates over the boundary thimbles. For a systematic investigation of the boundary points in Picard-Lefschetz theory, we refer to \cite{DelabaereHowls2002}.}:
    
    \begin{align}
        I 
          = \int\displaylimits_{\mathclap{\mathcal{C} \subset \mathbb{C}^N}} \e^{\im \phi(\vec{z})/\hbar} \d \vec{z} 
          = \sum_\sigma n_\sigma 
           \int\displaylimits_{\mathclap{\mathcal{T}_\sigma \subset \mathbb{C}^N}} 
             \e^{\im \phi(\vec{z})/\hbar} \, \d \vec{z}
        \label{eq:PL-integrals}
    \end{align}
    The value of the integral is preserved at any intermediate stage $\lambda$ of the downwards flow. 
    Notably, the expressions \eq{eq:PL-integral} and \eq{eq:PL-integrals} are thus strict equalities and independent of the dimension $N$, since all we have done so far is a contour deformation. 
    However, for a non-degenerate critical point and in the asymptotic limit $\hbar \rightarrow 0$ the integral along the thimble can be approximated by a Gaussian (as in \fig{fig:PL-schematics}(c)).
    This well-known approximation is the saddle-point method, as explained in e.g. \cite{bleistein1975asymptotic}:
    \begin{equation}
      \int\displaylimits_{\mathclap{\mathcal{T}_\sigma \subset \mathbb{C}^N}} 
             \e^{\im \phi(\vec{z})/\hbar} \, \d \vec{z}
             \approx 
             \sqrt{\frac{2 \pi \hbar}{ \mathrm{det}\left(  \phi''(\vec{z}_\sigma) \right) }} \e^{\im \phi(\vec{z}_\sigma) /\hbar}
      \label{eq:SPM-short}
    \end{equation}
    leading to the saddle-point approximation of the integral $I$,
    \begin{equation}
        I \approx \sum_{\sigma} n_\sigma \sqrt{\frac{2 \pi \hbar}{ \mathrm{det}\left(  \phi''(\vec{z}_\sigma) \right) }} \e^{\im \phi(\vec{z}_\sigma) /\hbar} \,.
    \end{equation}
    The saddle-point approximation inherits the relevance of the saddle point $\vec{z}_\sigma$ through the intersection $n_\sigma$.
    

    For the ionisation amplitude shown in \fig{fig:ATI-integrand-contours} this implies that if the main goal was (only) the efficient evaluation of the full integral we could terminate the flow procedure at any intermediate flow step $i$ and integrate along the obtained contour. 
    For the converged contour (panel (d)) the contribution of each saddle point can be assumed Gaussian, such that the total value of the integral can be approximated by a sum of Gaussians around the two relevant $t_s$ as shown in \eq{eq:ATI-sum}.

    With the reliable methods of evaluating an integral of type (\ref{eq:PL-integral}) by using either the downwards flow of the integration contour or, alternatively, determining relevant saddle points via the upwards flow, we can study the integral upon changes of external parameters of the phase function $\phi$.
    In the case of strong-field physics, where the phase function is the semi-classical action of the electron in the continuum, those external parameters might be specifics of the driving laser field, for example a phase shift in a two-colour field configuration, or -- in the case of HHG -- the energy of the observed photon, i.e., the harmonic order.

    Upon a continuous and smooth scan over such external parameters the saddle points $\vec{z}_\sigma$ vary smoothly in the complex $\vec{z}$ space, but may abruptly change their respective intersection number $n_\sigma$ at Stokes transitions.
    Moreover, saddle points might coalesce into higher-order critical points, at which the conventional saddle-point approximation \eq{eq:SPM-short} breaks down because the second derivative $\phi''$ in the denominator vanishes.
    To resolve the resulting caustics in the total integral we can use the downwards flow as it is in itself agnostic of the nature of the critical points.
    Evaluated across a range of external parameters it gives an exact representation of the integral regardless of possible `complications' in the saddle-point landscape.

    In the following sections we will describe aspects of Picard-Lefschetz theory that are relevant to its application in attosecond physics.
    After giving a more detailed description of the inner working of the deformation of the integration contour in \sect{sec:pl-maths}, we will present two possible approaches (and their numerical methods) to simplify the conditionally convergent highly-oscillatory integral in \eq{eq:PL-integral}:
    (a) ``The downwards flow'': Transforming the integration contour 
    according to \eq{eq:downwards-flow} 
    and evaluating the integral along this new contour by numerical quadrature. 
    And (b) ``The necklace algorithm'':
    Determining relevant critical points of $\phi(\vec{x})$ 
    by checking if there exists a steepest-ascent connection to the original integration domain
    and then evaluating their integral on the thimble, or by their Gaussian approximation.
    Equipped with those robust and novel techniques to evaluate whole families of integrals, in \sect{sec:pl-caustics} we will address the appearance of caustics in the space of external parameters.


  \subsection{On the deformation of the integration contour} \label{sec:pl-maths}
    
\begin{figure}[t]
      \includegraphics[width=\linewidth]{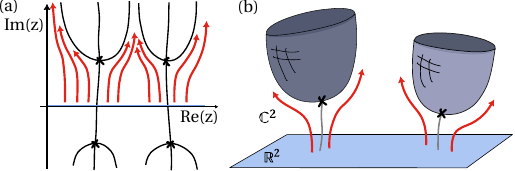}
      \caption{
      For (a) one- and (b) two-dimensional path integrals the downwards flow (directions indicated by red arrows) transform the original, real-valued integration domain (light blue) into the complex domain, ultimately towards the steepest descent contours (``thimbles'', grey) attached to the critical points (cross markers).
      }
      \label{fig:flow-sketch}
    \end{figure}

    Given that the exponent of the integrand in \eq{eq:PL-integral} is meromorphic 
    it can be written as
    \begin{equation}
      \im \phi(\vec{z}) / \hbar = h(\vec{z}) + \im H(\vec{z})
    \end{equation}
    where $h$ controls the amplitude of the integrand as 
    $|\e^{\im \phi}| = \e^h$, while $H$ controls the oscillations.%
    \footnote{We use the notation $h = \real(\phi)$ and $H=\imag(\phi)$ which is standard in the context of Picard-Lefschetz theory.}
    That is, to localise the integrand we are seeking a contour along which the value of $h$ decreases most rapidly and $H$ is constant.
    As we analytically continue $\im \phi(\vec{z})$ into the complex plane, it fulfils the Cauchy-Riemann equations 
    \begin{equation}
      \frac{\partial h}{\partial \real(\vec{z})} 
      = \frac{\partial H}{\partial \imag(\vec{z})} 
      \quad \mathrm{and} \quad 
      \frac{\partial h}{\partial \imag(\vec{z})} = - \frac{\partial H}{\partial \real(\vec{z})} \,.
      \label{eq:cauchy-riemann}
    \end{equation}
    Hence, contours of constant phase $H$ are those along which $\e^h$ vanishes (or increases) most rapidly, i.e., contours of steepest descent (ascent). 
    We can therefore find an optimal integration contour by deforming the integration path into the direction of decreasing $h$, using the downwards flow \eq{eq:downwards-flow} shown above%
    \footnote{Alternative names in other research areas are gradient, Morse or holomorphic flow .}
     and sketched in \fig{fig:flow-sketch}.
    Along the flow the value of $H$ remains constant as
    \begin{align}
      \frac{\partial \im \phi}{\partial \lambda} 
      = \frac{\partial \im \phi}{\partial \vec{z}} \frac{\partial \vec{z}}{\partial \lambda}
      = \frac{\partial \im \phi}{\partial \vec{z}} \left( - \frac{\partial \im \phi}{\partial \vec{z}} \right)^*
      = - \left| \frac{\partial \im \phi}{\partial \vec{z}}\right|^2
      \label{eq:constant-H}
    \end{align}
    which means that indeed $\imag \left(  \frac{\partial \im \phi}{\partial \lambda}  \right) =  \frac{\partial H}{\partial \lambda} =0$, while $h$ decreases most rapidly: 
    $\real \left(  \frac{\partial \im \phi}{\partial \lambda}  \right) =  \frac{\partial h}{\partial \lambda} < 0$.
    Flowing the entire original integration domain $\mathcal{C}_0$ into the complex plane converges the contour to a set of several disconnected Lefschetz thimbles \eq{eq:thimbles}.
    Each thimble is a $N$-dimensional manifold embedded in $\mathbb{C}^N$ (i.e., $2N$ real dimensions) and attached to a critical points $\vec{z}_\sigma$, as mentioned above and visualised in \fig{fig:flow-sketch} for $N=1$ and $N=2$ respectively.
    As $H(\vec{z}_\sigma)$ is constant along each of the thimbles $\mathcal{T}_\sigma \subset \mathbb{C}^N$, in the expression for the total integral \eq{eq:PL-integrals} it acts as a weighing factor for each contribution while the integration only needs to be carried out across $\e^{h(\vec{z})}$:
    \begin{equation}
      I = \sum_\sigma n_\sigma \, \e^{\im H(\vec{z}_\sigma)}  \int\displaylimits_{\mathclap{\mathcal{T}_\sigma}} \e^{h(\vec{z})} \d \vec{z}.
    \end{equation}

\begin{figure}[t]
      \includegraphics[width=\linewidth]{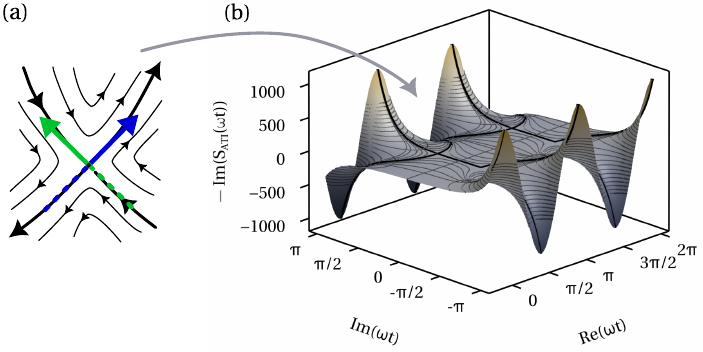}
      \caption{
      For an analytical function zeros of the first derivative constitute saddle points $\omega t_s$ in the complex plane, visible in the contour for $-\imag(\Sati)$ shown in (b).
      Around the saddle point, level lines of $\imag(\Sati(\omega t)) = \imag(\Sati(\omega t_s))$ (black) are locally orthogonal, as shown in (a), and define directions of steepest descent (blue) and steepest ascent (green) of $\real(\Sati(\omega t))$.
      }
      \label{fig:saddle-points}
    \end{figure}

      In analogy to the downwards flow, the \emph{upwards flow} is given by 
      \begin{equation}
        \frac{\mathrm{d} z_{\lambda,i}}{\mathrm{d} \lambda} =
        + \left(\frac{\partial \im \phi}{\partial z_{\lambda,i}} \right)^*
         \label{eq:upwards-flow}
      \end{equation}
      and shows the direction of steepest \emph{ascent} of $h$, while preserving $H$.
      The steepest ascent manifold of a saddle point $\vec{z}_\sigma$ is known as the
      \emph{dual} thimble $\mathcal{K}_\sigma \in \mathbb{C}^N$%
      \footnote{Also referred to as the ``anti'' or the ``unstable'' thimble \cite{bharathkumar2020lefschetz, alexandru2016monte}.}.
      The manifolds $\mathcal{T}_\sigma$ and $\mathcal{K}_\sigma$ intersect (and are locally orthogonal to each other) only in exactly one point: the critical point $\vec{z}_\sigma$, as it a stationary solution to both upwards and downwards flow.
      This is visualised in \fig{fig:saddle-points}(a) where steepest-descent (blue) and steepest-ascent (green) contours are locally orthogonal lines, intersecting at the saddle point.
      In the traditional treatment of saddle-point methods, the dual thimbles are often ignored.
      However, following Picard-Lefschetz theory, the dual thimble $\mathcal{K}_\sigma$ actually governs the relevance (meaning the contribution to the integral \eq{eq:PL-integrals}) of the thimble $\mathcal{T}_\sigma$ through its intersections with the original integration domain $\mathcal{C}_0 \in \mathbb{R}^N$.
      Those intersections are counted by the intersection number
      \begin{equation}
        n_\sigma = \langle {\mathcal{K}_\sigma}, \mathcal{C}_0 \rangle, \quad n_\sigma \in \mathbb{Z} \,,
      \end{equation}
      where the intersection operator $\langle \cdot, \cdot \rangle$ is rigorously defined in relative homology. 
      If the dual thimble $\mathcal{K}_\sigma$ attached to a critical point $\vec{z}_\sigma$ intersects the original integration domain ($n_\sigma = +1$), its thimble $\mathcal{T}_\sigma$ is part of the converged integration contour $\mathcal{T}$.
      If they don't intersect ($n_\sigma = 0$), the respective thimble has to be neglected.


  \subsection{Numerical methods for one- and two-dimensional integrals} \label{sec:pl-num-methods}

    The fact that the integration of a highly-oscillatory, conditionally convergent integral can be localised into contributions from a set of thimbles allows for a variety of computational approaches, e.g. finding the thimbles by means of Monte-Carlo sampling when $N$ is large \cite{cristoforetti2013monte,alexandru2016monte,nishimura2017combining}. 
    In the following we describe the two methods that are well-suited for the one- and two-dimensional integrals to describe strong-field ionisation and HHG, respectively. 
    First, the ``downwards flow'', which yields a discretised contour along which the integration can be carried out more efficiently.
    And secondly, the ``necklace algorithm'', which we developed to determine the intersection number $n_\sigma$ of a given saddle point, irrespective of any `classification' of saddle points (as one would usually do in attosecond science).

  \subsubsection{The downwards flow method}
    \label{sec:downwards-flow}



    The goal of the downwards flow method is to deform the integration contour according to the downwards flow \eq{eq:downwards-flow}.
    The numerical algorithm (and its description) are based on J.F.'s open-source implementation in C++, available at \cite{feldbruggepicardlefschetz}.
  
\begin{figure}[t]
      \includegraphics[width=\linewidth]{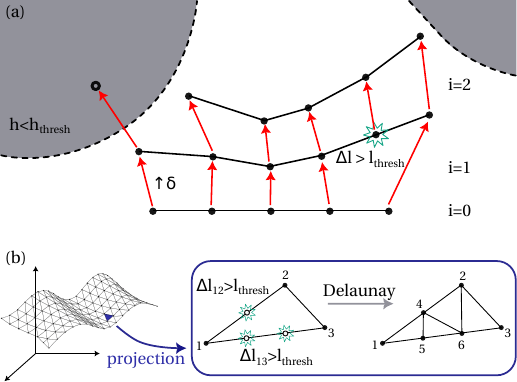}
      \caption{Sketch of the implementation schemes of the downwards flow procedure.
        (a) For a one-dimensional integration, the contour is discretised as a set of connected points that are iteratively (iteration step $i$) flowed, possibly subdivided (newly inserted point on the right-hand side) and discarded (empty circle on the left-hand side).
        (b) The two-dimensional integration follows the same procedure, but points are connected to triangles to form a surface. 
        For the subdivision each triangle is considered in its projected plane, vertices are subdivided and the set of old and newly inserted points are connected to triangles by Delaunay triangulation.
      }
      \label{fig:flow-algorithm-sketches}
    \end{figure}

    We will explain the algorithm for the case of a one-dimensional integral first, which is shown in \fig{fig:flow-algorithm-sketches}(a). 
    The initial step is the discretisation of the (original) integration domain.
    We discretise the real axis as a list of points, which are connected to line segments.
    Then, we iteratively apply the downwards flow to each of the points, moving it into the complex plane using a first-order Euler method
    \begin{equation}
      z \mapsto z - \delta_\mathrm{flow} \left( \frac{\partial h}{\partial z} \right)^*
      \label{eq:flow-step}
    \end{equation}
    with the small parameter $\delta_\mathrm{flow}$.
    Note that it is sufficient to consider the gradient of $h$ (rather than $\im \phi$) as $H$ remains constant along the flow anyway (see \eq{eq:constant-H}).
    We use an adaptive grid in the sense that as soon as two neighbouring points are further than a threshold distance $l_\mathrm{thresh}$ apart, we insert a new point in the middle (see on the right-hand side in \fig{fig:flow-algorithm-sketches}(a)).
    Furthermore, points are turned `inactive' (i.e., they are not moved any more) as soon as their $h$ value drops below a certain threshold, say $h_\mathrm{thresh}$, indicated as grey regions and empty points in \fig{fig:flow-algorithm-sketches}(a).
    This will eventually break up the integration contour (which was just the real line) into disconnected parts, as e.g.\, in \fig{fig:ATI-integrand-contours}(d).
    The deformation of the integration contour converges to the Lefschetz thimble as $\frac{\partial h}{\partial z}$ vanishes on the thimble.
    To avoid `overshooting' this zero-gradient contour of steepest descent, we normalise the gradient as soon as its magnitude drops below a certain threshold.
    The algorithm is terminated after a fixed number of flow steps $i=i_\mathrm{max}$, when the number of active points remains constant and the shape of the contour does not change for subsequent iteration steps.

    In the case of a two-dimensional integral the algorithm technically follows the same procedure. 
    However, each point now has two coordinates (each of them being a complex number!) and the integration `contour' is a surface, embedded in four real dimensions.
    That is, rather than using line segments we have discretised our integration domain into triangles now.
    For the subdivision we use the routine sketched in \fig{fig:flow-algorithm-sketches}(b):
    Each triangle is considered in its plane. 
    For edges exceeding $l_\mathrm{thresh}$, we insert $\left\lfloor \tfrac{\Delta l}{l_\mathrm{thresh}} \right\rfloor $ new points and then meshing the original triangle using a Delaunay triangulation%
    \footnote{Coincidentally, the eponymous Boris Delaunay is the father of Nikolai Delone that gave the ``D'' in the Ammosov–Delone–Krainov (ADK) ionization rates in strong-field physics.} of all points.

    Ultimately, for the evaluation of the integral we use a numerical quadrature of the obtained meshed surface \cite{bartholomew1959numerical}.

  \subsubsection{The necklace algorithm}
    \label{sec:necklace}


    Whether a given thimble $\mathcal{T}_\sigma$ contributes to \eq{eq:PL-integrals} or not is dictated by the intersection number $n_\sigma$ which counts the intersections between the dual thimble $\mathcal{K}_\sigma$ (i.e., the steepest-ascent manifold attached to a critical point) and the original integration domain $\mathcal{C}_0$.
    As the value of $H$ is constant along the steepest-ascent manifold, for a one-dimensional integral ($N=1$) finding the thimble and dual thimble attached to the critical point $z_\sigma \in \mathbb{C}$ corresponds to finding the respective contour level lines where $H(z) = H(z_\sigma)$, drawn as heavy black lines in \fig{fig:saddle-points}.
    This can easily be done numerically with e.g., a marching squares algorithm. 
    The arrows on the level lines in the inset panel in \fig{fig:saddle-points} indicate the direction of descending $h$.
    At the saddle point, these directions of maximised gradient are given by the (orthogonal) eigenvectors of the Hessian, one pointing in direction of steepest ascent (green), and one points in direction of steepest descent (blue).
    To find the dual thimble we therefore simply pick the contour level lines along which $h$ increases away from the saddle point in the steepest-ascent direction. 
    Locally, these lines coincide with the one of the eigenvectors and its inverse (dotted green vectors) of the Hessian.
    That is, if $h$ is ascending along a level line away from $z_\sigma$ and eventually connects to the real axis (the original integral domain), then $n_\sigma =1$ and the critical point contributes to the integral.
    For example, in \fig{fig:saddle-points} for the saddle points with a positive imaginary part the steepest-ascent lines intersect the real axis, whereas for those saddle points with a negative imaginary part they don't.
    This can also be concluded immediately from the fact that for saddles points with $\imag(z_\sigma) < 0$ we have $h(z_\sigma) > 0$, such that there is no way `uphill' from $z_\sigma$ to the real axis where $h=\real(\im \phi(\real(z))) = 0$. 

    For the case of a two-dimensional integral, finding the dual thimbles is more complicated.
    As mentioned above, they are now two-dimensional manifolds (i.e., surfaces) embedded in the four-dimensional space ($\real(z_1), \imag(z_1), \real(z_2), \imag(z_2)$).
    That means, tracing the contour levels $H(\vec{z}) = H(\vec{z}_\sigma)$ for a critical point $\vec{z}_\sigma$ will now yield contour level \emph{surfaces} embedded in 4D which is computationally more advanced.

    \begin{figure}[t]
      \includegraphics[width=\linewidth]{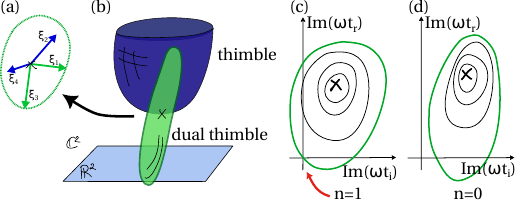}
      \caption{The necklace algorithm: 
      Thimble and dual thimble are surfaces embedded in 4D (illustrative 3D projection in panel (b)), locally spanned by the eigenvectors $\xi_\alpha$ of the Hessian at the saddle point (panel (a)).
      The brim of the dual thimble -- the necklace -- is initialised as a circle using the steepest-ascent eigenvectors (green), and then flowed upwards.
      (c) and (d) show projections of example necklaces in the $(\imag(\omega \ti), \imag(\omega \tr))$, where an intersection with $(0,0)$ implies an intersection with the real plane.}
      \label{fig:necklace}
    \end{figure}

    Here we present a novel technique -- which we call the `necklace algorithm' -- to determine the intersection number of a given saddle point for a two-dimensional integral, 
    a problem which has so far remained open,%
    \footnote{A similar approach has been used in \cite{han2021spinfoam,alexandru2022complex} to approximate the (steepest-descent) thimbles.}
    and for which tentative general solutions were only proposed very recently~\cite{shoji2025stable}.
    The basic idea, shown in \fig{fig:necklace}, is to initialise the `tip' of the dual thimble in the closest vicinity of the saddle point, and then use the upwards flow for its further construction `slice by slice'.
    We terminate the upwards flow as soon as each point reaches $h=0$, and then check for the intersection with the original integration domain.
    As we are only interested in this (potential) intersection, it is sufficient to consider the `brim' of the thimble, which~--~ as it is a discretised closed loop~--~we dub the necklace.
    This approach is guaranteed to identify \emph{all} possible intersections of the steepest-ascent manifold with the original integration domain.

    Let us briefly explain the procedure on the one-dimensional example.
    In the vicinity of the saddle point the direction of steepest descent and ascent can be found by linearising the flow.
    That is, we calculate the second derivatives w.r.t.\ to both real and imaginary part of the integration variable $z$ and identify the $2 \times 2$ real-valued Hessian matrix, the
    eigenvectors of which pointing in the direction of maximised gradient.
    The eigenvector corresponding to the negative eigenvalue points in direction of decreasing $h$ away from the saddle point, drawn as a blue vector in \fig{fig:saddle-points}, whereas the eigenvector corresponding to the positive eigenvalue increases $h$ away from the saddle point (drawn in green). 
    To obtain the full steepest-ascent manifold (a line), we flow the end point of the steepest-ascent eigenvector and its inverse (drawn as a dashed vector in the opposite direction) until it eventually reaches $h=0$.%
    \footnote{For the one-dimensional integral, of course, this was not necessary as we simply pick the respective contour line.}
    If either of the end points of the line hits the real axis, the intersection number $n_\sigma$ counts $+1$.

    Now, for the two-dimensional integral we follow the same procedure.
    We assume that the upwards flow \eq{eq:upwards-flow} in a small region around the critical point $\vec{z}_\sigma$ is linear in $h$ w.r.t.\ each of the four real-valued dimensions.
    That is, we use $\vec{z} = (\real(z_1),\imag(z_1), \real(z_2), \imag(z_2))$ and write
    \begin{align}
      \frac{d z_\alpha}{d \lambda} &= \left( \frac{\partial^2 h}{\partial z_\alpha \partial z_\beta} \right)^* \bigg|_{\vec{z} = \vec{z}_\sigma} ({z_\beta}^* - {z_{\sigma,\beta}}^*) 
      \nonumber \\ 
      &= {\mathcal{H}_{\alpha \beta}} ({z_\beta}^* - {z_{\sigma,\beta}}^*)
      \label{eq:linearised-flow}
    \end{align}
    where $\alpha, \beta = 1 \dots 4$, such that $\mathcal{H}$ is the real-valued, symmetric $4 \times 4$ Hessian of $h$.
    The solutions to the respective eigensystem
    \begin{equation}
      \mathcal{H} \tilde{\vec{v}} = \xi \tilde{\vec{v}}
      \label{eq:eigensystem}
    \end{equation}
    yields four eigenvalues, coming in pairs, where
    $ \xi_2 = - \xi_1$ and $\xi_4 = -\xi_3$.
    Analogously, for the corresponding eigenvectors we find 
    $\tilde{\vec{v}}_2 =  -\im \tilde{\vec{v}_1}$ and $\tilde{\vec{v}}_4 =  -\im \tilde{\vec{v}_3}$.
    Their linear combination solves \eq{eq:linearised-flow} and hence, defines the directions of constant $H$ around the saddle point $\vec{z}_\sigma$.
    The two vectors $\tilde{\vec{v}}$ with the smaller eigenvalues point towards the steepest descent of $h$, and the two vectors with larger eigenvalues point in the direction of steepest ascent of $h$, and they are drawn as blue and green vectors in \fig{fig:necklace}(a) respectively.
    By re-writing the eigenvectors into complex form a la
    $\vec{v}_\alpha = (\tilde{v}_{\alpha,1} + \im \tilde{v}_{\alpha,2}, \tilde{v}_{\alpha,3} + \im \tilde{v}_{\alpha,4})$, 
    and assuming $\xi_1$ and $\xi_3$ to be the two positive eigenvalues, we can therefore define the directions of steepest ascent,
     \begin{equation}
       \vec{z}(\lambda) - \vec{z}_\sigma = a_1 \vec{v}_1 \e^{\xi_1 \lambda} + a_3 \vec{v}_3 \e^{\xi_3 \lambda}\,. 
     \end{equation}
     with arbitrary coefficients $a_1$ and $a_3$.
     We initialise the dual thimble's brim by constructing a vanishing cycle (viz.\ a ``loop'' of constant $H$) around the saddle point.
      For that we use the two steepest-ascent vectors and draw the circle
      \begin{equation}
         \vec{z}(\gamma, \lambda = 0) = \vec{z}_\sigma + \epsilon \left(\cos\gamma \,  \vec{v_1} + \sin\gamma \, \vec{v_3} \right)
      \end{equation} 
      for $\gamma \in [0,2 \pi)$ and a small value $\epsilon$, as shown in \fig{fig:necklace}(a).


    Once the necklace has been initialised, we discretise it and apply the upwards flow \eq{eq:upwards-flow} to each resulting bead of this necklace, making use of the procedure described for the downwards flow in one dimension, and sketched in \fig{fig:flow-algorithm-sketches}(a).
    With that we construct the dual thimble `slice by slice' (or rather `ring by ring') until all beads reach $h=0$.

    Finally, it remains indeed to check whether the found brim of the dual thimble intersects the original integration domain $\mathcal{C}_0$.
    For that we look at the $\imag(\vec{z})$ projection of the necklace and check if the line crosses the origin $\imag(\vec{z})=(0,0)$.
    This is shown schematically in \fig{fig:necklace}(c) and (d), where in panel (c) we find the necklace intersects the real axis once, making $n_\sigma =1$, whereas in panel (d) there is no intersection.

    For any given individual saddle point, the necklace algorithm allows us to determine whether its thimble is a relevant contributor to the integral \eq{eq:PL-integrals}.
    As the necklace algorithm essentially traces contours of constant $H$ value, limitations naturally arise in cases where two neighbouring critical points have similar $H$ values.
    The upwards flow then might accidentally `slip' into (parts of) the dual thimble brim of the other critical point.
    A decision about the relevance of the individual critical points can then be made by identifying the parameters for the Stokes transitions, as will be laid out in the next section.

    Once we determined which critical points constitute relevant thimbles, finding their contribution to the integral is straightforward.
    Either we apply the standard saddle-point method and approximate the integral across the thimble to be of Gaussian shape as shown in \eq{eq:SPM-short}.
    Or, if we want an exact representation of the integral, we find the thimble attached to each critical point.
    For that, we initialise a small vanishing cycle around the critical point in the directions of steepest descent (i.e., using the eigenvectors with smaller eigenvalues from \eq{eq:eigensystem}) and then construct the thimble by applying the \emph{downwards} flow to it `slice by slice', analogously to flowing the necklace to obtain the dual thimble.
    The value of the integral can finally be found by evaluating the integrand along the thimble using a standard quadrature routine.




  \subsection{Evaluating integrals across ranges of external parameters -- Stokes transitions, caustics and catastrophes}
    \label{sec:pl-caustics}
\begin{figure}[t]
      \includegraphics[width=\linewidth]{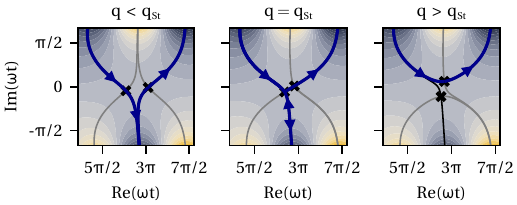}
      \caption{Topological change of steepest-descent contours (black) and the resulting integration contour (heavy dark blue line) around a Stokes transition (at $q=q_\mathrm{St}$) between two saddle points upon changing an external parameter $q$ (left to right panel). 
      }
      \label{fig:stokes}
    \end{figure}

    Equipped with tools to solve integrals like \eq{eq:PL-integral} with arbitrary phase functions $\phi$, we can study how they depend on external parameters.
    Upon a continuous and smooth scan over such external parameters, the saddle points $\vec{z}_\sigma$ vary smoothly in the complex $\vec{z}$ space.
    Their intersection number $n_\sigma$, however, may change abruptly at so-called \emph{Stokes transitions}, causing the total number of contributing saddle points (or rather, thimbles) to change~\cite{lando1997geometry, figueirademorissonfaria2002highorder, milosevic2002role, pisanty2020imaginary, Chipperfield2005conditions}.
    For this to happen there must be a topological change in the course of the steepest-descent integration contour.
    This is often caused by two (or more) critical points in close proximity, as shown in \fig{fig:stokes} for an integration contour depending on the external parameter $q$.
    In the left-hand panel where $q<q_\mathrm{St}$ the two critical points are both part of the converged deformed integration contour $\mathcal{T}$ (heavy blue line) and contribute separately to the integral via their thimbles $\mathcal{T}_{\sigma1}$ and $\mathcal{T}_{\sigma2}$.
    In the centre panel, at $q=q_\mathrm{St}$, their steepest-descent contours coincide and the deformed contour contains both critical points -- this value of the external parameter marks the Stokes transition.
    For a further increase of the external parameter where $q>q_\mathrm{St}$, the steepest-descent contours separate and $\mathcal{T}$ only contains one of the saddle points.

    A necessary condition for the Stokes transition between two critical points $\vec{z}_{\sigma1}$ and $\vec{z}_{\sigma2}$ is that $H(\vec{z}_{\sigma1}) = H(\vec{z}_{\sigma2})$, such that their steepest-descent contours may connect directly to each other.
    In simple examples, e.g., when $\phi(\vec{z})$ is a polynomial with two external parameters, Stokes transitions can be analytically solved for and yield lines in parameter space \cite{wright1980stokes,berry1990stokes, feldbrugge2023oscillatory}.
    For more complicated $\phi(\vec{z})$, where there is no closed form expression for $\vec{z}_\sigma$, candidate Stokes transition can be found numerically by identifying where in parameter space pairs of critical points assume the same value of $H$.
    Generally, for a phase function with $K$ external `control' parameters the Stokes transitions are $(K-1)$-dimensional manifolds in the $K$ dimensional parameter space.
    They are topological features of this parameter space, as any change of number of relevant saddle points is indubitably linked to a Stokes transition.
    Vice versa, Stokes transitions define regions in parameter space with a certain number of contributors to the integral.
    Consequently, if they can be calculated a priori it is unnecessary to calculate each critical point's relevance individually.

    More generally, as we evaluate the total integral across ranges of external parameters we find Stokes transitions and caustics. 
    The latter are the pronounced features that arise whenever multiple saddle points are in close proximity and ultimately coalesce. 
    Caustics can be observed in everyday life, e.g., the rippled bright features at the bottom of a swimming pool or the cusp-like structure in a coffee cup, as well as in more involved physical problems like real-time path integrals, lensing (both optical and gravitational) \cite{feldbrugge2023multiplane,feldbrugge2017lorentzian}, the formation of large-scale structures of our universe \cite{feldbrugge2018caustic}, and, of course, attosecond science \cite{raz2012spectral,birulia2019spectral,facciala2016probe,raab2025xuv}.
    From a mathematical point of view, caustics are best analysed in terms of catastrophe theory.
    Assuming the phase function $\phi(\vec{z})$ is a $(K+2)$-order polynomial with $K$ external `control' parameters dictates that there can be at most $K+1$ saddle points coalescing to one higher-order critical point, the catastrophe point.
    Evaluating the integral over a range of external parameters that involves such a coalescence of saddle points will yield characteristic patterns, depending on the codimension $K$, known as canonical diffraction patterns ~[\citealp{poston1978catastrophe, saunders1980introduction};\,\citealp{NIST_handbook}, \href{http://dlmf.nist.gov/#1}{\S36.3}].
    Correspondingly, the contribution of a critical point in the vicinity of a coalescence cannot be taken as Gaussian, but has to be modelled using a uniform approximation that involves the canonical diffraction integrals, which is a challenging task \cite{chester1957extension,berry1989uniform,schomerus1997bifurcations,stamnes1983evaluation}.
    For example, the diffraction pattern for the fold catastrophe ($K=1$), for which two saddle saddle points coalesce, is the Airy function. 
    Hence the Airy function is used to model the HHG spectrum (which can be interpreted as a scan over the control parameter harmonic order) around the cutoff at which the saddle point solutions for short and long trajectories (almost) coalesce \cite{figueirademorissonfaria2002highorder, pisanty2020imaginary}.

    The downwards flow procedure elegantly circumvents these problems as it is agnostic to the critical points. 
    With that, it offers the unique capability to evaluate the integral \eq{eq:PL-integral} exactly across parameters ranges and to naturally resolve the appearing caustic structures.

    Having laid out the ideas and resulting numerical methods of Picard-Lefschetz theory, we now turn back to attosecond science and strong-field physics.
    In the following chapter we apply the methods introduced above to the SFA integral for HHG, namely \eq{eq:hhg-integral}, and show how it allows us a quantum-orbit based evaluation for scenarios with arbitrary driving waveforms.


\section{HHG driven by two-colour laser fields} \label{sec:hhg_driven_by_two_colour_laser_fields}

  The harmonic response of an atom subjected to a strong laser field can (within the SFA framework \cite{lewenstein1994theory,corkum1993plasma}) be calculated in terms of the two-dimensional integral \eq{eq:hhg-integral} over ionisation and recombination times of the involved electronic wave packet, $\ti$ and $\tr$, respectively.
  This double integral is often rewritten in terms of contributions of separate quantum orbits \cite{salieres2001feynmans,milosevic2006quantumorbit}, i.e. semi-classical electron paths defined by a discrete ionisation and recombination time, which allows for an intuitive understanding of the process as the associated trajectories have different properties in e.g. the spatial divergences in the far-field \cite{he2010interference,chipperfield2006tracking}.
  The quantum orbits are pairs $(\ti, \tr)$ for which the semi-classical action $\Shhg(\ti, \tr)$ is stationary, i.e., saddle points in the complex plane defined by \eq{eq:speqs}.
  Notably, there are typically far more solutions to \eq{eq:speqs} than relevant quantum orbits to the process.
  So far, the existing heuristics to decide whether a given saddle point solution is a relevant quantum orbit rely on a classification of the solutions and dynamic symmetries of the driving field.

  For generic driving fields, however, those heuristics fail.
  In the following we demonstrate how the methods of Picard-Lefschetz theory described in \sect{sec:pl-num-methods} can be utilised to compute the harmonic dipole integral \eq{eq:hhg-integral}.
  The central insight is that the integral can be evaluated along a different contour $\mathcal{C}$ in the complex time planes and ultimately expressed as a sum over contributions from separate \mbox{thimbles}~$\mathcal{T}_s$:
  \begin{align}
    \vec{D}(q \omega) &= \int_{\mathcal{C}_0} \dots \d \vec{t} = \int_{\mathcal{C} \in \mathbb{C}^2} \dots \d \vec{t} \nonumber \\ &=
    \im \sum_s n_s
    \int\limits_{\mathclap{\mathcal{T}_s \in \mathbb{C}^2}} \d \vec{t} \, 
    \vec{d}\left( \vec{p}_s(\ti, \tr) + \vec{A}(\tr)\right) \nonumber \\
    & \qquad \vec{\Upsilon}\left( \vec{p}_s(\ti, \tr) + \vec{A}(\ti)\right) \nonumber \\ 
    & \qquad \left( \frac{2 \pi}{\im (\tr - \ti)} \right)^{3/2}
    \e^{-\im \Shhg(\ti, \tr)}
     \,. \label{eq:hhg-PL-sum}
  \end{align}
  Here, we notate $\vec{t} = (\ti,\tr)$ and the original integration domain is \mbox{$\mathcal{C}_0 = \{ (\ti, \tr) \in \mathbb{R}^2 \mid \tr > \ti \}$}.
  Each thimble $\mathcal{T}_s$ is attached to a critical point $(\tis, \trs)$ defined by \eq{eq:speqs}, and only contributes for non-vanishing intersection numbers $n_s$. 
  The key difference to and striking advantage over \eq{eq:hhg-sum-spm} is, that here we still have an equality, as we haven't made any assumption on the shape of the integrand around the critical points.

  Note that the presented integration methods ``only'' address the two-dimensional temporal integration and therefore hold for any definition of prefactors, waveforms etc.
  With that, the exponentiated phase factor is $-\im \Shhg$, which we consider in its dependency on ionisation and recombination times only, $\Shhg = \Shhg(\ti,\tr)$ as in \eq{eq:Shhg}. 


  As an example, within this manuscript we choose to consider HHG driven by a collinear and co-polarised two-colour field that consists of a fundamental laser field with frequency $\omega$, superimposed with its second harmonic (frequency $2 \omega$).
  A generic expression for the electric field then reads
  \begin{equation}
    \vec{E}(t) = E_1 \cos(\omega t) + E_2 \cos(2 \omega t + \varphi) 
    \label{eq:tc-field}
  \end{equation}
  with the field amplitudes $E_1$ and $E_2$ and the phase delay $\varphi$ between the two field components.
  For pulses longer than a few cycles, it is a good approximation to restrict our considerations to one cycle of the fundamental frequency, the period $T = 2 \pi/\omega$.
  These types of driving fields are ubiquitous in attosecond science, in both experiment and theory.
  They allow to probe the inner workings of the process of strong-field light-matter interaction itself, as well as to tailor the properties of the harmonic spectrum and/or the created attosecond pulse 
  \cite{shafir2012resolving, pedatzur2015attosecond, kneller2022look, raab2025xuv,mitra2020suppression,mauritsson2009subcycle,mansten2008spectral}.

  The following results are obtained using
  $\Ip = \SI{15.8}{eV}$ and 
  $I_0 = E_0^2 = 0.92\times 10^{14} \SI{}{W/cm^2}$ 
  ($ E_0 = \SI{0.05}{\au})$, 
  $\lambda = \SI{1030}{nm}$ 
  ($\omega = \SI{0.044}{\au}$), and we use atomic units ($\SI{}{\au}$) unless stated otherwise.
    \begin{figure}[t]
      \includegraphics[width=0.75\linewidth]{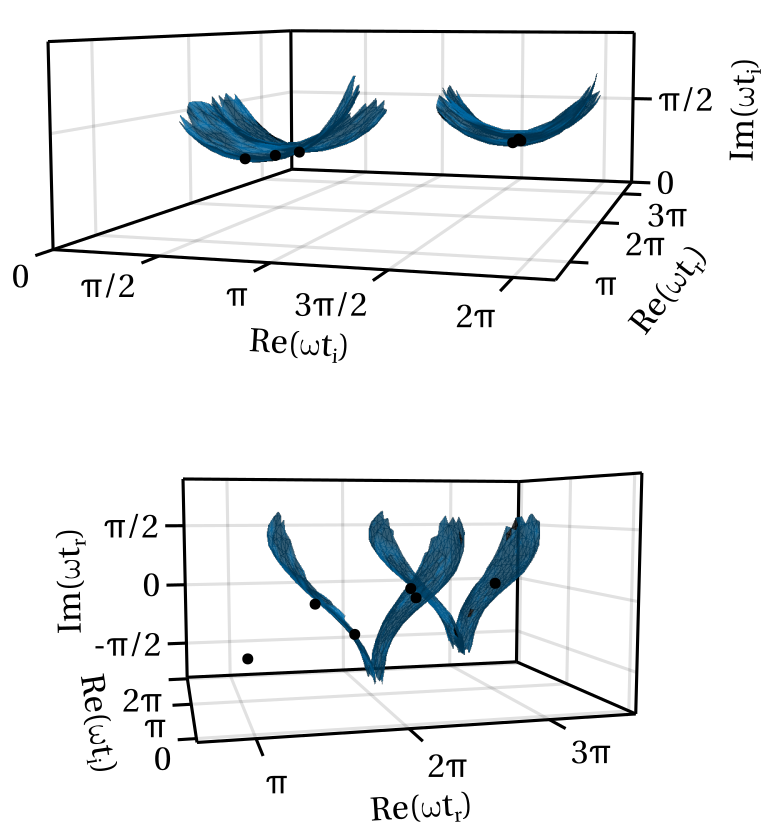}
      \caption{Two projections of the HHG thimble for a monochromatic driving field and 
      $q=25$: (a) Ionisation-time projection $(\real(\ti), \imag(\ti), \real(\tr))$ and recombination-time projection $(\real(\tr), \imag(\tr), \real(\ti))$.
      Saddle points are marked as dots.
    }
      \label{fig:HHG-thimble}
    \end{figure}

\subsection{Applying Picard-Lefschetz methods}

  \subsubsection{Using the downwards flow to deform the integration contour towards Lefschetz thimbles }

    \begin{figure*}[ht]
      \includegraphics[width=\textwidth]{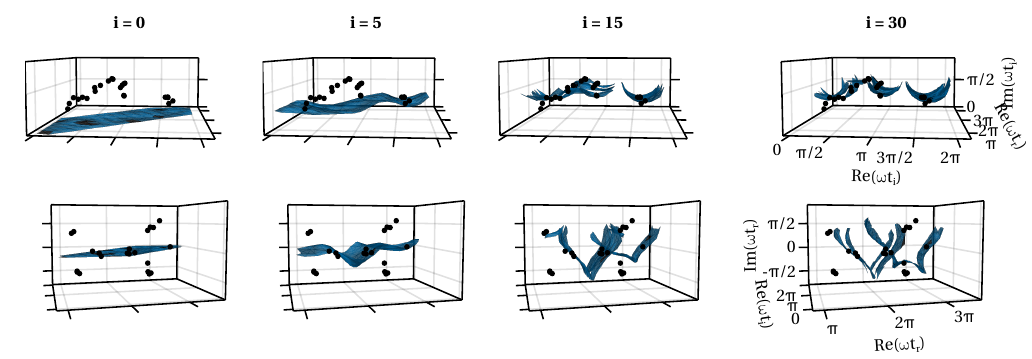}
      \caption{Several steps of the deformation of the integration domain towards the Lefschetz thimble, for HHG driven by the two-colour field shown in \fig{fig:ATI-landscapes}(b), harmonic order $q=25$.}
      \label{fig:HHG-thimbles-flow}
    \end{figure*}

    In the previous chapter we showed that there exists a continuous deformation of the original integration domain into a contour that minimises the integrand oscillations, which then allows for a more efficient numerical evaluation of the integrand along that new contour. 
    We demonstrate how this downward flow method is applied to the integration contour of the HHG integral \eq{eq:hhg-integral}.
    
    We restrict ionisation to one cycle, so $0 \leq \real(\ti) < T$, and recombination to happen after that, with travel times limited to one cycle as contributions from orbits with longer travel times are lower. 
    This original integration domain is then deformed into the complex domain for both $\ti$ and $\tr$, i.e., into the four-dimensional space ($\real(\ti)$, $\imag(\ti)$, $\real(\tr)$, $\imag(\tr)$). 
    As described in \sect{sec:pl-num-methods}, the deformation of the domain follows a simple first-order scheme for either variable, 
    \begin{equation}
      \ti \mapsto \ti + \delta_\mathrm{flow} \, \frac{\mathrm{d} \ti}{\mathrm{d} \lambda} 
      \quad \text{and} \quad
      \tr \mapsto \tr + \delta_\mathrm{flow} \, \frac{\mathrm{d} \tr}{\mathrm{d} \lambda}\,,
      \label{eq:flow-steps-HHG}
    \end{equation}
    with a small factor $\delta$.
    The direction for every flow step $\lambda$ is dictated by the downwards flow \eq{eq:downwards-flow}, recast for the case of HHG:%
    \footnote{We have dropped the $\lambda$ indices for better readability.}
     \begin{equation}
      \frac{\mathrm{d} \ti}{\mathrm{d} \lambda} = - \left(\frac{\partial \Shhg}{\partial \ti} \right)^*
       \quad \text{and} \quad
       \frac{\mathrm{d} \tr}{\mathrm{d} \lambda} = - \left(\frac{\partial \Shhg}{\partial \tr} \right)^*    
       \label{eq:downwards-flow-HHG}
    \end{equation}
    This routine 
    continuously deforms the (discretised) original integration domain into a two-dimensional steepest-descent surface embedded in 4D space and converges to the Lefschetz thimbles.

    The resulting thimbles for the simple case of a monochromatic driver 
    $E(t) = E_0 \cos(\omega t)$ (shown in \fig{fig:ATI-landscapes}(a))
    and harmonic order $q = 25$
    are shown in \fig{fig:HHG-thimble} in two projections, as well as the saddle point solutions (black markers). 
    In the ``ionisation projection'' $(\real(\ti), \imag(\ti), \real(\tr))$ (top), we observe the deformation into four disconnected surfaces corresponding to the ionisation windows around each maximum of the electric field. 
    In the ``recombination projection'' $(\real(\tr), \imag(\tr), \real(\ti))$ (bottom), we identify four separate surfaces, corresponding to the expected two pairs of ``short'' and ``long'' quantum orbits within each half cycle. 
    Each surface is the steepest descent manifold (thimble) of a relevant saddle point.
    Saddle points that are not included in the surface are irrelevant.

    For comparison, and in order to aid the understanding of the flow method, we show a range intermediate steps of the continuous deformation for a more complicated situation in \fig{fig:HHG-thimbles-flow}.
    The driving field is the two-colour field as in \fig{fig:ATI-landscapes}(b),
    and we show how the downwards flow deforms the integration domain towards the Lefschetz thimbles (increasing iteration steps from left to right)
    for harmonic order 25, using the same projections described above.
    From the ionisation-time projection (top row) we can make out separate ionisation windows, albeit not as distinct as in the monochromatic case.

    Finally, the harmonic dipole \eq{eq:hhg-integral} can be calculated with a simple quadrature along these discretised Lefschetz thimbles.
    This can be done at any intermediate step of the deformation (as the integral remains unchanged for just a change of contour, see \eq{eq:PL-integrals}).
    Note that this deformation of the integration domain is not an efficient method for the detection of all relevant saddle points.

  \subsubsection{Using the necklace algorithm to determine relevant quantum orbits}
    Saddle points of the action correspond to the quantum orbits that interfere when creating the harmonic dipole response. 
    The properties of those several electron trajectories, e.g.\ the spread of the wave packet and the recollision angle, imprint on the dipole and therewith on the properties of the emitted radiation~\cite{strelkov2012highorder, Pisanty2017, Pisanty2018, shafir2012resolving, dudovich2006measuring, Itatani2004}.
    Moreover, the contributions of the various quantum paths behave differently upon propagation, and give rise to distinct patterns in the far-field spectra measurement~\cite{zair2008quantum, Hoffmann2014, RoscamAbbing2020, Brugnera2011}.
    Phenomenologically, it is therefore interesting to understand which quantum orbits are at play for the creation of a certain dipole, i.e., to understand which saddle points are relevant contributors to the sum \eq{eq:hhg-PL-sum}.
    A given saddle point is a relevant contributor if and only if its attached steepest-ascent contour (the dual thimble) connects back to the original integration domain. 
    We find this possible intersection by propagating the brim of the dual thimble upwards until $h=0$ and then checking for an intersection of this brim with the real plane; 
    this is the ``necklace'' algorithm introduced in \sect{sec:necklace}.
    The necklace around the saddle point is initialised as a small circle in the plane of the two eigenvectors corresponding to the largest eigenvalues of the matrix
    \begin{equation}
      \mathcal{H}_{\alpha, \beta} = \frac{\partial^2}{\partial t_\alpha \partial t_\beta} \Shhg(t_1 + t_2\im, t_3 + t_4 \im)
    \end{equation}
    where $\alpha,\beta=1,\dots,4$, and where we have taken $\vec{t} = (\real(\ti), \imag(\ti), \real(\tr), \imag(\tr))$.
    Each bead of the (discretised) necklace then flows upwards in $h$, following
      \begin{equation}
      \frac{\mathrm{d} \ti}{\mathrm{d} \lambda} = +\left(\frac{\partial \Shhg}{\partial \ti} \right)^*
       \quad \text{and} \quad
       \frac{\mathrm{d} \tr}{\mathrm{d} \lambda} = +\left(\frac{\partial \Shhg}{\partial \tr} \right)^* \,
       \label{eq:upwards-flow-HHG}
    \end{equation} 
    until $h=0$ and $\Shhg$ becomes real. 
    If the converged necklace intersects our original integration domain, the given saddle point is relevant.

  \subsubsection{Comparison of the two methods: Harmonic spectra}
\begin{figure}[t]
        \includegraphics[width=\linewidth]{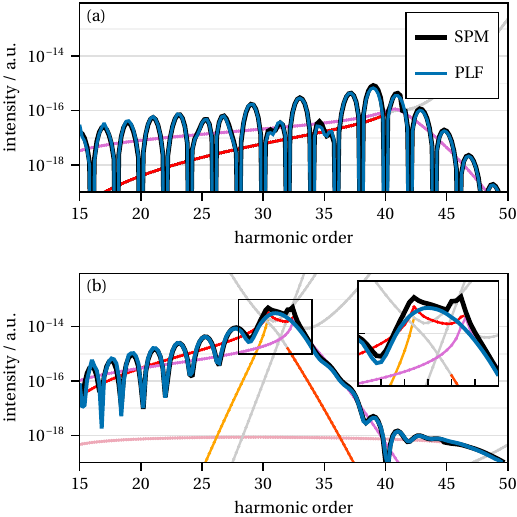}

        \caption{HHG spectrum for the two fields shown in \fig{fig:ATI-landscapes}, 
        ((a) for the monochromatic field in \fig{fig:ATI-landscapes}(a),
         (b) for the two-colour field in \fig{fig:ATI-landscapes}(d)),
          calculated as a sum of Gaussian contributions from relevant saddle points (SPM, black line), and as a quadrature of the deformed integration domain (PLF, blue line).
        Individual saddle points' contributions are shown as markers (coloured for contributing saddles, grey for non-contributing saddles).
        The bottom panel features an inset for the artificial discontinuity that arises for the SPM when saddle points are in close vicinity.}
        \label{fig:spectra}
      \end{figure}

    The resulting values of the integrals, in the form of spectral intensities \eq{eq:spectral-intensity} for a range of harmonic orders $q$, i.e., harmonic spectra, are shown in \fig{fig:spectra} for the monochromatic driving field (top panel) and the two-colour field (bottom panel) as in \fig{fig:HHG-saddles-in-cp}(a) and (b) respectively.
    We show the Gaussian contribution from each saddle point in coloured markers (for relevant saddle points; light grey for non-contributing saddles). 
    The coherent summation of relevant saddles' contribution is shown in black markers (saddle point-method, SPM, \eq{eq:hhg-sum-spm}), which is compared to the quadrature along the deformed integration contour (Picard-Lefschetz flow, PLF,~\eq{eq:hhg-PL-sum}) in blue. 
    For the monochromatic driving field (in panel (a)) we recognise the familiar structure of a typical HHG spectrum exhibiting quantum-path interference~\cite{zair2008quantum}.
    Throughout the spectrum there are two types of relevant contributions: from short and long trajectories (red and pink respectively), of which the former become non-relevant at the high-harmonic cutoff at $q_c = 42$.
    Note small deviations between the two integration approaches only occur around this Stokes transition where the two saddle points are in close vicinity and their contribution should not be modelled as Gaussian, but rather as an Airy-type integral \cite{pisanty2020imaginary,figueirademorissonfaria2002highorder,milosevic2002role}.

    For the complicated two-colour field (in panel (b)), the harmonic spectrum exhibits a more interesting structure, as we find more than only two types of trajectories.
    Throughout the first plateau (harmonic orders 15 to 32) we observe the expected interference structure from the two dominant trajectories marked in red and pink.
    However, we can identify a more interesting feature of the spectrum that we can now attribute to individual trajectories, and which is shown enlarged in the inset. 
    Around the first harmonic cutoff there are two other trajectories (yellow and orange) which contribute significantly to the integral and yield an overall spectral enhancement.
    This enhancement stems from the `cluster' of saddle points shown in \fig{fig:HHG-saddles-in-cp}(e) panel D and signifies the appearance of a caustic, which we address in more detail in the following section.
    In that region of the spectrum, the SPM exhibits both an artificial discontinuity and also the largest deviation from the PLF-based signal.
    Again, this is expected, as for saddle points in close vicinity the integrand along the steepest-descent contour does not resemble a Gaussian.
    For a proper analytical treatment within saddle-point based methods we would require a uniform approximation to smoothen out the discontinuities. 

    Within the PLF method the integrand is directly evaluated along the thimble and agnostic of any saddle (or higher order critical) points.
    As a result, the evaluated integral is naturally smooth throughout the spectrum and eliminates the need for a carefully constructed uniform approximation that connects different regimes of relevant saddle points.
    To this end, the PLF provides a unique tool that captures the exact value of the integral while still allowing for a separation into distinct contributions from the disconnected components of the thimble, where each of them may be identified with a specific electron trajectory.

\subsection{Spectral caustics}
    With the rigorous methods of Picard-Lefschetz theory, we can now study a phenomenon that has so far been inaccessible to semi-classical quantum orbit analysis: caustics.
    Caustics are the bright features that appear whenever multiple classical solutions of a quantum-mechanical system coincide, as for example the marbled pattern at the bottom of a swimming pool caused by multiple (classical) rays of sunlight that are bent by the curved water surface onto the same position. 
  In attosecond science, we find the same effect for specific shapes of the driving laser field's vector potential which causes the semi-classical electron trajectories to recombine at the same time and produce observable bright features in, for example, the HHG spectrum 
  \cite{dong2024caustic, 
  raz2012spectral, 
  yan2010lowEnergy, 
  uzan2020attosecond,
  birulia2019spectral, 
  goreslavskii2000tunneling, facciala2018highorder, facciala2016probe,
  raab2025xuv}.
  The most prominent example is the high-order harmonic cutoff, for which the saddle point solutions for short and long trajectories are very close (or even coalesce), see \fig{fig:spectra}(a) \cite{pisanty2020imaginary}, and which was first observed as a divergence in the simpler, fully classical, `simple man's' model~\cite{corkum1993plasma}.
  Mathematically, the appearance of caustics is linked to catastrophe theory (see \sect{sec:pl-caustics}) which relates the number of external parameters $K$ to the number of coalescing saddle points ($K+1$) and hence, types of caustic structures and the degree of enhancement. 
  In the context of HHG, the first `external parameter' is the harmonic order~$q$. 
  Across a spectrum we can therefore observe features related to a fold catastrophe ($K=1$) at which two saddle points coalesce \cite{pisanty2020imaginary}.

  As we increase the number of control parameters, e.g.,\ by adding a second driving field, we can observe higher-order diffraction patterns (cf.~\citenistsec{36.3}).
  Such bright features have been observed experimentally as intensity enhancements of specific harmonic orders in \cite{raz2012spectral} and \cite{raab2025xuv}, where they have been attributed to a swallowtail catastrophe diffraction pattern and coinciding classical trajectories, respectively. 
  Both experiments have slightly different parameters, but topologically they constitute the same situation:  
  a HHG setup with a collinear two-colour driver comprised of a fundamental field and a strong second harmonic component.
  The caustics can then be found by scanning over phase delays between the two constituent fields and measuring the harmonic spectra.
  They are related to a swallowtail catastrophe point ($K=3$), with the three `external' control parameters being harmonic order, phase delay and relative intensity.

\begin{figure}[t]
    \includegraphics[width=\linewidth]{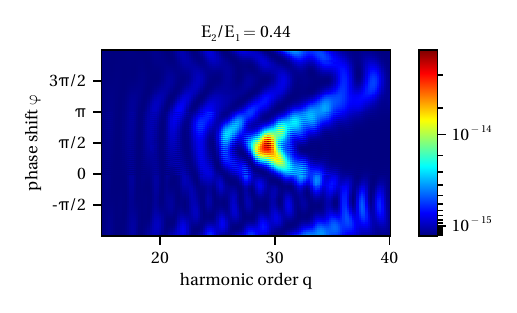}
    \caption{Harmonic intensity across a range of harmonic orders and phase shifts of the two-colour field \eq{eq:tc-field} with $E_2/E_1 = 0.44$, calculated using quadrature along the flowed integration domain (PLF method), resembling the canonical diffraction integral for a swallowtail catastrophe.}
    \label{fig:caustics}
    \end{figure}

  Here, we present the harmonic intensities for this setup across a scan over various phase delays $\varphi$ between the two constituent fields, calculated using the PLF method.
  That is, for every configuration of external parameters $(\varphi, q)$ we deform the integration domain using the downwards-flow algorithm to determine the Lefschetz thimbles and then evaluate the HHG dipole integral across the obtained thimble surface.
  In \fig{fig:caustics} we show this parameter scan for the electric field \eq{eq:tc-field} with the fixed relative intensity $E_2/E_1 = 0.44$ that reveals a part of the caustic pattern expected for a swallowtail catastrophe, and a significant enhancement around $q=30.$ and $\varphi = 0.59$.%
  \footnote{For comparison: in Fig. 2\ in \cite{raz2012spectral} the HHG yield was modelled as the respective canonical diffraction integral to demonstrate the expected pattern, and not actually calculated from the HHG integral.} 

\begin{figure}[t]
    \includegraphics[width=0.75\linewidth]{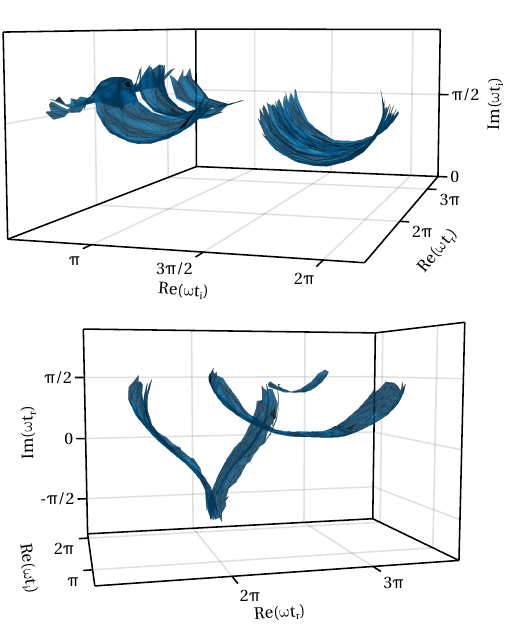}
    \caption{Thimble for the HHG swallowtail configuration ($E_2/E_1=0.44, q = 30., \varphi = 0.59$), the most intense point of \fig{fig:caustics}, in projections as in \fig{fig:HHG-thimbles-flow}.}
    \label{fig:thimble-sw}
    \end{figure}

  Lefschetz thimbles for the parameters with highest intensity 
  are shown in \fig{fig:thimble-sw}.
  While we can't see striking differences in the ionisation-time projection (top panel), we find 
  a rather flat surface in the recombination-time projection (bottom panel). 
  Comparing this to \fig{fig:HHG-thimbles-flow} where several saddle points' contributions yield `steep' and disconnected surfaces, this highlights the fact that around a swallowtail point multiple saddle point coalesce to one higher-order critical point for which the steepest-descent manifold covers a long range of recombination times.

  Most importantly, this demonstrates our capability to evaluate the HHG dipole integral exactly, even around scenarios exhibiting caustic features.
  The specific setup showcased here is of particular interest as it reveals a significant enhancement of the harmonic response. 
  However, for HHG driven by polychromatic fields it is not the only configuration where caustics appear, as -- loosely speaking -- each new frequency component of the driving field contributes new saddle-point solutions that may or may not coalesce (and hence produce caustics) in a specific parameter condition.
  One such caustic feature originating from the (near) coalescence of three saddle points (a cusp catastrophe point) appears in the scenario presented in the following chapter and is studied in more detail in Appendix \ref{sec:app:cusp}.


\subsection{The colour switchover}
\begin{figure}
    \includegraphics[width=\linewidth]{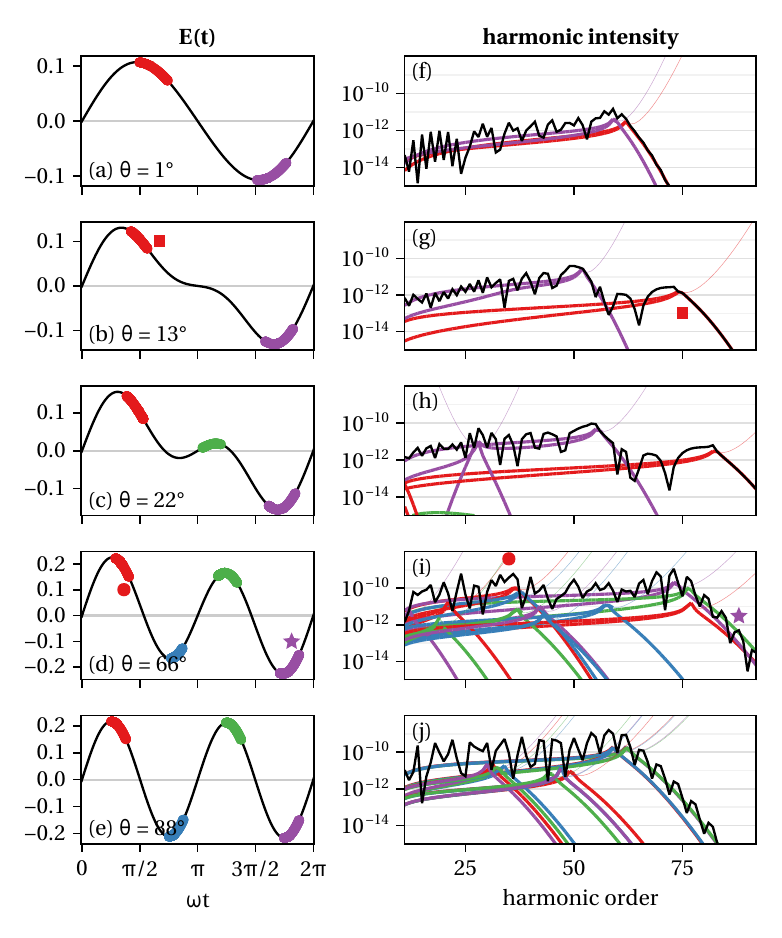}
    \caption{The colour switchover scheme: Electric fields $E(t)$ (in a.u.\,, left column) and respective HHG spectra $I(q\omega)$ (in a.u.\,, right column). 
    The contribution of the various trajectories are colour-coded based on their ionisation times marked in the electric field.
    The markers in (g) and (i) indicate the specific contributions corresponding to the trajectories in \fig{fig:cucs-trajectories}.
    }
    \label{fig:cucs-spectra}
  \end{figure}

    The necklace method introduced in \sect{sec:necklace} allows to determine relevant quantum orbits for any arbitrary driving field and independent of a classification of saddle point solutions.
    This now enables us to answer the broader question of how quantum orbits develop throughout any (arbitrary!) smooth change of parameters.
    As an example, here we address the smooth change from a monochromatic driver, via a two-colour field, to a monochromatic driver of second harmonic frequency -- a technique we term colour switchover (see evolution of field shapes in the left column of \fig{fig:cucs-spectra}), introduced in \cite{weber2025quantum}.

    It provides a framework to study two-colour driving fields with arbitrary intensity ratio (and phase shifts) that seamlessly connects perturbative setups (where the second colour field intensity is in the order of a few percent) to fully bichromatic setups (with two constituent fields of equal intensity).
    This transition beyond the perturbative regime has proven to bear interesting dynamics in an ionisation-only context already \cite{weber2025quantum,rook2022exploring}. 

    Here, we consider a colour switchover with constant ponderomotive energy $\Upond$,%
    \footnote{The ponderomotive energy is the time-averaged quiver energy of the free electron over one cycle of the driving field, $\Upond = \langle E(t) \rangle_T = \frac{1}{4} \left({A_1}^2 + {A_2}^2 \right)$.}
    i.e.\, constant total energy of the driving field.
    With that, the classical harmonic cutoff of the spectrum given by 
    $q_\mathrm{class} \approx \Ip + 3.17\Upond$
    remains constant throughout the switchover.
    We restrict our discussion to fields with zero phase delay ($\varphi = 0$) between the two components, such that the vector potential reads
    \begin{align}
      A(t) &= A_1 \cos(\omega t) + A_2 \cos(2 \omega t)  \\
      \quad \text{where} \quad 
      A_1 &= \cos(\theta)  E_0 / \omega
      \quad \text{and} \quad
      A_2 = \sin(\theta) E_0 / (2 \omega) \nonumber \,.
    \end{align} 
    The electric field is hence given by
    \begin{align}
      E(t) &= E_1 \sin(\omega t) + E_2 \sin(2 \omega t) \label{eq:cucs-field} \\
      \quad \text{where} \quad 
      E_1 &= E_0 \cos(\theta)
      \quad \text{and} \quad 
      E_2 = 2 E_0 \sin(\theta)\,.  \nonumber 
    \end{align}
    Here we have used the \emph{mixing angle} $\theta$ as a parameter to tune the amplitude ratio of the two constituent fields as $E_2/E_1 = 2 \tan \theta$. 
    For $\theta = 0 \degree$ the field \eq{eq:cucs-field} corresponds to a monochromatic field with frequency $\omega$, and for $\theta = 90 \degree$ to a monochromatic field with frequency $2 \omega$.
    For all intermediate values of $\theta$, \eq{eq:cucs-field} forms a two-colour field with varying amplitude ratio.

    On the left-hand column of \fig{fig:cucs-spectra} we show the total electric field $E(t)$ for the values $\theta = 1\degree, 13\degree, 22\degree, 67\degree$ and $88\degree$ (panels a-e) where we marked $\real(\tis)$ for contributing quantum orbits.
    This demonstrates the key intrigue of the scheme: 
    At the initial stage of the switchover ($\theta = 1\degree$) we have two (distinct and clearly separated) ionisation bursts within one cycle of the fundamental $T$, just after the maxima of the field at 
    $\omega t = \pi/2$ and $3 \pi/2$ (panel (a)).
    After the \emph{continuous} transition to $\theta = 88\degree$, however, we have four ionisation bursts in that same time frame, around $\omega t=0.6$, $2.2$, $3.8$ and $5.3$ (panel (e)).

    Especially with the understanding that ionisation events (and the associated recombination times, making them trajectories) correspond to saddle points, and saddle points are topologically stable features of analytic functions, this `jump' in number of ionisation events raises a set of questions \cite{weber2025quantum}:
      Where do the new ionisation events come from?
      Which of the `old' trajectories correspond to the `new' ones?
      When do the new trajectories start to become relevant for the harmonic spectrum?
    Picard-Lefschetz theory allow us to answer these questions.

  \subsubsection{Saddle-point dynamics in complex time}
    Tracking how the saddle points move in the complex plane throughout the colour switchover reveals rather complicated dynamics, which makes a consistent classification of trajectories tedious and non-unique, but still possible.
    For example, in the beginning of the switchover the saddle points can be classified as ``short'' and ``long'' trajectories as shown in \fig{fig:HHG-saddles-in-cp}(b) and (c).
    Upon the colour switchover the two `branches' may break up and reconnect with other -- newly emerging -- solutions, such that saddle points that would be classified as a short trajectories for the monochromatic $\omega$ driving field transition smoothly (!) into being long trajectories. 

    The new saddle point solutions mostly come in from high imaginary ionisation times (note e.g.\ the range of $\imag(\trs)$ in \fig{fig:HHG-saddles-in-cp}(e) panel for ionisation window B and C) and move down towards the real axis until they are equally spread out.
    Focussing on the ionisation times, we find that the typical structure around the first ionisation burst (around time $\omega \ti=1.9$, 
    structure as seen in \fig{fig:HHG-saddles-in-cp}(b)) is `ripped apart' by the newly incoming saddle points, which subsequently push some solutions to earlier times and some solutions to later times.
    During this process, trajectories frequently undergo pairwise (avoided) crossings similar to the one observed and extensively described for the high-harmonic cutoff in \cite{pisanty2020imaginary}. 
    As described therein, the full coalescence of two saddle points to a fold catastrophe point ($K=1$) renders the classification ambiguous and introduces branch cuts into the unified Riemann surface that solves the saddle-point equations.
    For the performed colour switchover considered here there are multiple instances of those full coalescences, we show one of them in \fig{fig:cucs-trajectories}(d) below.

  \subsubsection{Harmonic spectra throughout the switchover}
    Let us now highlight a few curious features that impact the observable harmonic spectrum.
    Therefore, in the right-hand side of \fig{fig:cucs-spectra} we present the harmonic spectra with their contributions from the individual saddle points for the respective stages of the colour switchover.
    We coloured the individual saddle points' contributions according to their ionisation window indicated on the left-hand side, and we show the resulting total intensity (SPM, from \eq{eq:hhg-sum-spm}) in black.
    The contributions of non-relevant saddle points are shown in faint lines.

    For the initial stage of the colour switchover, i.e., a purely monochromatic field with frequency $\omega$, the HHG spectrum looks like the one shown in \fig{fig:spectra}(a).
    The contributions from short and long trajectories from within in half cycle interfere with each other and due to the dynamical symmetry of the driving field result in the cancellation of the odd-order total intensities. 
    A slight perturbation to the driving field breaks this behaviour. 
    In panels (a) and (f) of \fig{fig:cucs-spectra} we show the field and the resulting spectrum shortly after the initial stage of the colour switchover ($\theta = 1\degree$), i.e.,\, a two-colour driving field with a weak $2\omega$ component.
    We see that the comb-like structure of \fig{fig:spectra}(a) is broken. %
     Moreover, the contributions of short-long pairs from within each half cycle start to separate, visible at e.g.\, the harmonic cutoff (here around order $60$).

    When increasing the strength of the second colour field (panels (g)-(j)), this spreading becomes more pronounced.
    The newly incoming saddle point solutions ultimately cause the harmonic cutoff in the spectrum
    to break up into two visible cutoffs (around $q=50$ and $q=75$ in (g)).
    For the later stage of the colour switchover the newly emergent trajectories produce multiple cutoffs, in panel (j) at $q \approx 32$, $47$ and $60$. 
    The former, however, do not impact the total shape of the harmonic spectrum as they stem from higher-order return pairs of saddle points, i.e., trajectories with longer travel times and hence weaker contribution \cite{chipperfield2006tracking}.

    Upon the full completion of the colour switchover we also restore the expected suppression of odd harmonic orders (of the $2 \omega$ driver) due to the symmetry of the driving field.


  \subsubsection{Spectral enhancements at the cusp catastrophe}

      \begin{figure}[t]
      \includegraphics[width=\linewidth]{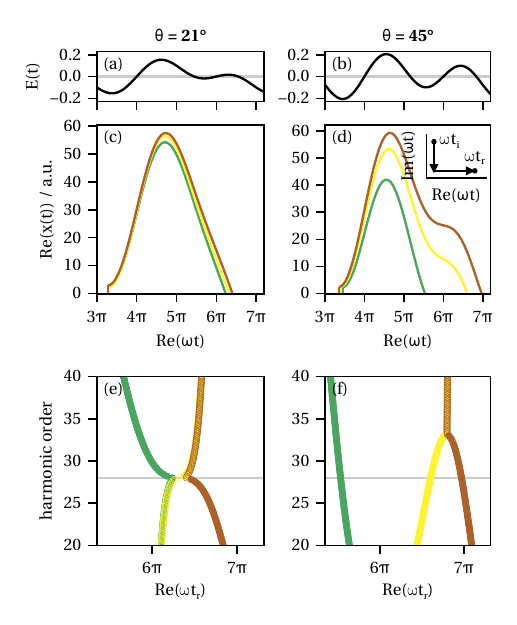}
      \caption{Electrical fields (panels (a) and (b)) and three according semi-classical electron trajectories (panels (c) and (d)) for three saddle points contributing to harmonic order $q=28$.
      The respective energy-time relations are shown in panel (e) and (f).
      The three assimilating trajectories for $\theta = 21\degree$ ($E_2/E_1 = 0.78$) on the left-hand side cause the enhancement seen in the respective spectrum \fig{fig:cucs-spectra}(h).
      The temporal contour along which the integral \eq{eq:trajectory} is evaluated is shown in the inset of panel (d).}
      \label{fig:cusp-trajs}
    \end{figure}

    The quantum-orbit based consideration of the colour switchover allows us to study the intriguing interplay of more than two trajectories. 
    For example, the colour switchover presented here entails an avoided crossing of three trajectories.
    Even if the chosen parameter scan does not include an exact coalesce%
    \footnote{In fact, the exact coalescence of multiple saddle points requires complex-valued external parameters, as is explained in full detail for the case of the coalescence of two saddle points at the high-harmonic cutoff in \cite{pisanty2020imaginary}.}
    it nevertheless results in an enhanced signal for specific harmonic orders, depending on the mixing angle.
    This can be seen in \fig{fig:cucs-spectra}(h) around $q=28$, where we find three contributing saddle points contributions in close vicinity (see the three purple lines) resulting in a noticeable increase of the total spectral intensity around this harmonic order.
    A more detailed analysis of this cusp catastrophe point is shown in Appendix \ref{sec:app:cusp}.

    In \fig{fig:cusp-trajs} we show the corresponding semi-classical electron trajectories for the respective three saddle points for this situation, as well as for a subsequent field configuration in the switchover, at $\theta=45\degree$.
    The trajectories are described by the displacement from the origin (the nucleus)
    \begin{equation}
      \vec{x}(t) = \int_{\tis}^{\trs} \left(\vec{p}_s + \vec{A}(t) \right) \d t \,,
      \label{eq:trajectory}
    \end{equation}
    where the temporal integration in \eq{eq:trajectory} starts from $\tis \in \mathbb{C}$ and goes down to $\real{\tis}$, then all the way to $\real{\trs}$ where it then terminates at $\trs \in \mathbb{C}$, shown as an inset in \fig{fig:cusp-trajs}. 
    The first leg of this contour can be interpreted as the trajectory `inside the tunnelling barrier' and it will give an imaginary-valued displacement $\vec{x}(t)$. 
    The second part of the time contour then describes the classical path of the electron under the influence of the driving laser field \cite{smirnova2013multielectron,Pisanty2017}.

    Phenomenologically, this is the same situation as in the swallowtail caustic described above.
    For most scenarios all the various electron trajectories interfere destructively and their contributions `counteract' each other.
    In situations like \fig{fig:cusp-trajs}(b), the depicted electron pathways lie so close together -- while still being individually relevant contributions -- that they interfere constructively as they all contribute with the same phase towards the total HHG dipole. 
    This results in a significant enhancement compared to other configurations.
    Technically, this is exactly the phenomenon that can be used to optimise the spectral yield of a desired harmonic order \cite{raab2025xuv}.

  \subsubsection{Individual quantum orbits throughout the switchover}
\begin{figure}[t]
      \includegraphics[width= \linewidth]{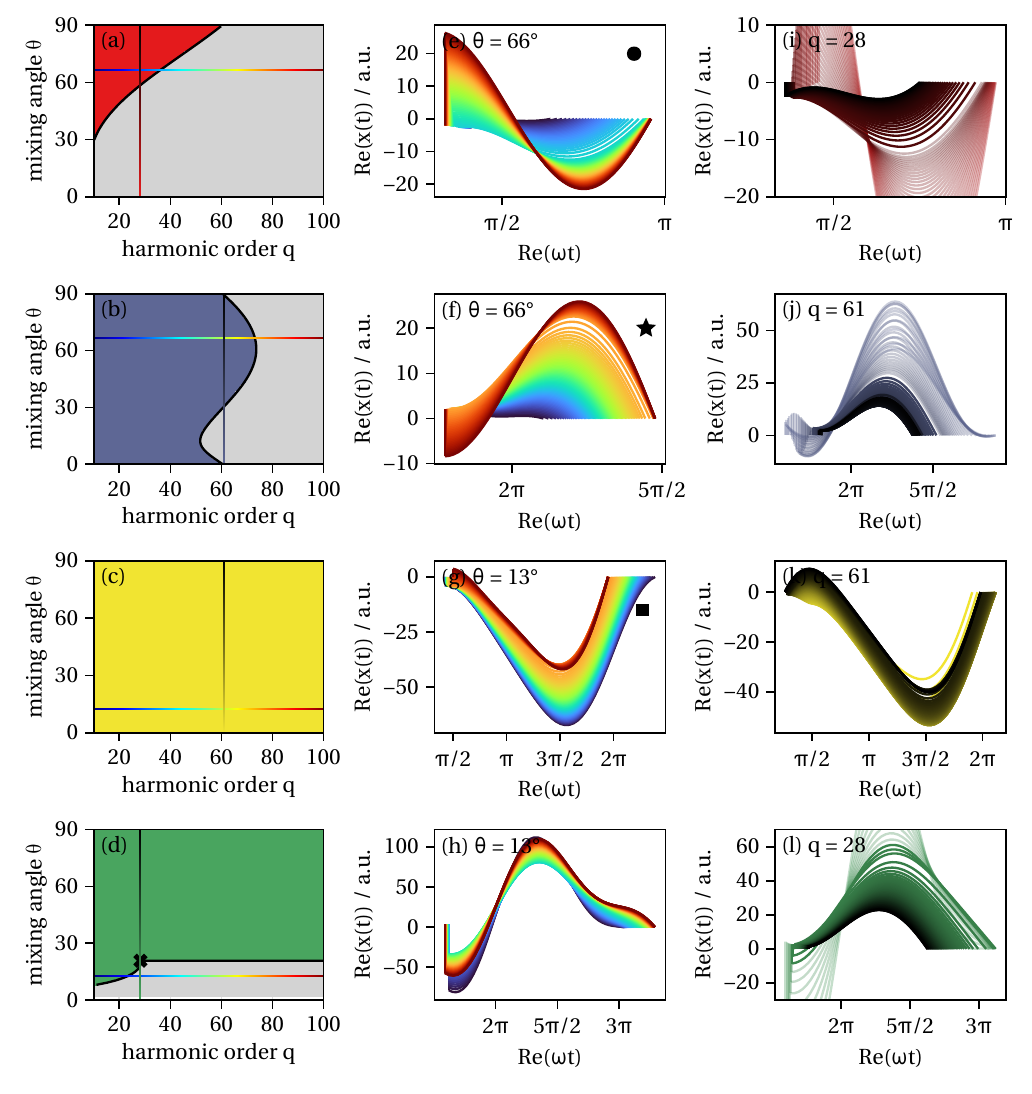}
      \caption{Tracking four different saddle points (rows) throughout the colour switchover and harmonic orders. 
      Left column: showing whether the saddle point is relevant (coloured) or not (grey).
      Centre column: Trajectories acc. to \eq{eq:trajectory} for the indicated $\theta$ value (horizontal line in the left panel), for all harmonic orders (lines coloured respectively).
      Right column: Trajectories for the indicated harmonic order (vertical line in the left panel), throughout the colour switchover (i.e., all values of $\theta$, colour shaded respectively).
      Markers in (e), (f) and (g) attribute contributions to the spectra in \fig{fig:cucs-spectra}(j) and (g).
      }
      \label{fig:cucs-trajectories}
    \end{figure}

    Apart from looking at the shape of the total harmonic response and how the several ionisation windows interfere, the necklace algorithm gives us the unique capability to follow individual quantum orbits throughout the colour switchover.
    This has been inaccessible within the existing understanding of saddle-point methods or their extension to uniform approximations.    
    In particular, tracking saddle points allows to examine how the respective electron trajectories and their relevance to the spectrum changes.
    In \fig{fig:cucs-trajectories} we showcase four (`types' of) saddle points in detail, as representative examples of common behaviours.
    In the left column we show whether the respective saddle point is relevant (coloured) or not (grey), as it is tracked for the range of harmonic orders and throughout the colour switchover. 
    Hence, the boundary of the coloured region of contributions are the Stokes lines in parameter space, drawn as a black line.
    In the second column we show the specific semi-classical electron trajectories as in \fig{fig:cusp-trajs}(b), for a fixed mixing angle (indicated with the rainbow-coloured horizontal line in the left panel) and a range of harmonic orders denoted with the respective colour.
    Similarly, the third column shows the trajectories for a fixed harmonic order (indicated by the vertical bar in the left panel) and across the colour switchover denoted with the colour gradient.
    For non-relevant trajectories the lines are drawn faint where it does not lead to confusion.

    The first row in \fig{fig:cucs-trajectories} shows what is eventually the most dominant short trajectory for the first ionisation window of the $2\omega$ field.
    That is, from panel (a) we find that for $\theta = 90\degree$ this saddle point is relevant up to harmonic order $q \approx 60$, which constitutes the cutoff of \fig{fig:cucs-spectra}(j).
    Prior to that (for $\theta < 90\degree$), this saddle point contributes only contributes for lower harmonic orders, or not at all.
    We find this behaviour particularly interesting, as this saddle points only starts contributing quite late in the colour switchover, but then in fact plays a prominent role for the spectrum of the fully $2\omega$ driving field.
    The trajectories shown in the centre panel (for all harmonic orders) correspond to the contribution marked with a circle in the harmonic spectrum in \fig{fig:cucs-spectra}(i).

    In contrast to that, the trajectory showcased in the second row contributes to (at least) the early plateau throughout the full colour switchover. 
    It is the first, and hence most dominant, short trajectory starting from ionisation burst four, marked in \fig{fig:cucs-spectra}(i) with a star.
    This trajectory remains one of the most dominant contributors to the spectrum throughout the colour switchover, so that its Stokes line marks a noticeable cutoff in the spectrum.
    The Stokes line in the left panel explains the shift in the high-order harmonic cutoff observed from orders $q=60$ for $\theta=0$ to $75, 82, 78$ and $60$ for $13\degree, 22\degree, 67\degree$ and $88\degree$ as seen in \fig{fig:cucs-spectra}(f-j) respectively.

    In the third row we show the first long trajectory of the second ionisation burst of the $2 \omega$ driver around $\omega \ti = 2.22$, marked in \fig{fig:cucs-spectra}(b) and (g) with a square.
    As a long trajectory it is relevant for all harmonic orders and remains so throughout the whole colour switchover.

    The trajectory shown in the last row is involved in the caustics mentioned to explain the enhancement around $q=28$ in \fig{fig:cucs-spectra}(h).
    Around the caustic there is a saddle-point coalescence that introduces ambiguity in the classification and shows up as a clear discontinuity in panel \fig{fig:cucs-trajectories}(d).
    For the beginning of the colour switchover this saddle point was a higher order return.
    After this branch cut, however, this trajectory eventually becomes the long trajectory of the fourth ionisation burst of the $2\omega$ field. 

    To conclude, \fig{fig:cucs-trajectories} demonstrates how the electron trajectories change smoothly upon parameter scans, but their relevance to the total spectrum may change abruptly. 
    In turn, tracking contributions from distinct quantum orbits throughout a parameter scan allows us to attribute the observable features of the harmonic spectra to these specific saddle-point dynamics.

\section{Outlook}
  This paper introduces the ideas of Picard-Lefschetz theory to attosecond science and strong-field physics.
  We presented two computational methods that utilise these concepts: the continuous downwards flow of the integration domain, and the (novel) necklace algorithm to determine the relevance of given saddle points. 
  Both of those methods have the flow of the discretised integration domain
  (Eqs. \ref{eq:downwards-flow} and \ref{eq:upwards-flow}, or \ref{eq:downwards-flow-HHG} and \ref{eq:upwards-flow-HHG} in the context of HHG) as a central algorithmic element.
  While these flows theoretically preserve $\imag(\phi(x))$, their discretised numerical implementation as a first-order Euler scheme can lead to numerical instabilities and hence, limitations in the usage of the methods.
  This issue becomes particularly important near Stokes transitions, where a precise treatment of $\imag(\phi(x))$ is essential by definition.
  Looking forward, a more rigorous way to identify and incorporate Stokes transitions can be developed.

  We have applied our new methods to calculate the HHG response across ranges of external parameters that contain Stokes transitions and hence, show caustics.
  Generally, the study of caustics is inherently linked to the framework of catastrophe theory. 
  By identifying parameters that cause the coalescence of saddle points once can hence classify the respective catastrophe.
  This allows us to compute the expected enhancement of the signal at the catastrophe point compared to the signal in its vicinity.
  The so-called `twinkling exponents' \cite{berry1977focusing} have been used to motivate the enhancements within a harmonic spectrum \cite{raz2012spectral,raab2025xuv}, but the rigorous derivation for arbitrary parameter scans is still missing in the context of attosecond experiments.

  Furthermore, the identification of catastrophe points allows to develop uniform approximations that smoothen the integral contribution of several saddle points in its parameter vicinity. 
  Realising these for specific parameters scans would allow a fully saddle-point based analytical approach without artificial discontinuities even in the case of coalescences of three or more saddle points. 

  Alternatively, however, the separate thimbles 
  can be evaluated individually using a standard quadrature of the surface elements.
  As shown by \eq{eq:hhg-PL-sum} this yields the exact integral for the HHG dipole in terms of distinct contributions, irrespective of the saddle points (and their vicinity).


\section{Conclusion}
  The description of strong-field induced processes like tunnel ionisation and the generation of high-order harmonics (HHG) is often linked to the intuitive picture of interfering semi-classical electron trajectories.
  Mathematically, this corresponds to making the saddle-point approximation to the integrals that describe the atomic response in the SFA formulation, such as the harmonic radiation dipole \eq{eq:hhg-integral}.
  The spectral intensity of a given harmonic order is expressed as a sum over contributions from discrete quantum orbits, i.e. a sum of Gaussians centred around saddle points.
  For a given laser field, however, there are far more solutions to the saddle point equations (\ref{eq:speqs}) than quantum orbits that contribute to the dynamics.
  Their selection has so far been based on heuristics and empirical rules, which fail for generic state-of-the-art lasers to drive the process.

  In this paper, we presented Picard-Lefschetz theory as a tool to rigorously and unambiguously evaluate the dipole response in a quantum-orbit based fashion for any arbitrary driving laser waveforms.
  For that, we understand the dipole response as a two-dimensional path integral over ionisation and recombination times.
  Continuously deforming the two-dimensional integration contour into the complex plane allows to rewrite the integral as a sum of contributions along so-called Lefschetz thimbles.
  These are steepest-descent contours (surfaces) attached to the saddle points.
  The continuos deformation of the contour towards the thimble is dictated by the `downwards flow' and preserves the value of the integral, such that any intermediate flow step is an equal, and hence exact, representation of the integral.
  Conversely, as an alternative approach, we can identify contributing saddle points by checking whether there is an `upwards flow' (a steepest-ascent contour, the dual thimble) that connects them back to the original integration domain -- the plane of real ionisation and recombination times.
  For the case of a one-dimensional integral the steepest-ascent contours are lines that connect back to the real axes for relevant saddle points -- a property that is computationally straight-forward to inspect.
  For the case of a two-dimensional integral the steepest-ascent contours are surfaces (embedded in four-dimensional real space) which cannot be readily determined.
  We therefore introduce a novel procedure, the ``necklace'' algorithm, 
  in which we only flow the brim of the dual thimble and indubitably identify all possible intersections with the real plane. 
  This allows us to systematically determine relevant saddle points to a two-dimensional integral.
  
  We apply these methods to strong-field phenomena that have so far been inaccessible to semi-classical analysis.
  One of them is the appearance of spectral caustics, where the close proximity (or even the full coalescence) of multiple saddle-point solutions (read: trajectories) causes a significant enhancement in the observed signal.
  In these scenarios, the correct analytical representation of the integral requires uniform approximations that account for the non-Gaussian shape around the saddle points.
  Evaluating the integrand along the deformed contour however, is independent of the nature of the critical points, and hence allows us to evaluate the SFA integral exactly even in the vicinity of saddle-point coalescences.

  The versatility of the introduced Picard-Lefschetz methods allows us furthermore to address questions of a new class of parameter scans: the colour switchover \cite{weber2025quantum}. 
  The gradual replacement of a monochromatic driving field with its second harmonic, via two-colour configurations of increasing amplitude ratio, connects the perturbative second-colour regime to fully bichromatic driving fields. 
  Using the necklace algorithm, we are able to identify the relevant quantum orbits throughout the full range of different driving field configurations.
  For the perturbative case, we can attribute the unfolding of the high-harmonic cutoff to the dominant pairs of trajectories from the respective half cycles.
  Increasing the relative strength of the second harmonic eventually leads to newly emerging ionisation bursts that produce topologically stable enhancements in the spectrum.
  These arise due to the unavoidable proximity to a three-fold saddle point coalescence (a cusp catastrophe), which is here demonstrated for the first time in the context of attosecond science. 
  Moreover, tracking individual saddle point solutions throughout the colour switchover allows to show how the electron trajectories react to the change of driving field.

  Ultimately, we have shown the rigorous link between the SFA integrals for both the ionisation amplitude (one-dimensional time integral) and the HHG response (two-dimensional), and its interpretation in terms of quantum orbits, for arbitrary driving fields.
  This opens up the possibility to analyse semi-classical trajectories for generic parameter scans and more complex wave forms, e.g.\ three-dimensionally structured light fields.
  Depending on the specific configuration, there might be other methods derived from Picard-Lefschetz theory that simplify the identification of relevant quantum orbits, or more generally, the evaluation of the integral. 
  More specifically, the relevance of quantum orbits and the occurrence of Stokes phenomena may be further illuminated by including higher-order corrections in the saddle point approximation using the mathematical theory of resurgence \cite{1991RSPSA.434..657B, 2019AnPhy.40967914D}.
  Furthermore, we look forward to seeing Picard-Lefschetz methods applied to other highly-oscillatory integrals within attosecond science as well, including 
  high-order ATI, with its description of rescattered electrons 
  \cite{milosevic2006quantumorbit, milosevic2019xray, jasarevic2020application}
  dynamic interference in ionisation stabilisation~\cite{Demekhin2012dynamic, Vismarra2025dynamic}, or
  attosecond streaking~\cite{Itatani2002attosecond}. 
  As the theoretical framework is independent of the dimensionality of the integral, it could be used to simplify the five-dimensional integrals arising in calculating the response of solid targets to strong laser field radiation~\cite{Moos2020intense}.

\section{Acknowledgements}
  We wish to acknowledge Iain Murray for encouraging this interdisciplinary collaboration between AW and JF.

  AW and EP acknowledge Royal Society funding under URF\textbackslash R1\textbackslash 211390 and RF\textbackslash ERE\textbackslash 210255.
  The work of JF is supported by the STFC Consolidated Grant ‘Particle Physics at the Higgs Centre,’ and, respectively, by a Higgs Fellowship at the University of Edinburgh. 

  For the purpose of open access, the authors have applied a Creative Commons Attribution (CC BY) license to any Author Accepted Manuscript version arising from this submission.



\appendix

\section{The cusp catastrophe point} \label{sec:app:cusp}
\begin{figure}[ht]
    \includegraphics[width=0.75\linewidth]{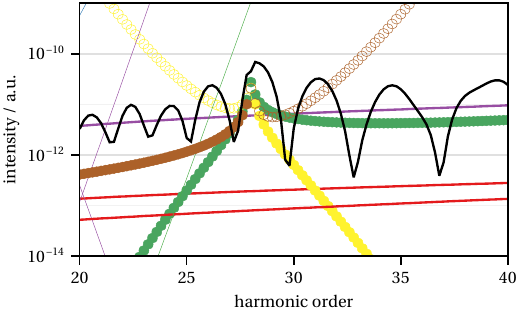}
    \caption{Zoom in on the spectrum of \fig{fig:cucs-spectra}(h), highlighting the three different contributions that cause the enhancement of harmonic order 28 due to the nearby cusp catastrophe point.}
    \label{fig:cusp-spectrum-zoom}
    \end{figure}

\begin{figure}[ht]
    \includegraphics[width=\linewidth]{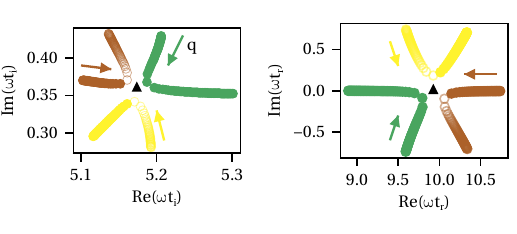}
    \caption{Saddle points in the complex plane for the spectrum shown in \fig{fig:cucs-spectra}(h), indicating a cusp catastrophe.}
    \label{fig:saddles-cusp}
    \end{figure}

    As a topological feature of the colour switchover we identify a cusp catastrophe point causing observable enhancements in the spectrum.
    For the exact coalescence of three saddle points, however, we require external parameters to assume complex values.
    The smaller those imaginary parts are, the larger is the effect of the catastrophe point on the observed quantity. 
    For the given phase shift $\varphi = 0$ presented in the main text of this paper we find the largest relative enhancement of this cusp point when $\theta = 21\degree$, see \fig{fig:cusp-spectrum-zoom}.
    The saddle points for $\theta = 21\degree$ ($E_2/E_1 = 0.78$), $\varphi = 0$ and for a range of $q = 20$ to $40$ are shown in \fig{fig:saddles-cusp}, indicating three `branches' of solutions in close proximity. 
    The exact coalescence for three solutions only happens in complex parameter space, at  
    $\theta = (21.28 + 0.03\im)\degree$, $\varphi=0$ and $q = 27.95 - 0.1\im
    $. 
    This cusp point is marked as a triangle in \fig{fig:saddles-cusp}, sitting in the centre of the three real-parameter saddle points.

\begin{figure}[t]
    \includegraphics[width=0.75\linewidth]{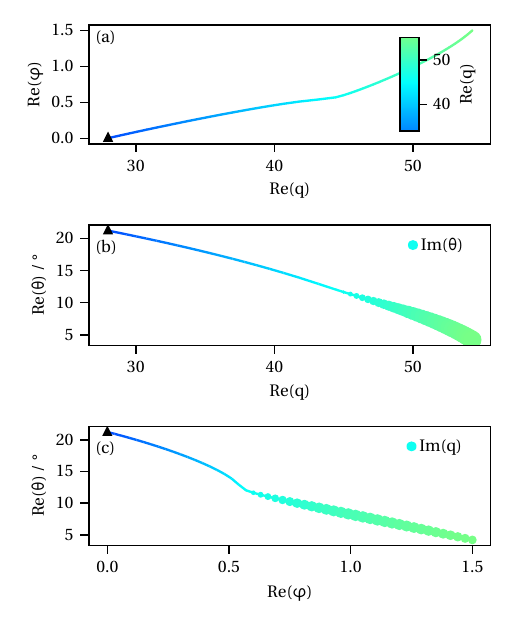}
    \caption{Projections of parameter combinations $(\theta, \varphi,q)$ for a cusp catastrophe point at which three saddle points coalesce. 
    For the exact coalescence we assume $\theta \in \mathbb{C}$ and $q \in \mathbb{C}$ and denote the respective imaginary parts as marker size in panels (b) and (c).
    In all three panels the colour indicates the real part of the harmonic order $q$.
    The cusp point reported in the main text and above is drawn as a triangle.
}
    \label{fig:cusp-parameter-space}
    \end{figure}

    In \fig{fig:cusp-parameter-space} we show how the exact location of this cusp point changes depending on the external parameters. 
    The $(\real(q), \real(\theta))$ projection in panel (b) indicates how the specific harmonic order which is enhanced by the coalescence changes as we perform the colour switchover.
    Panel (c) confirms that for colour switchovers with a different two-colour phase delay $\varphi$ there is still a cusp catastrophe point. 
    However, the increased imaginary parts of the external parameters (indicated by the marker size) suggest that it plays a subdominant role for the total spectrum.

\newpage

    \bibliographystyle{lancelot} 
    \bibliography{references-zotero-final,otherRefs}

\begin{thebibliography}{113}%
\makeatletter
\providecommand \@ifxundefined [1]{%
 \@ifx{#1\undefined}
}%
\providecommand \@ifnum [1]{%
 \ifnum #1\expandafter \@firstoftwo
 \else \expandafter \@secondoftwo
 \fi
}%
\providecommand \@ifx [1]{%
 \ifx #1\expandafter \@firstoftwo
 \else \expandafter \@secondoftwo
 \fi
}%
\providecommand \natexlab [1]{#1}%
\providecommand \enquote  [1]{``#1''}%
\providecommand \bibnamefont  [1]{#1}%
\providecommand \bibfnamefont [1]{#1}%
\providecommand \citenamefont [1]{#1}%
\providecommand \href@noop [0]{\@secondoftwo}%
\providecommand \href [0]{\begingroup \@sanitize@url \@href}%
\providecommand \@href[1]{\@@startlink{#1}\@@href}%
\providecommand \@@href[1]{\endgroup#1\@@endlink}%
\providecommand \@sanitize@url [0]{\catcode `\\12\catcode `\$12\catcode
  `\&12\catcode `\#12\catcode `\^12\catcode `\_12\catcode `\%12\relax}%
\providecommand \@@startlink[1]{}%
\providecommand \@@endlink[0]{}%
\providecommand \url  [0]{\begingroup\@sanitize@url \@url }%
\providecommand \@url [1]{\endgroup\@href {#1}{\urlprefix }}%
\providecommand \urlprefix  [0]{URL }%
\providecommand \Eprint [0]{\href }%
\providecommand \doibase [0]{https://doi.org/}%
\providecommand \selectlanguage [0]{\@gobble}%
\providecommand \bibinfo  [0]{\@secondoftwo}%
\providecommand \bibfield  [0]{\@secondoftwo}%
\providecommand \translation [1]{[#1]}%
\providecommand \BibitemOpen [0]{}%
\providecommand \bibitemStop [0]{}%
\providecommand \bibitemNoStop [0]{.\EOS\space}%
\providecommand \EOS [0]{\spacefactor3000\relax}%
\providecommand \BibitemShut  [1]{\csname bibitem#1\endcsname}%
\let\auto@bib@innerbib\@empty
\bibitem [{\citenamefont {Lewenstein}(1994)}]{lewenstein1994theory}%
  \BibitemOpen
  \bibfield  {author} {\bibinfo {author} {\bibfnamefont {M.}~\bibnamefont
  {Lewenstein}},\ }\bibfield  {title} {\bibinfo {title} {Theory of
  high-harmonic generation by low-frequency laser fields},\ }\href
  {https://doi.org/10.1103/PhysRevA.49.2117} {\bibfield  {journal} {\bibinfo
  {journal} {\emph {Phys. Rev. A}}\ }\textbf {\bibinfo {volume} {49}}\bibfield
  {number} {\bibinfo  {number} { no.~3},\ \bibinfo {pages} {pp.~2117--2132}}
  (\bibinfo {year} {1994})}\BibitemShut {NoStop}%
\bibitem [{\citenamefont {Keldysh}(1965)}]{keldysh1964ionization}%
  \BibitemOpen
  \bibfield  {author} {\bibinfo {author} {\bibfnamefont {L.}~\bibnamefont
  {Keldysh}},\ }\bibfield  {title} {\bibinfo {title} {Ionization in the field
  of a strong electromagnetic wave},\ }\href@noop {} {\bibfield  {journal}
  {\bibinfo  {journal} {\emph {Sov. Phys. {JETP}}}\ }\textbf {\bibinfo {volume}
  {20}}\bibfield  {number} {\bibinfo  {number} { no.~5},\ \bibinfo {pages}
  {p.~1307}} (\bibinfo {year} {1965})},\ \bibinfo {note}
  {[\href{http://www.jetp.ras.ru/cgi-bin/e/index/r/47/5/p1945?a=list}{\textit{Zh.
  Eksp. Teor. Fiz.} \textbf{47} no.~5, p.~1945 (1965)}]}\BibitemShut {NoStop}%
\bibitem [{\citenamefont {Perelomov}\ \emph {et~al.}(1967)\citenamefont
  {Perelomov}, \citenamefont {Popov},\ and\ \citenamefont
  {Terent'ev}}]{perelomov1967IonizationII}%
  \BibitemOpen
  \bibfield  {author} {\bibinfo {author} {\bibfnamefont {A.}~\bibnamefont
  {Perelomov}}, \bibinfo {author} {\bibfnamefont {V.}~\bibnamefont {Popov}},\
  and\ \bibinfo {author} {\bibfnamefont {M.}~\bibnamefont {Terent'ev}},\
  }\bibfield  {title} {\bibinfo {title} {Ionization of atoms in an alternating
  electric field: {II}},\ }\href@noop {} {\bibfield  {journal} {\bibinfo
  {journal} {\emph {Sov. Phys. JETP}}\ }\textbf {\bibinfo {volume}
  {24}}\bibfield  {number} {\bibinfo  {number} { no.~1},\ \bibinfo {pages}
  {p.~207}} (\bibinfo {year} {1967})},\ \bibinfo {note}
  {[\href{http://www.jetp.ac.ras/cgi-bin/e/index/r/51/1/p309?a=list}{\textit{Zh.
  Eksp. Teor. Fiz.}, \textbf{51} no.~1, p.~309 (1967)}]}\BibitemShut {NoStop}%
\bibitem [{\citenamefont {Becker}\ \emph {et~al.}(2002)\citenamefont {Becker},
  \citenamefont {Grasbon}, \citenamefont {Kopold}, \citenamefont {Milo{\v
  s}evi{\'c}}, \citenamefont {Paulus},\ and\ \citenamefont
  {Walther}}]{becker2002abovethreshold}%
  \BibitemOpen
  \bibfield  {author} {\bibinfo {author} {\bibfnamefont {W.}~\bibnamefont
  {Becker}}, \bibinfo {author} {\bibfnamefont {F.}~\bibnamefont {Grasbon}},
  \bibinfo {author} {\bibfnamefont {R.}~\bibnamefont {Kopold}}, \emph
  {et~al.},\ }\bibfield  {title} {\bibinfo {title} {Above-{{Threshold
  Ionization}}: {{From Classical Features}} to {{Quantum Effects}}},\ }in\
  \href {https://doi.org/10.1016/S1049-250X(02)80006-4} {\emph {\bibinfo
  {booktitle} {Advances {{In Atomic}}, {{Molecular}}, and {{Optical
  Physics}}}}},\ Vol.~\bibinfo {volume} {48},\ \bibinfo {editor} {edited by\
  \bibinfo {editor} {\bibfnamefont {B.}~\bibnamefont {Bederson}}\ and\ \bibinfo
  {editor} {\bibfnamefont {H.}~\bibnamefont {Walther}}}\ (\bibinfo  {publisher}
  {Academic Press},\ \bibinfo {year} {2002})\ pp.\ \bibinfo {pages}
  {35--98}\BibitemShut {NoStop}%
\bibitem [{\citenamefont {Kopold}\ \emph {et~al.}(2000)\citenamefont {Kopold},
  \citenamefont {Becker},\ and\ \citenamefont {Kleber}}]{kopold2000quantum}%
  \BibitemOpen
  \bibfield  {author} {\bibinfo {author} {\bibfnamefont {R.}~\bibnamefont
  {Kopold}}, \bibinfo {author} {\bibfnamefont {W.}~\bibnamefont {Becker}},\
  and\ \bibinfo {author} {\bibfnamefont {M.}~\bibnamefont {Kleber}},\
  }\bibfield  {title} {\bibinfo {title} {Quantum path analysis of high-order
  above-threshold ionization, {{Dedicated}} to {{Marlan O}}. {{Scully}} on the
  occasion of his 60th birthday},\ }\href
  {https://doi.org/10.1016/S0030-4018(99)00521-0} {\bibfield  {journal}
  {\bibinfo  {journal} {\emph {Optics Communications}}\ }\textbf {\bibinfo
  {volume} {179}}\bibfield  {number} {\bibinfo  {number} { no.~1},\ \bibinfo
  {pages} {pp.~39--50}} (\bibinfo {year} {2000})}\BibitemShut {NoStop}%
\bibitem [{\citenamefont {Sali{\`e}res}\ \emph {et~al.}(2001)\citenamefont
  {Sali{\`e}res}, \citenamefont {Carr{\'e}}, \citenamefont {Le~D{\'e}roff},
  \citenamefont {Grasbon}, \citenamefont {Paulus}, \citenamefont {Walther},
  \citenamefont {Kopold}, \citenamefont {Becker}, \citenamefont {Milo{\v
  s}evi{\'c}}, \citenamefont {Sanpera},\ and\ \citenamefont
  {Lewenstein}}]{salieres2001feynmans}%
  \BibitemOpen
  \bibfield  {author} {\bibinfo {author} {\bibfnamefont {P.}~\bibnamefont
  {Sali{\`e}res}}, \bibinfo {author} {\bibfnamefont {B.}~\bibnamefont
  {Carr{\'e}}}, \bibinfo {author} {\bibfnamefont {L.}~\bibnamefont
  {Le~D{\'e}roff}}, \emph {et~al.},\ }\bibfield  {title} {\bibinfo {title}
  {Feynman's {{Path-Integral Approach}} for {{Intense-Laser-Atom
  Interactions}}},\ }\href {https://doi.org/10.1126/science.108836} {\bibfield
  {journal} {\bibinfo  {journal} {\emph {Science}}\ }\textbf {\bibinfo {volume}
  {292}}\bibfield  {number} {\bibinfo  {number} { no.~5518},\ \bibinfo {pages}
  {pp.~902--905}} (\bibinfo {year} {2001})}\BibitemShut {NoStop}%
\bibitem [{\citenamefont {Lefschetz}(1924)}]{lefschetz1924lanalysis}%
  \BibitemOpen
  \bibfield  {author} {\bibinfo {author} {\bibfnamefont {S.}~\bibnamefont
  {Lefschetz}},\ }\href@noop {} {\emph {\bibinfo {title} {L'{{Analysis}} Situs
  et La G{\'e}om{\'e}trie Alg{\'e}brique}}}\ (\bibinfo  {publisher}
  {Gauthier-Villars et cie},\ \bibinfo {year} {1924})\BibitemShut {NoStop}%
\bibitem [{\citenamefont {Picard}\ and\ \citenamefont
  {Simart}(1897)}]{picard1897theorie}%
  \BibitemOpen
  \bibfield  {author} {\bibinfo {author} {\bibfnamefont {E.}~\bibnamefont
  {Picard}}\ and\ \bibinfo {author} {\bibfnamefont {G.}~\bibnamefont
  {Simart}},\ }\href@noop {} {\emph {\bibinfo {title} {{Th{\'e}orie des
  fonctions alg{\'e}briques de deux variables ind{\'e}pendantes}}}}\ (\bibinfo
  {publisher} {Paris, Gauthier-Villars},\ \bibinfo {year} {1897})\BibitemShut
  {NoStop}%
\bibitem [{\citenamefont {Pham}(1983)}]{pham1983vanishing}%
  \BibitemOpen
  \bibfield  {author} {\bibinfo {author} {\bibfnamefont {F.}~\bibnamefont
  {Pham}},\ }\bibinfo {title} {Vanishing homologies and the n variables
  saddlepoint method},\ in\ \href@noop {} {\emph {\bibinfo {booktitle}
  {Singularities, {{Part}} 2}}},\ \bibinfo {series and number} {\bibinfo
  {number} {2}}\ (\bibinfo  {publisher} {American Mathematical Soc.},\ \bibinfo
  {year} {1983})\ pp.\ \bibinfo {pages} {310--333}\BibitemShut {NoStop}%
\bibitem [{\citenamefont {Witten}(2010)}]{witten2010analytic}%
  \BibitemOpen
  \bibfield  {author} {\bibinfo {author} {\bibfnamefont {E.}~\bibnamefont
  {Witten}}.\EOS\ }
\newblock \bibinfo {title} {Analytic {{Continuation Of Chern-Simons
  Theory}}}.\EOS\
\newblock
  \href{https://arxiv.org/abs/1001.2933}{arXiv:\allowbreak{}1001.\allowbreak{}2933}
  (\bibinfo {year} {2010})\BibitemShut {NoStop}%
\bibitem [{\citenamefont {Feldbrugge}\ and\ \citenamefont
  {Turok}(2023)}]{feldbrugge2023existence}%
  \BibitemOpen
  \bibfield  {author} {\bibinfo {author} {\bibfnamefont {J.}~\bibnamefont
  {Feldbrugge}}\ and\ \bibinfo {author} {\bibfnamefont {N.}~\bibnamefont
  {Turok}},\ }\bibfield  {title} {\bibinfo {title} {Existence of real time
  quantum path integrals},\ }\href {https://doi.org/10.1016/j.aop.2023.169315}
  {\bibfield  {journal} {\bibinfo  {journal} {\emph {Annals Phys.}}\ }\textbf
  {\bibinfo {volume} {454}},\ \bibinfo {pages} {p.~169315} (\bibinfo {year}
  {2023})}\BibitemShut {NoStop}%
\bibitem [{\citenamefont {Raz}\ \emph {et~al.}(2012)\citenamefont {Raz},
  \citenamefont {Pedatzur}, \citenamefont {Bruner},\ and\ \citenamefont
  {Dudovich}}]{raz2012spectral}%
  \BibitemOpen
  \bibfield  {author} {\bibinfo {author} {\bibfnamefont {O.}~\bibnamefont
  {Raz}}, \bibinfo {author} {\bibfnamefont {O.}~\bibnamefont {Pedatzur}},
  \bibinfo {author} {\bibfnamefont {B.~D.}\ \bibnamefont {Bruner}},\ and\
  \bibinfo {author} {\bibfnamefont {N.}~\bibnamefont {Dudovich}},\ }\bibfield
  {title} {\bibinfo {title} {Spectral caustics in attosecond science},\ }\href
  {https://doi.org/10.1038/nphoton.2011.353} {\bibfield  {journal} {\bibinfo
  {journal} {\emph {Nat. Photonics}}\ }\textbf {\bibinfo {volume} {6}}\bibfield
   {number} {\bibinfo  {number} { no.~3},\ \bibinfo {pages} {pp.~170--173}}
  (\bibinfo {year} {2012})}\BibitemShut {NoStop}%
\bibitem [{\citenamefont {Birulia}\ and\ \citenamefont
  {Strelkov}(2019)}]{birulia2019spectral}%
  \BibitemOpen
  \bibfield  {author} {\bibinfo {author} {\bibfnamefont {V.~A.}\ \bibnamefont
  {Birulia}}\ and\ \bibinfo {author} {\bibfnamefont {V.~V.}\ \bibnamefont
  {Strelkov}},\ }\bibfield  {title} {\bibinfo {title} {Spectral caustic in
  two-color high-order harmonic generation: {{Role}} of {{Coulomb}} effects},\
  }\href {https://doi.org/10.1103/PhysRevA.99.043413} {\bibfield  {journal}
  {\bibinfo  {journal} {\emph {Phys. Rev. A}}\ }\textbf {\bibinfo {volume}
  {99}}\bibfield  {number} {\bibinfo  {number} { no.~4},\ \bibinfo {pages}
  {p.~043413}} (\bibinfo {year} {2019})}\BibitemShut {NoStop}%
\bibitem [{\citenamefont {Raab}\ \emph {et~al.}(2025)\citenamefont {Raab},
  \citenamefont {Redon}, \citenamefont {Abbing}, \citenamefont {Fang},
  \citenamefont {Guo}, \citenamefont {Smorenburg}, \citenamefont {Mauritsson},
  \citenamefont {Viotti}, \citenamefont {L'Huillier},\ and\ \citenamefont
  {Arnold}}]{raab2025xuv}%
  \BibitemOpen
  \bibfield  {author} {\bibinfo {author} {\bibfnamefont {A.-K.}\ \bibnamefont
  {Raab}}, \bibinfo {author} {\bibfnamefont {M.}~\bibnamefont {Redon}},
  \bibinfo {author} {\bibfnamefont {S.~R.}\ \bibnamefont {Abbing}}, \emph
  {et~al.},\ }\bibfield  {title} {\bibinfo {title} {{{XUV}} yield optimization
  of two-color high-order harmonic generation in gases},\ }\bibfield  {journal}
  {\emph {\bibinfo  {journal} {Nanophotonics}}\ }\href
  {https://doi.org/10.1515/nanoph-2024-0579} {10.1515/nanoph-2024-0579}
  (\bibinfo {year} {2025})\BibitemShut {NoStop}%
\bibitem [{\citenamefont {Dong}\ \emph {et~al.}(2024)\citenamefont {Dong},
  \citenamefont {Xia},\ and\ \citenamefont {Liu}}]{dong2024caustic}%
  \BibitemOpen
  \bibfield  {author} {\bibinfo {author} {\bibfnamefont {F.}~\bibnamefont
  {Dong}}, \bibinfo {author} {\bibfnamefont {Q.}~\bibnamefont {Xia}},\ and\
  \bibinfo {author} {\bibfnamefont {J.}~\bibnamefont {Liu}},\ }\bibfield
  {title} {\bibinfo {title} {Caustic effects on high-order harmonic generation
  in graphene},\ }\href {https://doi.org/10.1103/PhysRevA.109.L041102}
  {\bibfield  {journal} {\bibinfo  {journal} {\emph {Phys. Rev. A}}\ }\textbf
  {\bibinfo {volume} {109}}\bibfield  {number} {\bibinfo  {number} { no.~4},\
  \bibinfo {pages} {p.~L041102}} (\bibinfo {year} {2024})}\BibitemShut
  {NoStop}%
\bibitem [{\citenamefont {Faccial{\`a}}\ \emph {et~al.}(2018)\citenamefont
  {Faccial{\`a}}, \citenamefont {Pabst}, \citenamefont {Bruner}, \citenamefont
  {Ciriolo}, \citenamefont {Devetta}, \citenamefont {Negro}, \citenamefont
  {Geetha}, \citenamefont {Pusala}, \citenamefont {Soifer}, \citenamefont
  {Dudovich}, \citenamefont {Stagira},\ and\ \citenamefont
  {Vozzi}}]{facciala2018highorder}%
  \BibitemOpen
  \bibfield  {author} {\bibinfo {author} {\bibfnamefont {D.}~\bibnamefont
  {Faccial{\`a}}}, \bibinfo {author} {\bibfnamefont {S.}~\bibnamefont {Pabst}},
  \bibinfo {author} {\bibfnamefont {B.~D.}\ \bibnamefont {Bruner}}, \emph
  {et~al.},\ }\bibfield  {title} {\bibinfo {title} {High-order harmonic
  generation spectroscopy by recolliding electron caustics},\ }\href
  {https://doi.org/10.1088/1361-6455/aac351} {\bibfield  {journal} {\bibinfo
  {journal} {\emph {J. Phys. B: At. Mol. Opt. Phys.}}\ }\textbf {\bibinfo
  {volume} {51}}\bibfield  {number} {\bibinfo  {number} { no.~13},\ \bibinfo
  {pages} {p.~134002}} (\bibinfo {year} {2018})}\BibitemShut {NoStop}%
\bibitem [{\citenamefont {Faccial{\`a}}\ \emph {et~al.}(2016)\citenamefont
  {Faccial{\`a}}, \citenamefont {Pabst}, \citenamefont {Bruner}, \citenamefont
  {Ciriolo}, \citenamefont {De~Silvestri}, \citenamefont {Devetta},
  \citenamefont {Negro}, \citenamefont {Soifer}, \citenamefont {Stagira},
  \citenamefont {Dudovich},\ and\ \citenamefont {Vozzi}}]{facciala2016probe}%
  \BibitemOpen
  \bibfield  {author} {\bibinfo {author} {\bibfnamefont {D.}~\bibnamefont
  {Faccial{\`a}}}, \bibinfo {author} {\bibfnamefont {S.}~\bibnamefont {Pabst}},
  \bibinfo {author} {\bibfnamefont {B.~D.}\ \bibnamefont {Bruner}}, \emph
  {et~al.},\ }\bibfield  {title} {\bibinfo {title} {Probe of {{Multielectron
  Dynamics}} in {{Xenon}} by {{Caustics}} in {{High-Order Harmonic
  Generation}}},\ }\href {https://doi.org/10.1103/PhysRevLett.117.093902}
  {\bibfield  {journal} {\bibinfo  {journal} {\emph {Phys. Rev. Letters}}\
  }\textbf {\bibinfo {volume} {117}}\bibfield  {number} {\bibinfo  {number} {
  no.~9},\ \bibinfo {pages} {p.~093902}} (\bibinfo {year} {2016})}\BibitemShut
  {NoStop}%
\bibitem [{\citenamefont {Corkum}(1993)}]{corkum1993plasma}%
  \BibitemOpen
  \bibfield  {author} {\bibinfo {author} {\bibfnamefont {P.~B.}\ \bibnamefont
  {Corkum}},\ }\bibfield  {title} {\bibinfo {title} {Plasma perspective on
  strong field multiphoton ionization},\ }\href
  {https://doi.org/10.1103/PhysRevLett.71.1994} {\bibfield  {journal} {\bibinfo
   {journal} {\emph {Phys. Rev. Letters}}\ }\textbf {\bibinfo {volume}
  {71}}\bibfield  {number} {\bibinfo  {number} { no.~13},\ \bibinfo {pages}
  {pp.~1994--1997}} (\bibinfo {year} {1993})}\BibitemShut {NoStop}%
\bibitem [{\citenamefont {Kulander}\ \emph {et~al.}(1993)\citenamefont
  {Kulander}, \citenamefont {Schafer},\ and\ \citenamefont
  {Krause}}]{kulander1993superintense}%
  \BibitemOpen
  \bibfield  {author} {\bibinfo {author} {\bibfnamefont {K.~C.}\ \bibnamefont
  {Kulander}}, \bibinfo {author} {\bibfnamefont {K.~J.}\ \bibnamefont
  {Schafer}},\ and\ \bibinfo {author} {\bibfnamefont {J.~L.}\ \bibnamefont
  {Krause}},\ }\bibfield  {title} {\bibinfo {title} {Super-intense laser-atom
  physics},\ }\href@noop {} {\bibfield  {journal} {\bibinfo  {journal} {\emph
  {NATO Advanced Science Institutes Series}}\ }\textbf {\bibinfo {volume}
  {316}} (\bibinfo {year} {1993})}\BibitemShut {NoStop}%
\bibitem [{\citenamefont {Jin}\ \emph {et~al.}(2014)\citenamefont {Jin},
  \citenamefont {Wang}, \citenamefont {Wei}, \citenamefont {Le},\ and\
  \citenamefont {Lin}}]{jin2014waveforms}%
  \BibitemOpen
  \bibfield  {author} {\bibinfo {author} {\bibfnamefont {C.}~\bibnamefont
  {Jin}}, \bibinfo {author} {\bibfnamefont {G.}~\bibnamefont {Wang}}, \bibinfo
  {author} {\bibfnamefont {H.}~\bibnamefont {Wei}}, \emph {et~al.},\ }\bibfield
   {title} {\bibinfo {title} {Waveforms for optimal sub-{{keV}} high-order
  harmonics with synthesized two- or three-colour laser fields},\ }\href
  {https://doi.org/10.1038/ncomms5003} {\bibfield  {journal} {\bibinfo
  {journal} {\emph {Nat. Commun.}}\ }\textbf {\bibinfo {volume} {5}}\bibfield
  {number} {\bibinfo  {number} { no.~1},\ \bibinfo {pages} {p.~4003}} (\bibinfo
  {year} {2014})}\BibitemShut {NoStop}%
\bibitem [{\citenamefont {Cirmi}\ \emph {et~al.}(2023)\citenamefont {Cirmi},
  \citenamefont {Mainz}, \citenamefont {{Silva-Toledo}}, \citenamefont
  {Scheiba}, \citenamefont {{\c C}ankaya}, \citenamefont {Kubullek},
  \citenamefont {Rossi},\ and\ \citenamefont {K{\"a}rtner}}]{cirmi2023optical}%
  \BibitemOpen
  \bibfield  {author} {\bibinfo {author} {\bibfnamefont {G.}~\bibnamefont
  {Cirmi}}, \bibinfo {author} {\bibfnamefont {R.~E.}\ \bibnamefont {Mainz}},
  \bibinfo {author} {\bibfnamefont {M.~A.}\ \bibnamefont {{Silva-Toledo}}},
  \emph {et~al.},\ }\bibfield  {title} {\bibinfo {title} {Optical {{Waveform
  Synthesis}} and {{Its Applications}}},\ }\href
  {https://doi.org/10.1002/lpor.202200588} {\bibfield  {journal} {\bibinfo
  {journal} {\emph {Laser \& Photonics Reviews}}\ }\textbf {\bibinfo {volume}
  {17}}\bibfield  {number} {\bibinfo  {number} { no.~4},\ \bibinfo {pages}
  {p.~2200588}} (\bibinfo {year} {2023})}\BibitemShut {NoStop}%
\bibitem [{\citenamefont {Mitra}\ \emph {et~al.}(2020)\citenamefont {Mitra},
  \citenamefont {Biswas}, \citenamefont {Sch{\"o}tz}, \citenamefont {Pisanty},
  \citenamefont {F{\"o}rg}, \citenamefont {Kavuri}, \citenamefont {Burger},
  \citenamefont {Okell}, \citenamefont {H{\"o}gner}, \citenamefont {Pupeza},
  \citenamefont {Pervak}, \citenamefont {Lewenstein}, \citenamefont {Wnuk},\
  and\ \citenamefont {Kling}}]{mitra2020suppression}%
  \BibitemOpen
  \bibfield  {author} {\bibinfo {author} {\bibfnamefont {S.}~\bibnamefont
  {Mitra}}, \bibinfo {author} {\bibfnamefont {S.}~\bibnamefont {Biswas}},
  \bibinfo {author} {\bibfnamefont {J.}~\bibnamefont {Sch{\"o}tz}}, \emph
  {et~al.},\ }\bibfield  {title} {\bibinfo {title} {Suppression of individual
  peaks in two-colour high harmonic generation},\ }\href
  {https://doi.org/10.1088/1361-6455/ab859c} {\bibfield  {journal} {\bibinfo
  {journal} {\emph {J. Phys. B: At. Mol. Opt. Phys.}}\ }\textbf {\bibinfo
  {volume} {53}}\bibfield  {number} {\bibinfo  {number} { no.~13},\ \bibinfo
  {pages} {p.~134004}} (\bibinfo {year} {2020})}\BibitemShut {NoStop}%
\bibitem [{\citenamefont {Mansten}\ \emph {et~al.}(2008)\citenamefont
  {Mansten}, \citenamefont {Dahlstr{\"o}m}, \citenamefont {Johnsson},
  \citenamefont {Swoboda}, \citenamefont {L'Huillier},\ and\ \citenamefont
  {Mauritsson}}]{mansten2008spectral}%
  \BibitemOpen
  \bibfield  {author} {\bibinfo {author} {\bibfnamefont {E.}~\bibnamefont
  {Mansten}}, \bibinfo {author} {\bibfnamefont {J.~M.}\ \bibnamefont
  {Dahlstr{\"o}m}}, \bibinfo {author} {\bibfnamefont {P.}~\bibnamefont
  {Johnsson}}, \emph {et~al.},\ }\bibfield  {title} {\bibinfo {title} {Spectral
  shaping of attosecond pulses using two-colour laser fields},\ }\href
  {https://doi.org/10.1088/1367-2630/10/8/083041} {\bibfield  {journal}
  {\bibinfo  {journal} {\emph {New Journal of Physics}}\ }\textbf {\bibinfo
  {volume} {10}}\bibfield  {number} {\bibinfo  {number} { no.~8},\ \bibinfo
  {pages} {p.~083041}} (\bibinfo {year} {2008})}\BibitemShut {NoStop}%
\bibitem [{\citenamefont {Chipperfield}\ \emph {et~al.}(2009)\citenamefont
  {Chipperfield}, \citenamefont {Robinson}, \citenamefont {Tisch},\ and\
  \citenamefont {Marangos}}]{chipperfield2009ideal}%
  \BibitemOpen
  \bibfield  {author} {\bibinfo {author} {\bibfnamefont {L.~E.}\ \bibnamefont
  {Chipperfield}}, \bibinfo {author} {\bibfnamefont {J.~S.}\ \bibnamefont
  {Robinson}}, \bibinfo {author} {\bibfnamefont {J.~W.~G.}\ \bibnamefont
  {Tisch}},\ and\ \bibinfo {author} {\bibfnamefont {J.~P.}\ \bibnamefont
  {Marangos}},\ }\bibfield  {title} {\bibinfo {title} {Ideal {{Waveform}} to
  {{Generate}} the {{Maximum Possible Electron Recollision Energy}} for {{Any
  Given Oscillation Period}}},\ }\href
  {https://doi.org/10.1103/PhysRevLett.102.063003} {\bibfield  {journal}
  {\bibinfo  {journal} {\emph {Phys. Rev. Letters}}\ }\textbf {\bibinfo
  {volume} {102}}\bibfield  {number} {\bibinfo  {number} { no.~6},\ \bibinfo
  {pages} {p.~063003}} (\bibinfo {year} {2009})}\BibitemShut {NoStop}%
\bibitem [{\citenamefont {Kneller}\ \emph {et~al.}(2022)\citenamefont
  {Kneller}, \citenamefont {Azoury}, \citenamefont {Federman}, \citenamefont
  {Kr{\"u}ger}, \citenamefont {Uzan}, \citenamefont {Orenstein}, \citenamefont
  {Bruner}, \citenamefont {Smirnova}, \citenamefont {Patchkovskii},
  \citenamefont {Ivanov},\ and\ \citenamefont {Dudovich}}]{kneller2022look}%
  \BibitemOpen
  \bibfield  {author} {\bibinfo {author} {\bibfnamefont {O.}~\bibnamefont
  {Kneller}}, \bibinfo {author} {\bibfnamefont {D.}~\bibnamefont {Azoury}},
  \bibinfo {author} {\bibfnamefont {Y.}~\bibnamefont {Federman}}, \emph
  {et~al.},\ }\bibfield  {title} {\bibinfo {title} {A look under the tunnelling
  barrier via attosecond-gated interferometry},\ }\href
  {https://doi.org/10.1038/s41566-022-00955-7} {\bibfield  {journal} {\bibinfo
  {journal} {\emph {Nat. Photonics}}\ }\textbf {\bibinfo {volume}
  {16}}\bibfield  {number} {\bibinfo  {number} { no.~4},\ \bibinfo {pages}
  {pp.~304--310}} (\bibinfo {year} {2022})}\BibitemShut {NoStop}%
\bibitem [{\citenamefont {He}\ \emph {et~al.}(2010)\citenamefont {He},
  \citenamefont {Dahlstr{\"o}m}, \citenamefont {Rakowski}, \citenamefont
  {Heyl}, \citenamefont {Persson}, \citenamefont {Mauritsson},\ and\
  \citenamefont {L'Huillier}}]{he2010interference}%
  \BibitemOpen
  \bibfield  {author} {\bibinfo {author} {\bibfnamefont {X.}~\bibnamefont
  {He}}, \bibinfo {author} {\bibfnamefont {J.~M.}\ \bibnamefont
  {Dahlstr{\"o}m}}, \bibinfo {author} {\bibfnamefont {R.}~\bibnamefont
  {Rakowski}}, \emph {et~al.},\ }\bibfield  {title} {\bibinfo {title}
  {Interference effects in two-color high-order harmonic generation},\ }\href
  {https://doi.org/10.1103/PhysRevA.82.033410} {\bibfield  {journal} {\bibinfo
  {journal} {\emph {Phys. Rev. A}}\ }\textbf {\bibinfo {volume} {82}}\bibfield
  {number} {\bibinfo  {number} { no.~3},\ \bibinfo {pages} {p.~033410}}
  (\bibinfo {year} {2010})}\BibitemShut {NoStop}%
\bibitem [{\citenamefont {Shafir}\ \emph {et~al.}(2012)\citenamefont {Shafir},
  \citenamefont {Soifer}, \citenamefont {Bruner}, \citenamefont {Dagan},
  \citenamefont {Mairesse}, \citenamefont {Patchkovskii}, \citenamefont
  {Ivanov}, \citenamefont {Smirnova},\ and\ \citenamefont
  {Dudovich}}]{shafir2012resolving}%
  \BibitemOpen
  \bibfield  {author} {\bibinfo {author} {\bibfnamefont {D.}~\bibnamefont
  {Shafir}}, \bibinfo {author} {\bibfnamefont {H.}~\bibnamefont {Soifer}},
  \bibinfo {author} {\bibfnamefont {B.~D.}\ \bibnamefont {Bruner}}, \emph
  {et~al.},\ }\bibfield  {title} {\bibinfo {title} {Resolving the time when an
  electron exits a tunnelling barrier},\ }\href
  {https://doi.org/10.1038/nature11025} {\bibfield  {journal} {\bibinfo
  {journal} {\emph {Nature}}\ }\textbf {\bibinfo {volume} {485}}\bibfield
  {number} {\bibinfo  {number} { no.~7398},\ \bibinfo {pages} {pp.~343--346}}
  (\bibinfo {year} {2012})}\BibitemShut {NoStop}%
\bibitem [{\citenamefont {Dudovich}\ \emph {et~al.}(2006)\citenamefont
  {Dudovich}, \citenamefont {Smirnova}, \citenamefont {Levesque}, \citenamefont
  {Mairesse}, \citenamefont {Ivanov}, \citenamefont {Villeneuve},\ and\
  \citenamefont {Corkum}}]{dudovich2006measuring}%
  \BibitemOpen
  \bibfield  {author} {\bibinfo {author} {\bibfnamefont {N.}~\bibnamefont
  {Dudovich}}, \bibinfo {author} {\bibfnamefont {O.}~\bibnamefont {Smirnova}},
  \bibinfo {author} {\bibfnamefont {J.}~\bibnamefont {Levesque}}, \emph
  {et~al.},\ }\bibfield  {title} {\bibinfo {title} {Measuring and controlling
  the birth of attosecond {{XUV}} pulses},\ }\href
  {https://doi.org/10.1038/nphys434} {\bibfield  {journal} {\bibinfo  {journal}
  {\emph {Nat. Phys.}}\ }\textbf {\bibinfo {volume} {2}}\bibfield  {number}
  {\bibinfo  {number} { no.~11},\ \bibinfo {pages} {pp.~781--786}} (\bibinfo
  {year} {2006})}\BibitemShut {NoStop}%
\bibitem [{\citenamefont {Zhao}\ and\ \citenamefont
  {Lein}(2013)}]{zhao2013determination}%
  \BibitemOpen
  \bibfield  {author} {\bibinfo {author} {\bibfnamefont {J.}~\bibnamefont
  {Zhao}}\ and\ \bibinfo {author} {\bibfnamefont {M.}~\bibnamefont {Lein}},\
  }\bibfield  {title} {\bibinfo {title} {Determination of {{Ionization}} and
  {{Tunneling Times}} in {{High-Order Harmonic Generation}}},\ }\href
  {https://doi.org/10.1103/PhysRevLett.111.043901} {\bibfield  {journal}
  {\bibinfo  {journal} {\emph {Phys. Rev. Letters}}\ }\textbf {\bibinfo
  {volume} {111}}\bibfield  {number} {\bibinfo  {number} { no.~4},\ \bibinfo
  {pages} {p.~043901}} (\bibinfo {year} {2013})}\BibitemShut {NoStop}%
\bibitem [{\citenamefont {Eicke}\ and\ \citenamefont
  {Lein}(2019)}]{eicke2019Attoclock}%
  \BibitemOpen
  \bibfield  {author} {\bibinfo {author} {\bibfnamefont {N.}~\bibnamefont
  {Eicke}}\ and\ \bibinfo {author} {\bibfnamefont {M.}~\bibnamefont {Lein}},\
  }\bibfield  {title} {\bibinfo {title} {Attoclock with counter-rotating
  bicircular laser fields},\ }\href
  {https://doi.org/10.1103/PhysRevA.99.031402} {\bibfield  {journal} {\bibinfo
  {journal} {\emph {Phys. Rev. A}}\ }\textbf {\bibinfo {volume} {99}}\bibfield
  {number} {\bibinfo  {number} { no.~3},\ \bibinfo {pages} {p.~031402}}
  (\bibinfo {year} {2019})}\BibitemShut {NoStop}%
\bibitem [{\citenamefont {Mauritsson}\ \emph {et~al.}(2009)\citenamefont
  {Mauritsson}, \citenamefont {Dahlstr{\"o}m}, \citenamefont {Mansten},\ and\
  \citenamefont {Fordell}}]{mauritsson2009subcycle}%
  \BibitemOpen
  \bibfield  {author} {\bibinfo {author} {\bibfnamefont {J.}~\bibnamefont
  {Mauritsson}}, \bibinfo {author} {\bibfnamefont {J.~M.}\ \bibnamefont
  {Dahlstr{\"o}m}}, \bibinfo {author} {\bibfnamefont {E.}~\bibnamefont
  {Mansten}},\ and\ \bibinfo {author} {\bibfnamefont {T.}~\bibnamefont
  {Fordell}},\ }\bibfield  {title} {\bibinfo {title} {Sub-cycle control of
  attosecond pulse generation using two-colour laser fields},\ }\href
  {https://doi.org/10.1088/0953-4075/42/13/134003} {\bibfield  {journal}
  {\bibinfo  {journal} {\emph {J. Phys. B: At. Mol. Opt. Phys.}}\ }\textbf
  {\bibinfo {volume} {42}}\bibfield  {number} {\bibinfo  {number} { no.~13},\
  \bibinfo {pages} {p.~134003}} (\bibinfo {year} {2009})}\BibitemShut {NoStop}%
\bibitem [{\citenamefont {Ruiz}\ \emph {et~al.}(2009)\citenamefont {Ruiz},
  \citenamefont {Hoffmann}, \citenamefont {Torres}, \citenamefont
  {Chipperfield},\ and\ \citenamefont {Marangos}}]{ruiz2009control}%
  \BibitemOpen
  \bibfield  {author} {\bibinfo {author} {\bibfnamefont {C.}~\bibnamefont
  {Ruiz}}, \bibinfo {author} {\bibfnamefont {D.~J.}\ \bibnamefont {Hoffmann}},
  \bibinfo {author} {\bibfnamefont {R.}~\bibnamefont {Torres}}, \emph
  {et~al.},\ }\bibfield  {title} {\bibinfo {title} {Control of the polarization
  of attosecond pulses using a two-color field},\ }\href
  {https://doi.org/10.1088/1367-2630/11/11/113045} {\bibfield  {journal}
  {\bibinfo  {journal} {\emph {New Journal of Physics}}\ }\textbf {\bibinfo
  {volume} {11}}\bibfield  {number} {\bibinfo  {number} { no.~11},\ \bibinfo
  {pages} {p.~113045}} (\bibinfo {year} {2009})}\BibitemShut {NoStop}%
\bibitem [{\citenamefont {Roscam~Abbing}\ \emph
  {et~al.}(2020{\natexlab{a}})\citenamefont {Roscam~Abbing}, \citenamefont
  {Campi}, \citenamefont {Sajjadian}, \citenamefont {Lin}, \citenamefont
  {Smorenburg},\ and\ \citenamefont {Kraus}}]{roscamabbing2020divergence}%
  \BibitemOpen
  \bibfield  {author} {\bibinfo {author} {\bibfnamefont {S.}~\bibnamefont
  {Roscam~Abbing}}, \bibinfo {author} {\bibfnamefont {F.}~\bibnamefont
  {Campi}}, \bibinfo {author} {\bibfnamefont {F.~S.}\ \bibnamefont
  {Sajjadian}}, \emph {et~al.},\ }\bibfield  {title} {\bibinfo {title}
  {Divergence {{Control}} of {{High-Harmonic Generation}}},\ }\href
  {https://doi.org/10.1103/PhysRevApplied.13.054029} {\bibfield  {journal}
  {\bibinfo  {journal} {\emph {Phys. Rev. Applied}}\ }\textbf {\bibinfo
  {volume} {13}}\bibfield  {number} {\bibinfo  {number} { no.~5},\ \bibinfo
  {pages} {p.~054029}} (\bibinfo {year} {2020}{\natexlab{a}})}\BibitemShut
  {NoStop}%
\bibitem [{\citenamefont {Haessler}\ \emph {et~al.}(2014)\citenamefont
  {Haessler}, \citenamefont {Bal{\v c}iunas}, \citenamefont {Fan},
  \citenamefont {Andriukaitis}, \citenamefont {Pug{\v z}lys}, \citenamefont
  {Baltu{\v s}ka}, \citenamefont {Witting}, \citenamefont {Squibb},
  \citenamefont {Za{\"i}r}, \citenamefont {Tisch}, \citenamefont {Marangos},\
  and\ \citenamefont {Chipperfield}}]{haessler2014optimization}%
  \BibitemOpen
  \bibfield  {author} {\bibinfo {author} {\bibfnamefont {S.}~\bibnamefont
  {Haessler}}, \bibinfo {author} {\bibfnamefont {T.}~\bibnamefont {Bal{\v
  c}iunas}}, \bibinfo {author} {\bibfnamefont {G.}~\bibnamefont {Fan}}, \emph
  {et~al.},\ }\bibfield  {title} {\bibinfo {title} {Optimization of {{Quantum
  Trajectories Driven}} by {{Strong-Field Waveforms}}},\ }\href
  {https://doi.org/10.1103/PhysRevX.4.021028} {\bibfield  {journal} {\bibinfo
  {journal} {\emph {Phys. Rev. X}}\ }\textbf {\bibinfo {volume} {4}}\bibfield
  {number} {\bibinfo  {number} { no.~2},\ \bibinfo {pages} {p.~021028}}
  (\bibinfo {year} {2014})}\BibitemShut {NoStop}%
\bibitem [{\citenamefont {Baykusheva}\ and\ \citenamefont
  {W{\"o}rner}(2018)}]{baykusheva2018chiral}%
  \BibitemOpen
  \bibfield  {author} {\bibinfo {author} {\bibfnamefont {D.}~\bibnamefont
  {Baykusheva}}\ and\ \bibinfo {author} {\bibfnamefont {H.~J.}\ \bibnamefont
  {W{\"o}rner}},\ }\bibfield  {title} {\bibinfo {title} {Chiral
  {{Discrimination}} through {{Bielliptical High-Harmonic Spectroscopy}}},\
  }\href {https://doi.org/10.1103/PhysRevX.8.031060} {\bibfield  {journal}
  {\bibinfo  {journal} {\emph {Phys. Rev. X}}\ }\textbf {\bibinfo {volume}
  {8}}\bibfield  {number} {\bibinfo  {number} { no.~3},\ \bibinfo {pages}
  {p.~031060}} (\bibinfo {year} {2018})}\BibitemShut {NoStop}%
\bibitem [{\citenamefont {Ayuso}\ \emph {et~al.}(2019)\citenamefont {Ayuso},
  \citenamefont {Neufeld}, \citenamefont {Ordonez}, \citenamefont {Decleva},
  \citenamefont {Lerner}, \citenamefont {Cohen}, \citenamefont {Ivanov},\ and\
  \citenamefont {Smirnova}}]{ayuso2019synthetic}%
  \BibitemOpen
  \bibfield  {author} {\bibinfo {author} {\bibfnamefont {D.}~\bibnamefont
  {Ayuso}}, \bibinfo {author} {\bibfnamefont {O.}~\bibnamefont {Neufeld}},
  \bibinfo {author} {\bibfnamefont {A.~F.}\ \bibnamefont {Ordonez}}, \emph
  {et~al.},\ }\bibfield  {title} {\bibinfo {title} {Synthetic chiral light for
  efficient control of chiral light--matter interaction},\ }\href
  {https://doi.org/10.1038/s41566-019-0531-2} {\bibfield  {journal} {\bibinfo
  {journal} {\emph {Nat. Photonics}}\ }\textbf {\bibinfo {volume}
  {13}}\bibfield  {number} {\bibinfo  {number} { no.~12},\ \bibinfo {pages}
  {pp.~866--871}} (\bibinfo {year} {2019})}\BibitemShut {NoStop}%
\bibitem [{\citenamefont {Feldbrugge}\ \emph
  {et~al.}(2017{\natexlab{a}})\citenamefont {Feldbrugge}, \citenamefont
  {Lehners},\ and\ \citenamefont {Turok}}]{feldbrugge2017lorentzian}%
  \BibitemOpen
  \bibfield  {author} {\bibinfo {author} {\bibfnamefont {J.}~\bibnamefont
  {Feldbrugge}}, \bibinfo {author} {\bibfnamefont {J.-L.}\ \bibnamefont
  {Lehners}},\ and\ \bibinfo {author} {\bibfnamefont {N.}~\bibnamefont
  {Turok}},\ }\bibfield  {title} {\bibinfo {title} {Lorentzian quantum
  cosmology},\ }\href {https://doi.org/10.1103/PhysRevD.95.103508} {\bibfield
  {journal} {\bibinfo  {journal} {\emph {Phys. Rev. D}}\ }\textbf {\bibinfo
  {volume} {95}}\bibfield  {number} {\bibinfo  {number} { no.~10},\ \bibinfo
  {pages} {p.~103508}} (\bibinfo {year} {2017}{\natexlab{a}})}\BibitemShut
  {NoStop}%
\bibitem [{\citenamefont {Tanizaki}\ and\ \citenamefont
  {Koike}(2014)}]{tanizaki2014realtime}%
  \BibitemOpen
  \bibfield  {author} {\bibinfo {author} {\bibfnamefont {Y.}~\bibnamefont
  {Tanizaki}}\ and\ \bibinfo {author} {\bibfnamefont {T.}~\bibnamefont
  {Koike}},\ }\bibfield  {title} {\bibinfo {title} {Real-time {{Feynman}} path
  integral with {{Picard}}--{{Lefschetz}} theory and its applications to
  quantum tunneling},\ }\href {https://doi.org/10.1016/j.aop.2014.09.003}
  {\bibfield  {journal} {\bibinfo  {journal} {\emph {Annals of Physics}}\
  }\textbf {\bibinfo {volume} {351}},\ \bibinfo {pages} {pp.~250--274}
  (\bibinfo {year} {2014})}\BibitemShut {NoStop}%
\bibitem [{\citenamefont {Bharathkumar}\ and\ \citenamefont
  {Joseph}(2020)}]{bharathkumar2020lefschetz}%
  \BibitemOpen
  \bibfield  {author} {\bibinfo {author} {\bibfnamefont {R.}~\bibnamefont
  {Bharathkumar}}\ and\ \bibinfo {author} {\bibfnamefont {A.}~\bibnamefont
  {Joseph}},\ }\bibfield  {title} {\bibinfo {title} {Lefschetz thimbles and
  quantum phases in zero-dimensional bosonic models},\ }\href
  {https://doi.org/10.1140/epjc/s10052-020-08493-8} {\bibfield  {journal}
  {\bibinfo  {journal} {\emph {The European Physical Journal C}}\ }\textbf
  {\bibinfo {volume} {80}}\bibfield  {number} {\bibinfo  {number} { no.~10},\
  \bibinfo {pages} {p.~923}} (\bibinfo {year} {2020})}\BibitemShut {NoStop}%
\bibitem [{\citenamefont {Feldbrugge}\ \emph {et~al.}(2023)\citenamefont
  {Feldbrugge}, \citenamefont {Pen},\ and\ \citenamefont
  {Turok}}]{feldbrugge2023oscillatory}%
  \BibitemOpen
  \bibfield  {author} {\bibinfo {author} {\bibfnamefont {J.}~\bibnamefont
  {Feldbrugge}}, \bibinfo {author} {\bibfnamefont {U.-L.}\ \bibnamefont
  {Pen}},\ and\ \bibinfo {author} {\bibfnamefont {N.}~\bibnamefont {Turok}},\
  }\bibfield  {title} {\bibinfo {title} {Oscillatory path integrals for radio
  astronomy},\ }\href {https://doi.org/10.1016/j.aop.2023.169255} {\bibfield
  {journal} {\bibinfo  {journal} {\emph {Annals of Physics}}\ }\textbf
  {\bibinfo {volume} {451}},\ \bibinfo {pages} {p.~169255} (\bibinfo {year}
  {2023})}\BibitemShut {NoStop}%
\bibitem [{\citenamefont {Feldbrugge}\ and\ \citenamefont
  {Jones}(2025)}]{feldbrugge2025efficient}%
  \BibitemOpen
  \bibfield  {author} {\bibinfo {author} {\bibfnamefont {J.}~\bibnamefont
  {Feldbrugge}}\ and\ \bibinfo {author} {\bibfnamefont {J.~Y.~L.}\ \bibnamefont
  {Jones}},\ }\bibfield  {title} {\bibinfo {title} {Efficient evaluation of
  real-time path integrals},\ }\href
  {https://doi.org/10.1103/PhysRevD.111.083524} {\bibfield  {journal} {\bibinfo
   {journal} {\emph {Phys. Rev. D}}\ }\textbf {\bibinfo {volume}
  {111}}\bibfield  {number} {\bibinfo  {number} { no.~8},\ \bibinfo {pages}
  {p.~083524}} (\bibinfo {year} {2025})}\BibitemShut {NoStop}%
\bibitem [{\citenamefont {Weber}\ \emph {et~al.}(2025)\citenamefont {Weber},
  \citenamefont {Khokhlova},\ and\ \citenamefont {Pisanty}}]{weber2025quantum}%
  \BibitemOpen
  \bibfield  {author} {\bibinfo {author} {\bibfnamefont {A.}~\bibnamefont
  {Weber}}, \bibinfo {author} {\bibfnamefont {M.}~\bibnamefont {Khokhlova}},\
  and\ \bibinfo {author} {\bibfnamefont {E.}~\bibnamefont {Pisanty}},\
  }\bibfield  {title} {\bibinfo {title} {Quantum tunneling without a barrier},\
  }\href {https://doi.org/10.1103/PhysRevA.111.043103} {\bibfield  {journal}
  {\bibinfo  {journal} {\emph {Phys. Rev. A}}\ }\textbf {\bibinfo {volume}
  {111}}\bibfield  {number} {\bibinfo  {number} { no.~4},\ \bibinfo {pages}
  {p.~043103}} (\bibinfo {year} {2025})}\BibitemShut {NoStop}%
\bibitem [{\citenamefont {Weber}(2025)}]{figuremaker-code}%
  \BibitemOpen
  \bibfield  {author} {\bibinfo {author} {\bibfnamefont {A.}~\bibnamefont
  {Weber}},\ }\href {https://doi.org/10.5281/zenodo.17298777} {\bibinfo {title}
  {Figure-maker notebooks and code for {Picard- Lefschetz} methods for {ATI}
  and {HHG} integrals}},\ \bibinfo {howpublished} {Zenodo,
  \href{https://doi.org/10.5281/zenodo.17298777}{doi:\allowbreak{}10.5281/\allowbreak{}zenodo.\allowbreak{}17298777}}
  (\bibinfo {year} {2025})\BibitemShut {NoStop}%
\bibitem [{\citenamefont {Le}\ \emph {et~al.}(2016)\citenamefont {Le},
  \citenamefont {Wei}, \citenamefont {Jin},\ and\ \citenamefont
  {Lin}}]{le2016strongfield}%
  \BibitemOpen
  \bibfield  {author} {\bibinfo {author} {\bibfnamefont {A.-T.}\ \bibnamefont
  {Le}}, \bibinfo {author} {\bibfnamefont {H.}~\bibnamefont {Wei}}, \bibinfo
  {author} {\bibfnamefont {C.}~\bibnamefont {Jin}},\ and\ \bibinfo {author}
  {\bibfnamefont {C.~D.}\ \bibnamefont {Lin}},\ }\bibfield  {title} {\bibinfo
  {title} {Strong-field approximation and its extension for high-order harmonic
  generation with mid-infrared lasers},\ }\href
  {https://doi.org/10.1088/0953-4075/49/5/053001} {\bibfield  {journal}
  {\bibinfo  {journal} {\emph {J. Phys. B: At. Mol. Opt. Phys.}}\ }\textbf
  {\bibinfo {volume} {49}}\bibfield  {number} {\bibinfo  {number} { no.~5},\
  \bibinfo {pages} {p.~053001}} (\bibinfo {year} {2016})}\BibitemShut {NoStop}%
\bibitem [{\citenamefont {Popruzhenko}(2014)}]{popruzhenko2014keldysh}%
  \BibitemOpen
  \bibfield  {author} {\bibinfo {author} {\bibfnamefont {S.~V.}\ \bibnamefont
  {Popruzhenko}},\ }\bibfield  {title} {\bibinfo {title} {Keldysh theory of
  strong field ionization: History, applications, difficulties and
  perspectives},\ }\href {https://doi.org/10.1088/0953-4075/47/20/204001}
  {\bibfield  {journal} {\bibinfo  {journal} {\emph {J. Phys. B: At. Mol. Opt.
  Phys.}}\ }\textbf {\bibinfo {volume} {47}}\bibfield  {number} {\bibinfo
  {number} { no.~20},\ \bibinfo {pages} {p.~204001}} (\bibinfo {year}
  {2014})}\BibitemShut {NoStop}%
\bibitem [{\citenamefont {Smirnova}\ and\ \citenamefont
  {Ivanov}(2013)}]{smirnova2013multielectron}%
  \BibitemOpen
  \bibfield  {author} {\bibinfo {author} {\bibfnamefont {O.}~\bibnamefont
  {Smirnova}}\ and\ \bibinfo {author} {\bibfnamefont {M.}~\bibnamefont
  {Ivanov}}.\EOS\ }
\newblock \bibinfo {title} {Multielectron {{High Harmonic Generation}}: Simple
  man on a complex plane}.\EOS\
\newblock
  \href{https://arxiv.org/abs/1304.2413}{arXiv:\allowbreak{}1304.\allowbreak{}2413}
  (\bibinfo {year} {2013})\BibitemShut {NoStop}%
\bibitem [{\citenamefont {Amini}\ \emph {et~al.}(2019)\citenamefont {Amini},
  \citenamefont {Biegert}, \citenamefont {Calegari}, \citenamefont
  {Chac{\'o}n}, \citenamefont {Ciappina}, \citenamefont {Dauphin},
  \citenamefont {Efimov}, \citenamefont {Figueira De Morisson~Faria},
  \citenamefont {Giergiel}, \citenamefont {Gniewek}, \citenamefont {Landsman},
  \citenamefont {Lesiuk}, \citenamefont {Mandrysz}, \citenamefont {Maxwell},
  \citenamefont {Moszy{\'n}ski}, \citenamefont {Ortmann}, \citenamefont
  {{Antonio P{\'e}rez-Hern{\'a}ndez}}, \citenamefont {Pic{\'o}n}, \citenamefont
  {Pisanty}, \citenamefont {{Prauzner-Bechcicki}}, \citenamefont {Sacha},
  \citenamefont {Su{\'a}rez}, \citenamefont {Za{\"i}r}, \citenamefont
  {Zakrzewski},\ and\ \citenamefont {Lewenstein}}]{amini2019symphony}%
  \BibitemOpen
  \bibfield  {author} {\bibinfo {author} {\bibfnamefont {K.}~\bibnamefont
  {Amini}}, \bibinfo {author} {\bibfnamefont {J.}~\bibnamefont {Biegert}},
  \bibinfo {author} {\bibfnamefont {F.}~\bibnamefont {Calegari}}, \emph
  {et~al.},\ }\bibfield  {title} {\bibinfo {title} {Symphony on strong field
  approximation},\ }\href {https://doi.org/10.1088/1361-6633/ab2bb1} {\bibfield
   {journal} {\bibinfo  {journal} {\emph {Reports on Progress in Physics}}\
  }\textbf {\bibinfo {volume} {82}}\bibfield  {number} {\bibinfo  {number} {
  no.~11},\ \bibinfo {pages} {p.~116001}} (\bibinfo {year} {2019})}\BibitemShut
  {NoStop}%
\bibitem [{\citenamefont {Bleistein}\ and\ \citenamefont
  {Handelsman}(1975)}]{bleistein1975asymptotic}%
  \BibitemOpen
  \bibfield  {author} {\bibinfo {author} {\bibfnamefont {N.}~\bibnamefont
  {Bleistein}}\ and\ \bibinfo {author} {\bibfnamefont {R.~A.}\ \bibnamefont
  {Handelsman}},\ }\href@noop {} {\emph {\bibinfo {title} {Asymptotic
  Expansions of Integrals}}}\ (\bibinfo  {publisher} {Ardent Media},\ \bibinfo
  {year} {1975})\BibitemShut {NoStop}%
\bibitem [{\citenamefont {Wong}(2020)}]{wong2020asymptotic}%
  \BibitemOpen
  \bibfield  {author} {\bibinfo {author} {\bibfnamefont {R.}~\bibnamefont
  {Wong}},\ }\href {https://doi.org/10.1201/9781003072584} {\emph {\bibinfo
  {title} {Asymptotic and {{Computational Analysis}}: {{Conference}} in
  {{Honor}} of {{Frank W}}.j. {{Olver}}'s 65th {{Birthday}}}}}\ (\bibinfo
  {publisher} {CRC Press},\ \bibinfo {address} {Boca Raton},\ \bibinfo {year}
  {2020})\BibitemShut {NoStop}%
\bibitem [{\citenamefont {Ja{\v s}arevi{\'c}}\ \emph
  {et~al.}(2020)\citenamefont {Ja{\v s}arevi{\'c}}, \citenamefont
  {Hasovi{\'c}}, \citenamefont {Kopold}, \citenamefont {Becker},\ and\
  \citenamefont {Milo{\v s}evi{\'c}}}]{jasarevic2020application}%
  \BibitemOpen
  \bibfield  {author} {\bibinfo {author} {\bibfnamefont {A.}~\bibnamefont
  {Ja{\v s}arevi{\'c}}}, \bibinfo {author} {\bibfnamefont {E.}~\bibnamefont
  {Hasovi{\'c}}}, \bibinfo {author} {\bibfnamefont {R.}~\bibnamefont {Kopold}},
  \emph {et~al.},\ }\bibfield  {title} {\bibinfo {title} {Application of the
  saddle-point method to strong-laser-field ionization},\ }\href
  {https://doi.org/10.1088/1751-8121/ab749b} {\bibfield  {journal} {\bibinfo
  {journal} {\emph {Journal of Physics A: Mathematical and Theoretical}}\
  }\textbf {\bibinfo {volume} {53}}\bibfield  {number} {\bibinfo  {number} {
  no.~12},\ \bibinfo {pages} {p.~125201}} (\bibinfo {year} {2020})}\BibitemShut
  {NoStop}%
\bibitem [{\citenamefont {Milo{\v s}evi{\'c}}\ and\ \citenamefont
  {Becker}(2002)}]{milosevic2002role}%
  \BibitemOpen
  \bibfield  {author} {\bibinfo {author} {\bibfnamefont {D.~B.}\ \bibnamefont
  {Milo{\v s}evi{\'c}}}\ and\ \bibinfo {author} {\bibfnamefont
  {W.}~\bibnamefont {Becker}},\ }\bibfield  {title} {\bibinfo {title} {Role of
  long quantum orbits in high-order harmonic generation},\ }\href
  {https://doi.org/10.1103/PhysRevA.66.063417} {\bibfield  {journal} {\bibinfo
  {journal} {\emph {Phys. Rev. A}}\ }\textbf {\bibinfo {volume} {66}}\bibfield
  {number} {\bibinfo  {number} { no.~6},\ \bibinfo {pages} {p.~063417}}
  (\bibinfo {year} {2002})}\BibitemShut {NoStop}%
\bibitem [{\citenamefont {{Figueira de Morisson Faria}}\ \emph
  {et~al.}(2000)\citenamefont {{Figueira de Morisson Faria}}, \citenamefont
  {Milo{\v s}evi{\'c}},\ and\ \citenamefont
  {Paulus}}]{figueirademorissonfaria2000phasedependent}%
  \BibitemOpen
  \bibfield  {author} {\bibinfo {author} {\bibfnamefont {C.}~\bibnamefont
  {{Figueira de Morisson Faria}}}, \bibinfo {author} {\bibfnamefont {D.~B.}\
  \bibnamefont {Milo{\v s}evi{\'c}}},\ and\ \bibinfo {author} {\bibfnamefont
  {G.~G.}\ \bibnamefont {Paulus}},\ }\bibfield  {title} {\bibinfo {title}
  {Phase-dependent effects in bichromatic high-order harmonic generation},\
  }\href {https://doi.org/10.1103/PhysRevA.61.063415} {\bibfield  {journal}
  {\bibinfo  {journal} {\emph {Phys. Rev. A}}\ }\textbf {\bibinfo {volume}
  {61}}\bibfield  {number} {\bibinfo  {number} { no.~6},\ \bibinfo {pages}
  {p.~063415}} (\bibinfo {year} {2000})}\BibitemShut {NoStop}%
\bibitem [{\citenamefont {{Figueira de Morisson Faria}}\ \emph
  {et~al.}(2002)\citenamefont {{Figueira de Morisson Faria}}, \citenamefont
  {Schomerus},\ and\ \citenamefont
  {Becker}}]{figueirademorissonfaria2002highorder}%
  \BibitemOpen
  \bibfield  {author} {\bibinfo {author} {\bibfnamefont {C.}~\bibnamefont
  {{Figueira de Morisson Faria}}}, \bibinfo {author} {\bibfnamefont
  {H.}~\bibnamefont {Schomerus}},\ and\ \bibinfo {author} {\bibfnamefont
  {W.}~\bibnamefont {Becker}},\ }\bibfield  {title} {\bibinfo {title}
  {High-order above-threshold ionization: {{The}} uniform approximation and the
  effect of the binding potential},\ }\href
  {https://doi.org/10.1103/PhysRevA.66.043413} {\bibfield  {journal} {\bibinfo
  {journal} {\emph {Phys. Rev. A}}\ }\textbf {\bibinfo {volume} {66}}\bibfield
  {number} {\bibinfo  {number} { no.~4},\ \bibinfo {pages} {p.~043413}}
  (\bibinfo {year} {2002})}\BibitemShut {NoStop}%
\bibitem [{\citenamefont {Pisanty}\ \emph {et~al.}(2020)\citenamefont
  {Pisanty}, \citenamefont {Ciappina},\ and\ \citenamefont
  {Lewenstein}}]{pisanty2020imaginary}%
  \BibitemOpen
  \bibfield  {author} {\bibinfo {author} {\bibfnamefont {E.}~\bibnamefont
  {Pisanty}}, \bibinfo {author} {\bibfnamefont {M.~F.}\ \bibnamefont
  {Ciappina}},\ and\ \bibinfo {author} {\bibfnamefont {M.}~\bibnamefont
  {Lewenstein}},\ }\bibfield  {title} {\bibinfo {title} {The imaginary part of
  the high-harmonic cutoff},\ }\href {https://doi.org/10.1088/2515-7647/ab8f1e}
  {\bibfield  {journal} {\bibinfo  {journal} {\emph {Journal of Physics:
  Photonics}}\ }\textbf {\bibinfo {volume} {2}}\bibfield  {number} {\bibinfo
  {number} { no.~3},\ \bibinfo {pages} {p.~034013}} (\bibinfo {year}
  {2020})}\BibitemShut {NoStop}%
\bibitem [{\citenamefont {Milo{\v
  s}evi{\'c}}(2025)}]{milosevic2025application}%
  \BibitemOpen
  \bibfield  {author} {\bibinfo {author} {\bibfnamefont {D.~B.}\ \bibnamefont
  {Milo{\v s}evi{\'c}}},\ }\bibfield  {title} {\bibinfo {title} {Application of
  the uniform approximation to integrals occurring in ionization by a strong
  elliptically polarized laser field},\ }\href
  {https://doi.org/10.1103/PhysRevA.111.053105} {\bibfield  {journal} {\bibinfo
   {journal} {\emph {Phys. Rev. A}}\ }\textbf {\bibinfo {volume}
  {111}}\bibfield  {number} {\bibinfo  {number} { no.~5},\ \bibinfo {pages}
  {p.~053105}} (\bibinfo {year} {2025})}\BibitemShut {NoStop}%
\bibitem [{\citenamefont {Habibovi{\'c}}\ and\ \citenamefont {Milo{\v
  s}evi{\'c}}(2025)}]{habibovic2025complete}%
  \BibitemOpen
  \bibfield  {author} {\bibinfo {author} {\bibfnamefont {D.}~\bibnamefont
  {Habibovi{\'c}}}\ and\ \bibinfo {author} {\bibfnamefont {D.~B.}\ \bibnamefont
  {Milo{\v s}evi{\'c}}},\ }\bibfield  {title} {\bibinfo {title} {Complete
  classification and additional saddle-point solutions for high-order
  above-threshold ionization induced by a strong laser field},\ }\href
  {https://doi.org/10.1103/PhysRevA.111.023103} {\bibfield  {journal} {\bibinfo
   {journal} {\emph {Phys. Rev. A}}\ }\textbf {\bibinfo {volume}
  {111}}\bibfield  {number} {\bibinfo  {number} { no.~2},\ \bibinfo {pages}
  {p.~023103}} (\bibinfo {year} {2025})}\BibitemShut {NoStop}%
\bibitem [{\citenamefont {Gibbs}\ \emph {et~al.}(2024)\citenamefont {Gibbs},
  \citenamefont {Hewett},\ and\ \citenamefont
  {Huybrechs}}]{gibbs2024numerical}%
  \BibitemOpen
  \bibfield  {author} {\bibinfo {author} {\bibfnamefont {A.}~\bibnamefont
  {Gibbs}}, \bibinfo {author} {\bibfnamefont {D.~P.}\ \bibnamefont {Hewett}},\
  and\ \bibinfo {author} {\bibfnamefont {D.}~\bibnamefont {Huybrechs}},\
  }\bibfield  {title} {\bibinfo {title} {Numerical evaluation of oscillatory
  integrals via automated steepest descent contour deformation},\ }\href
  {https://doi.org/10.1016/j.jcp.2024.112787} {\bibfield  {journal} {\bibinfo
  {journal} {\emph {Journal of Computational Physics}}\ }\textbf {\bibinfo
  {volume} {501}},\ \bibinfo {pages} {p.~112787} (\bibinfo {year}
  {2024})}\BibitemShut {NoStop}%
\bibitem [{\citenamefont {Shanin}\ \emph {et~al.}(2022)\citenamefont {Shanin},
  \citenamefont {Korolkov},\ and\ \citenamefont {Kniazeva}}]{shanin2022saddle}%
  \BibitemOpen
  \bibfield  {author} {\bibinfo {author} {\bibfnamefont {A.~V.}\ \bibnamefont
  {Shanin}}, \bibinfo {author} {\bibfnamefont {A.~I.}\ \bibnamefont
  {Korolkov}},\ and\ \bibinfo {author} {\bibfnamefont {K.~S.}\ \bibnamefont
  {Kniazeva}},\ }\bibfield  {title} {\bibinfo {title} {Saddle {{Point Method}}
  for {{Transient Processes}} in {{Waveguides}}},\ }\href
  {https://doi.org/10.1142/S2591728521500183} {\bibfield  {journal} {\bibinfo
  {journal} {\emph {Journal of Theoretical and Computational Acoustics}}\
  }\textbf {\bibinfo {volume} {30}}\bibfield  {number} {\bibinfo  {number} {
  no.~04},\ \bibinfo {pages} {p.~2150018}} (\bibinfo {year}
  {2022})}\BibitemShut {NoStop}%
\bibitem [{\citenamefont {Faisal}(1973)}]{faisal1973multiple}%
  \BibitemOpen
  \bibfield  {author} {\bibinfo {author} {\bibfnamefont {F.~H.~M.}\
  \bibnamefont {Faisal}},\ }\bibfield  {title} {\bibinfo {title} {Multiple
  absorption of laser photons by atoms},\ }\href
  {https://doi.org/10.1088/0022-3700/6/4/011} {\bibfield  {journal} {\bibinfo
  {journal} {\emph {Journal of Physics B: Atomic and Molecular Physics}}\
  }\textbf {\bibinfo {volume} {6}}\bibfield  {number} {\bibinfo  {number} {
  no.~4},\ \bibinfo {pages} {pp.~L89--L92}} (\bibinfo {year}
  {1973})}\BibitemShut {NoStop}%
\bibitem [{\citenamefont {Reiss}(1980)}]{reiss1980effect}%
  \BibitemOpen
  \bibfield  {author} {\bibinfo {author} {\bibfnamefont {H.~R.}\ \bibnamefont
  {Reiss}},\ }\bibfield  {title} {\bibinfo {title} {Effect of an intense
  electromagnetic field on a weakly bound system},\ }\href
  {https://doi.org/10.1103/PhysRevA.22.1786} {\bibfield  {journal} {\bibinfo
  {journal} {\emph {Phys. Rev. A}}\ }\textbf {\bibinfo {volume} {22}}\bibfield
  {number} {\bibinfo  {number} { no.~5},\ \bibinfo {pages} {pp.~1786--1813}}
  (\bibinfo {year} {1980})}\BibitemShut {NoStop}%
\bibitem [{\citenamefont {Nayak}\ \emph {et~al.}(2019)\citenamefont {Nayak},
  \citenamefont {Dumergue}, \citenamefont {K{\"u}hn}, \citenamefont {Mondal},
  \citenamefont {Csizmadia}, \citenamefont {Harshitha}, \citenamefont
  {F{\"u}le}, \citenamefont {Upadhyay~Kahaly}, \citenamefont {Farkas},
  \citenamefont {Major}, \citenamefont {{Szaszk{\'o}-Bog{\'a}r}}, \citenamefont
  {F{\"o}ldi}, \citenamefont {Majorosi}, \citenamefont {Tsatrafyllis},
  \citenamefont {Skantzakis}, \citenamefont {Neori{\v c}i{\'c}}, \citenamefont
  {Shirozhan}, \citenamefont {Vampa}, \citenamefont {Varj{\'u}}, \citenamefont
  {Tzallas}, \citenamefont {Sansone}, \citenamefont {Charalambidis},\ and\
  \citenamefont {Kahaly}}]{nayak2019saddle}%
  \BibitemOpen
  \bibfield  {author} {\bibinfo {author} {\bibfnamefont {A.}~\bibnamefont
  {Nayak}}, \bibinfo {author} {\bibfnamefont {M.}~\bibnamefont {Dumergue}},
  \bibinfo {author} {\bibfnamefont {S.}~\bibnamefont {K{\"u}hn}}, \emph
  {et~al.},\ }\bibfield  {title} {\bibinfo {title} {Saddle point approaches in
  strong field physics and generation of attosecond pulses},\ }\href
  {https://doi.org/10.1016/j.physrep.2019.10.002} {\bibfield  {journal}
  {\bibinfo  {journal} {\emph {Physics Reports}}\ }\textbf {\bibinfo {volume}
  {833}},\ \bibinfo {pages} {pp.~1--52} (\bibinfo {year} {2019})}\BibitemShut
  {NoStop}%
\bibitem [{\citenamefont {Milo{\v s}evi{\'c}}\ \emph
  {et~al.}(2006)\citenamefont {Milo{\v s}evi{\'c}}, \citenamefont {Bauer},\
  and\ \citenamefont {Becker}}]{milosevic2006quantumorbit}%
  \BibitemOpen
  \bibfield  {author} {\bibinfo {author} {\bibfnamefont {D.~B.}\ \bibnamefont
  {Milo{\v s}evi{\'c}}}, \bibinfo {author} {\bibfnamefont {D.}~\bibnamefont
  {Bauer}},\ and\ \bibinfo {author} {\bibfnamefont {W.}~\bibnamefont
  {Becker}},\ }\bibfield  {title} {\bibinfo {title} {Quantum-orbit theory of
  high-order atomic processes in intense laser fields},\ }\href
  {https://doi.org/10.1080/09500340500186099} {\bibfield  {journal} {\bibinfo
  {journal} {\emph {J. Mod. Opt.}}\ }\textbf {\bibinfo {volume} {53}}\bibfield
  {number} {\bibinfo  {number} { no.~1-2},\ \bibinfo {pages} {pp.~125--134}}
  (\bibinfo {year} {2006})}\BibitemShut {NoStop}%
\bibitem [{\citenamefont {Milo{\v s}evi{\'c}}\ and\ \citenamefont
  {Becker}(2019)}]{milosevic2019xray}%
  \BibitemOpen
  \bibfield  {author} {\bibinfo {author} {\bibfnamefont {D.~B.}\ \bibnamefont
  {Milo{\v s}evi{\'c}}}\ and\ \bibinfo {author} {\bibfnamefont
  {W.}~\bibnamefont {Becker}},\ }\bibfield  {title} {\bibinfo {title} {X-ray
  harmonic generation by orthogonally polarized two-color fields: {{Spectral}}
  shape and polarization},\ }\href
  {https://doi.org/10.1103/PhysRevA.100.031401} {\bibfield  {journal} {\bibinfo
   {journal} {\emph {Phys. Rev. A}}\ }\textbf {\bibinfo {volume}
  {100}}\bibfield  {number} {\bibinfo  {number} { no.~3},\ \bibinfo {pages}
  {p.~031401}} (\bibinfo {year} {2019})}\BibitemShut {NoStop}%
\bibitem [{\citenamefont {Arnold}\ \emph {et~al.}(2012)\citenamefont {Arnold},
  \citenamefont {Varchenko},\ and\ \citenamefont
  {{Gusein-Zade}}}]{arnold2012singularities}%
  \BibitemOpen
  \bibfield  {author} {\bibinfo {author} {\bibfnamefont {V.~I.}\ \bibnamefont
  {Arnold}}, \bibinfo {author} {\bibfnamefont {A.~N.}\ \bibnamefont
  {Varchenko}},\ and\ \bibinfo {author} {\bibfnamefont {S.~M.}\ \bibnamefont
  {{Gusein-Zade}}},\ }\href@noop {} {\emph {\bibinfo {title} {Singularities of
  {{Differentiable Maps}}: {{Volume II Monodromy}} and {{Asymptotic
  Integrals}}}}}\ (\bibinfo  {publisher} {Springer Science \& Business Media},\
  \bibinfo {year} {2012})\BibitemShut {NoStop}%
\bibitem [{\citenamefont {Feldbrugge}\ \emph
  {et~al.}(2017{\natexlab{b}})\citenamefont {Feldbrugge}, \citenamefont
  {Lehners},\ and\ \citenamefont {Turok}}]{feldbrugge2017no}%
  \BibitemOpen
  \bibfield  {author} {\bibinfo {author} {\bibfnamefont {J.}~\bibnamefont
  {Feldbrugge}}, \bibinfo {author} {\bibfnamefont {J.-L.}\ \bibnamefont
  {Lehners}},\ and\ \bibinfo {author} {\bibfnamefont {N.}~\bibnamefont
  {Turok}},\ }\bibfield  {title} {\bibinfo {title} {No smooth beginning for
  spacetime},\ }\href {https://doi.org/10.1103/PhysRevLett.119.171301}
  {\bibfield  {journal} {\bibinfo  {journal} {\emph {Phys. Rev. Lett.}}\
  }\textbf {\bibinfo {volume} {119}}\bibfield  {number} {\bibinfo  {number} {
  no.~17},\ \bibinfo {pages} {p.~171301}} (\bibinfo {year}
  {2017}{\natexlab{b}})}\BibitemShut {NoStop}%
\bibitem [{\citenamefont {Feldbrugge}(2023)}]{feldbrugge2023multiplane}%
  \BibitemOpen
  \bibfield  {author} {\bibinfo {author} {\bibfnamefont {J.}~\bibnamefont
  {Feldbrugge}},\ }\bibfield  {title} {\bibinfo {title} {Multiplane lensing in
  wave optics},\ }\href {https://doi.org/10.1093/mnras/stad349} {\bibfield
  {journal} {\bibinfo  {journal} {\emph {Mon. Not. Roy. Astron. Soc.}}\
  }\textbf {\bibinfo {volume} {520}}\bibfield  {number} {\bibinfo  {number} {
  no.~2},\ \bibinfo {pages} {pp.~2995--3006}} (\bibinfo {year}
  {2023})}\BibitemShut {NoStop}%
\bibitem [{\citenamefont {{Bonga}}\ \emph {et~al.}(2025)\citenamefont
  {{Bonga}}, \citenamefont {{Feldbrugge}},\ and\ \citenamefont
  {{Metidieri}}}]{2025PhRvD.111f3061B}%
  \BibitemOpen
  \bibfield  {author} {\bibinfo {author} {\bibfnamefont {B.}~\bibnamefont
  {{Bonga}}}, \bibinfo {author} {\bibfnamefont {J.}~\bibnamefont
  {{Feldbrugge}}},\ and\ \bibinfo {author} {\bibfnamefont {A.~R.}\ \bibnamefont
  {{Metidieri}}},\ }\bibfield  {title} {\bibinfo {title} {{Wave optics for
  rotating stars}},\ }\href {https://doi.org/10.1103/PhysRevD.111.063061}
  {\bibfield  {journal} {\bibinfo  {journal} {\emph {\prd}}\ }\textbf {\bibinfo
  {volume} {111}}\bibfield  {number} {\bibinfo  {number} { no.~6},\ \bibinfo
  {eid} {063061}} (\bibinfo {year} {2025})}\BibitemShut {NoStop}%
\bibitem [{\citenamefont {Fujii}\ \emph {et~al.}(2013)\citenamefont {Fujii},
  \citenamefont {Honda}, \citenamefont {Kato}, \citenamefont {Kikukawa},
  \citenamefont {Komatsu},\ and\ \citenamefont {Sano}}]{fujii2013hybrid}%
  \BibitemOpen
  \bibfield  {author} {\bibinfo {author} {\bibfnamefont {H.}~\bibnamefont
  {Fujii}}, \bibinfo {author} {\bibfnamefont {D.}~\bibnamefont {Honda}},
  \bibinfo {author} {\bibfnamefont {M.}~\bibnamefont {Kato}}, \emph {et~al.},\
  }\bibfield  {title} {\bibinfo {title} {Hybrid {{Monte Carlo}} on {{Lefschetz
  Thimbles}} -- {{A}} study of the residual sign problem},\ }\href
  {https://doi.org/10.1007/JHEP10(2013)147} {\bibfield  {journal} {\bibinfo
  {journal} {\emph {Journal of High Energy Physics}}\ }\textbf {\bibinfo
  {volume} {2013}}\bibfield  {number} {\bibinfo  {number} { no.~10},\ \bibinfo
  {pages} {p.~147}} (\bibinfo {year} {2013})}\BibitemShut {NoStop}%
\bibitem [{\citenamefont {Cristoforetti}\ \emph {et~al.}(2012)\citenamefont
  {Cristoforetti}, \citenamefont {Renzo},\ and\ \citenamefont
  {Scorzato}}]{collaboration2012high}%
  \BibitemOpen
  \bibfield  {author} {\bibinfo {author} {\bibfnamefont {M.}~\bibnamefont
  {Cristoforetti}}, \bibinfo {author} {\bibfnamefont {F.~D.}\ \bibnamefont
  {Renzo}},\ and\ \bibinfo {author} {\bibfnamefont {L.}~\bibnamefont
  {Scorzato}} (\bibinfo {collaboration} {Collaboration, AuroraScience}),\
  }\bibfield  {title} {\bibinfo {title} {High density {{QCD}} on a
  {{Lefschetz}} thimble?},\ }\href {https://doi.org/10.1103/PhysRevD.86.074506}
  {\bibfield  {journal} {\bibinfo  {journal} {\emph {Phys. Rev. D}}\ }\textbf
  {\bibinfo {volume} {86}}\bibfield  {number} {\bibinfo  {number} { no.~7},\
  \bibinfo {pages} {p.~074506}} (\bibinfo {year} {2012})}\BibitemShut {NoStop}%
\bibitem [{\citenamefont {Cristoforetti}\ \emph {et~al.}(2013)\citenamefont
  {Cristoforetti}, \citenamefont {Renzo}, \citenamefont {Mukherjee},\ and\
  \citenamefont {Scorzato}}]{cristoforetti2013monte}%
  \BibitemOpen
  \bibfield  {author} {\bibinfo {author} {\bibfnamefont {M.}~\bibnamefont
  {Cristoforetti}}, \bibinfo {author} {\bibfnamefont {F.~D.}\ \bibnamefont
  {Renzo}}, \bibinfo {author} {\bibfnamefont {A.}~\bibnamefont {Mukherjee}},\
  and\ \bibinfo {author} {\bibfnamefont {L.}~\bibnamefont {Scorzato}},\
  }\bibfield  {title} {\bibinfo {title} {Monte {{Carlo}} simulations on the
  {{Lefschetz}} thimble: Taming the sign problem},\ }\href
  {https://doi.org/10.1103/PhysRevD.88.051501} {\bibfield  {journal} {\bibinfo
  {journal} {\emph {Phys. Rev. D}}\ }\textbf {\bibinfo {volume} {88}}\bibfield
  {number} {\bibinfo  {number} { no.~5},\ \bibinfo {pages} {p.~051501}}
  (\bibinfo {year} {2013})}\BibitemShut {NoStop}%
\bibitem [{\citenamefont {{Feldbrugge}}\ and\ \citenamefont
  {{Pen}}(2025)}]{feldbrugge2025realtime}%
  \BibitemOpen
  \bibfield  {author} {\bibinfo {author} {\bibfnamefont {J.}~\bibnamefont
  {{Feldbrugge}}}\ and\ \bibinfo {author} {\bibfnamefont {U.-L.}\ \bibnamefont
  {{Pen}}}.\EOS\ }
\newblock \bibinfo {title} {{The real-time Feynman path integral for step
  potentials}}.\EOS\
\newblock
  \href{https://arxiv.org/abs/2508.17578}{arXiv:\allowbreak{}2508.\allowbreak{}17578}
  (\bibinfo {year} {2025})\BibitemShut {NoStop}%
\bibitem [{\citenamefont {{Feldbrugge}}\ \emph {et~al.}(2025)\citenamefont
  {{Feldbrugge}}, \citenamefont {{Jow}},\ and\ \citenamefont
  {{Pen}}}]{2025PhRvD.111h5027F}%
  \BibitemOpen
  \bibfield  {author} {\bibinfo {author} {\bibfnamefont {J.}~\bibnamefont
  {{Feldbrugge}}}, \bibinfo {author} {\bibfnamefont {D.~L.}\ \bibnamefont
  {{Jow}}},\ and\ \bibinfo {author} {\bibfnamefont {U.-L.}\ \bibnamefont
  {{Pen}}},\ }\bibfield  {title} {\bibinfo {title} {{Complex classical paths in
  quantum reflections and tunneling}},\ }\href
  {https://doi.org/10.1103/PhysRevD.111.085027} {\bibfield  {journal} {\bibinfo
   {journal} {\emph {\prd}}\ }\textbf {\bibinfo {volume} {111}}\bibfield
  {number} {\bibinfo  {number} { no.~8},\ \bibinfo {eid} {085027}} (\bibinfo
  {year} {2025})}\BibitemShut {NoStop}%
\bibitem [{\citenamefont {Delabaere}\ and\ \citenamefont
  {Howls}(2002)}]{DelabaereHowls2002}%
  \BibitemOpen
  \bibfield  {author} {\bibinfo {author} {\bibfnamefont {E.}~\bibnamefont
  {Delabaere}}\ and\ \bibinfo {author} {\bibfnamefont {C.~J.}\ \bibnamefont
  {Howls}},\ }\bibfield  {title} {\bibinfo {title} {Global asymptotics for
  multiple integrals with boundaries},\ }\href
  {https://doi.org/10.1215/S0012-9074-02-11221-6} {\bibfield  {journal}
  {\bibinfo  {journal} {\emph {Duke Mathematical Journal}}\ }\textbf {\bibinfo
  {volume} {112}}\bibfield  {number} {\bibinfo  {number} { no.~2},\ \bibinfo
  {pages} {pp.~251--291}} (\bibinfo {year} {2002})}\BibitemShut {NoStop}%
\bibitem [{\citenamefont {Alexandru}\ \emph {et~al.}(2016)\citenamefont
  {Alexandru}, \citenamefont {Ba{\c s}ar},\ and\ \citenamefont
  {Bedaque}}]{alexandru2016monte}%
  \BibitemOpen
  \bibfield  {author} {\bibinfo {author} {\bibfnamefont {A.}~\bibnamefont
  {Alexandru}}, \bibinfo {author} {\bibfnamefont {G.}~\bibnamefont {Ba{\c
  s}ar}},\ and\ \bibinfo {author} {\bibfnamefont {P.}~\bibnamefont {Bedaque}},\
  }\bibfield  {title} {\bibinfo {title} {Monte {{Carlo}} algorithm for
  simulating fermions on {{Lefschetz}} thimbles},\ }\href
  {https://doi.org/10.1103/PhysRevD.93.014504} {\bibfield  {journal} {\bibinfo
  {journal} {\emph {Phys. Rev. D}}\ }\textbf {\bibinfo {volume} {93}}\bibfield
  {number} {\bibinfo  {number} { no.~1},\ \bibinfo {pages} {p.~014504}}
  (\bibinfo {year} {2016})}\BibitemShut {NoStop}%
\bibitem [{\citenamefont {Nishimura}\ and\ \citenamefont
  {Shimasaki}(2017)}]{nishimura2017combining}%
  \BibitemOpen
  \bibfield  {author} {\bibinfo {author} {\bibfnamefont {J.}~\bibnamefont
  {Nishimura}}\ and\ \bibinfo {author} {\bibfnamefont {S.}~\bibnamefont
  {Shimasaki}},\ }\bibfield  {title} {\bibinfo {title} {Combining the complex
  {{Langevin}} method and the generalized {{Lefschetz-thimble}} method},\
  }\href {https://doi.org/10.1007/JHEP06(2017)023} {\bibfield  {journal}
  {\bibinfo  {journal} {\emph {Journal of High Energy Physics}}\ }\textbf
  {\bibinfo {volume} {2017}}\bibfield  {number} {\bibinfo  {number} { no.~6},\
  \bibinfo {pages} {p.~23}} (\bibinfo {year} {2017})}\BibitemShut {NoStop}%
\bibitem [{\citenamefont {Feldbrugge}\ \emph {et~al.}(2021)\citenamefont
  {Feldbrugge}, \citenamefont {Pen},\ and\ \citenamefont
  {Turok}}]{feldbruggepicardlefschetz}%
  \BibitemOpen
  \bibfield  {author} {\bibinfo {author} {\bibfnamefont {J.~L.}\ \bibnamefont
  {Feldbrugge}}, \bibinfo {author} {\bibfnamefont {U.-L.}\ \bibnamefont
  {Pen}},\ and\ \bibinfo {author} {\bibfnamefont {N.}~\bibnamefont {Turok}},\
  }\href@noop {} {\bibinfo {title} {Picard-{{Lefschetz Path Integrals}}}},\
  \bibinfo {howpublished} {GitHub, \url{https://p-lpi.github.io/}} (\bibinfo
  {year} {2021})\BibitemShut {NoStop}%
\bibitem [{\citenamefont {Bartholomew}(1959)}]{bartholomew1959numerical}%
  \BibitemOpen
  \bibfield  {author} {\bibinfo {author} {\bibfnamefont {G.~E.}\ \bibnamefont
  {Bartholomew}},\ }\bibfield  {title} {\bibinfo {title} {Numerical integration
  over the triangle},\ }\href
  {https://doi.org/10.1090/S0025-5718-1959-0107976-5} {\bibfield  {journal}
  {\bibinfo  {journal} {\emph {Mathematical Tables and Other Aids to
  Computation}}\ ,\ \bibinfo {pages} {pp.~295--298}} (\bibinfo {year}
  {1959})}\BibitemShut {NoStop}%
\bibitem [{\citenamefont {Han}\ \emph {et~al.}(2021)\citenamefont {Han},
  \citenamefont {Huang}, \citenamefont {Liu}, \citenamefont {Qu},\ and\
  \citenamefont {Wan}}]{han2021spinfoam}%
  \BibitemOpen
  \bibfield  {author} {\bibinfo {author} {\bibfnamefont {M.}~\bibnamefont
  {Han}}, \bibinfo {author} {\bibfnamefont {Z.}~\bibnamefont {Huang}}, \bibinfo
  {author} {\bibfnamefont {H.}~\bibnamefont {Liu}}, \emph {et~al.},\ }\bibfield
   {title} {\bibinfo {title} {Spinfoam on a {{Lefschetz}} thimble: {{Markov}}
  chain {{Monte Carlo}} computation of a {{Lorentzian}} spinfoam propagator},\
  }\href {https://doi.org/10.1103/PhysRevD.103.084026} {\bibfield  {journal}
  {\bibinfo  {journal} {\emph {Phys. Rev. D}}\ }\textbf {\bibinfo {volume}
  {103}}\bibfield  {number} {\bibinfo  {number} { no.~8},\ \bibinfo {pages}
  {p.~084026}} (\bibinfo {year} {2021})}\BibitemShut {NoStop}%
\bibitem [{\citenamefont {Alexandru}\ \emph {et~al.}(2022)\citenamefont
  {Alexandru}, \citenamefont {Ba{\c s}ar}, \citenamefont {Bedaque},\ and\
  \citenamefont {Warrington}}]{alexandru2022complex}%
  \BibitemOpen
  \bibfield  {author} {\bibinfo {author} {\bibfnamefont {A.}~\bibnamefont
  {Alexandru}}, \bibinfo {author} {\bibfnamefont {G.}~\bibnamefont {Ba{\c
  s}ar}}, \bibinfo {author} {\bibfnamefont {P.~F.}\ \bibnamefont {Bedaque}},\
  and\ \bibinfo {author} {\bibfnamefont {N.~C.}\ \bibnamefont {Warrington}},\
  }\bibfield  {title} {\bibinfo {title} {Complex paths around the sign
  problem},\ }\href {https://doi.org/10.1103/RevModPhys.94.015006} {\bibfield
  {journal} {\bibinfo  {journal} {\emph {Reviews of Modern Physics}}\ }\textbf
  {\bibinfo {volume} {94}}\bibfield  {number} {\bibinfo  {number} { no.~1},\
  \bibinfo {pages} {p.~015006}} (\bibinfo {year} {2022})}\BibitemShut {NoStop}%
\bibitem [{\citenamefont {Shoji}\ and\ \citenamefont
  {Trailović}(2025)}]{shoji2025stable}%
  \BibitemOpen
  \bibfield  {author} {\bibinfo {author} {\bibfnamefont {Y.}~\bibnamefont
  {Shoji}}\ and\ \bibinfo {author} {\bibfnamefont {K.}~\bibnamefont
  {Trailović}}.\EOS\ }
\newblock \bibinfo {title} {Stable {{Evaluation}} of {{Lefschetz Thimble
  Intersection Numbers}}: {{Towards Real-Time Path Integrals}}}.\EOS\
\newblock
  \href{https://arxiv.org/abs/2510.06334}{arXiv:\allowbreak{}2510.\allowbreak{}06334}
  (\bibinfo {year} {2025})\BibitemShut {NoStop}%
\bibitem [{\citenamefont {Lando}(1997)}]{lando1997geometry}%
  \BibitemOpen
  \bibfield  {author} {\bibinfo {author} {\bibfnamefont {S.~K.}\ \bibnamefont
  {Lando}},\ }\bibfield  {title} {\bibinfo {title} {Geometry of the stokes sets
  for families of functions of one variable},\ }\href
  {https://doi.org/10.1007/BF02434983} {\bibfield  {journal} {\bibinfo
  {journal} {\emph {Journal of Mathematical Sciences}}\ }\textbf {\bibinfo
  {volume} {83}}\bibfield  {number} {\bibinfo  {number} { no.~4},\ \bibinfo
  {pages} {pp.~534--538}} (\bibinfo {year} {1997})}\BibitemShut {NoStop}%
\bibitem [{\citenamefont {Chipperfield}\ \emph {et~al.}(2005)\citenamefont
  {Chipperfield}, \citenamefont {Gaier}, \citenamefont {Knight}, \citenamefont
  {Marangos},\ and\ \citenamefont {Tisch}}]{Chipperfield2005conditions}%
  \BibitemOpen
  \bibfield  {author} {\bibinfo {author} {\bibfnamefont {L.~E.}\ \bibnamefont
  {Chipperfield}}, \bibinfo {author} {\bibfnamefont {L.~N.}\ \bibnamefont
  {Gaier}}, \bibinfo {author} {\bibfnamefont {P.~L.}\ \bibnamefont {Knight}},
  \emph {et~al.},\ }\bibfield  {title} {\bibinfo {title} {Conditions for the
  reliable production of attosecond pulses using ultra-short laser-generated
  high harmonics},\ }\href {https://doi.org/10.1080/0950034042000275379}
  {\bibfield  {journal} {\bibinfo  {journal} {\emph {J. Mod. Opt.}}\ }\textbf
  {\bibinfo {volume} {52}}\bibfield  {number} {\bibinfo  {number} { no.~2--3},\
  \bibinfo {pages} {pp.~243--260}} (\bibinfo {year} {2005})}\BibitemShut
  {NoStop}%
\bibitem [{\citenamefont {Wright}(1980)}]{wright1980stokes}%
  \BibitemOpen
  \bibfield  {author} {\bibinfo {author} {\bibfnamefont {F.~J.}\ \bibnamefont
  {Wright}},\ }\bibfield  {title} {\bibinfo {title} {The {{Stokes}} set of the
  cusp diffraction catastrophe},\ }\href
  {https://doi.org/10.1088/0305-4470/13/9/018} {\bibfield  {journal} {\bibinfo
  {journal} {\emph {J. Phys. A: Math. Theor.}}\ }\textbf {\bibinfo {volume}
  {13}}\bibfield  {number} {\bibinfo  {number} { no.~9},\ \bibinfo {pages}
  {pp.~2913--2928}} (\bibinfo {year} {1980})}\BibitemShut {NoStop}%
\bibitem [{\citenamefont {Berry}\ and\ \citenamefont
  {Howls}(1990)}]{berry1990stokes}%
  \BibitemOpen
  \bibfield  {author} {\bibinfo {author} {\bibfnamefont {M.~V.}\ \bibnamefont
  {Berry}}\ and\ \bibinfo {author} {\bibfnamefont {C.~J.}\ \bibnamefont
  {Howls}},\ }\bibfield  {title} {\bibinfo {title} {Stokes surfaces of
  diffraction catastrophes with codimension three},\ }\href
  {https://doi.org/10.1088/0951-7715/3/2/003} {\bibfield  {journal} {\bibinfo
  {journal} {\emph {Nonlinearity}}\ }\textbf {\bibinfo {volume} {3}}\bibfield
  {number} {\bibinfo  {number} { no.~2},\ \bibinfo {pages} {pp.~281--291}}
  (\bibinfo {year} {1990})}\BibitemShut {NoStop}%
\bibitem [{\citenamefont {Feldbrugge}\ \emph {et~al.}(2018)\citenamefont
  {Feldbrugge}, \citenamefont {van~de Weygaert}, \citenamefont {Hidding},\ and\
  \citenamefont {Feldbrugge}}]{feldbrugge2018caustic}%
  \BibitemOpen
  \bibfield  {author} {\bibinfo {author} {\bibfnamefont {J.}~\bibnamefont
  {Feldbrugge}}, \bibinfo {author} {\bibfnamefont {R.}~\bibnamefont {van~de
  Weygaert}}, \bibinfo {author} {\bibfnamefont {J.}~\bibnamefont {Hidding}},\
  and\ \bibinfo {author} {\bibfnamefont {J.}~\bibnamefont {Feldbrugge}},\
  }\bibfield  {title} {\bibinfo {title} {Caustic {{Skeleton}} \& {{Cosmic
  Web}}},\ }\href {https://doi.org/10.1088/1475-7516/2018/05/027} {\bibfield
  {journal} {\bibinfo  {journal} {\emph {Journal of Cosmology and Astroparticle
  Physics}}\ }\textbf {\bibinfo {volume} {2018}}\bibfield  {number} {\bibinfo
  {number} { no.~05},\ \bibinfo {pages} {pp.~027--027}} (\bibinfo {year}
  {2018})}\BibitemShut {NoStop}%
\bibitem [{\citenamefont {Poston}\ and\ \citenamefont
  {Stewart}(1978)}]{poston1978catastrophe}%
  \BibitemOpen
  \bibfield  {author} {\bibinfo {author} {\bibfnamefont {T.}~\bibnamefont
  {Poston}}\ and\ \bibinfo {author} {\bibfnamefont {I.}~\bibnamefont
  {Stewart}},\ }\href@noop {} {\emph {\bibinfo {title} {Catastrophe {{Theory}}
  and {{Its Applications}}}}},\ \bibinfo {edition} {first edition}\ ed.,\
  \bibinfo {series} {Surveys and {{Reference Works}} in {{Mathematics}}}\
  No.~\bibinfo {number} {2}\ (\bibinfo  {publisher} {Pitman Publishing Ltd.},\
  \bibinfo {address} {Bath, UK},\ \bibinfo {year} {1978})\BibitemShut {NoStop}%
\bibitem [{\citenamefont {Saunders}(1980)}]{saunders1980introduction}%
  \BibitemOpen
  \bibfield  {author} {\bibinfo {author} {\bibfnamefont {P.~T.}\ \bibnamefont
  {Saunders}},\ }\href@noop {} {\emph {\bibinfo {title} {An {{Introduction}} to
  {{Catastrophe Theory}}}}}\ (\bibinfo  {publisher} {Cambridge University
  Press},\ \bibinfo {year} {1980})\BibitemShut {NoStop}%
\bibitem [{\citenamefont {Olver}\ \emph {et~al.}(2010)\citenamefont {Olver},
  \citenamefont {Lozier}, \citenamefont {Boisvert},\ and\ \citenamefont
  {Clark}}]{NIST_handbook}%
  \BibitemOpen
  \bibinfo {editor} {\bibfnamefont {F.~W.~J.}\ \bibnamefont {Olver}}, \bibinfo
  {editor} {\bibfnamefont {D.~W.}\ \bibnamefont {Lozier}}, \bibinfo {editor}
  {\bibfnamefont {R.~F.}\ \bibnamefont {Boisvert}},\ and\ \bibinfo {editor}
  {\bibfnamefont {C.~W.}\ \bibnamefont {Clark}},\ eds.,\ \href@noop {} {\emph
  {\bibinfo {title} {{NIST} Handbook of Mathematical Functions}}}\ (\bibinfo
  {publisher} {Cambridge University Press NIST},\ \bibinfo {address}
  {Cambridge, New York},\ \bibinfo {year} {2010})\ \bibinfo {note} {available
  online as the \href{http://dlmf.nist.gov/}{Digital Library of Mathematical
  Functions}.}\BibitemShut {Stop}%
\bibitem [{\citenamefont {Chester}\ \emph {et~al.}(1957)\citenamefont
  {Chester}, \citenamefont {Friedman},\ and\ \citenamefont
  {Ursell}}]{chester1957extension}%
  \BibitemOpen
  \bibfield  {author} {\bibinfo {author} {\bibfnamefont {C.}~\bibnamefont
  {Chester}}, \bibinfo {author} {\bibfnamefont {B.}~\bibnamefont {Friedman}},\
  and\ \bibinfo {author} {\bibfnamefont {F.}~\bibnamefont {Ursell}},\
  }\bibfield  {title} {\bibinfo {title} {An extension of the method of steepest
  descents},\ }\href {https://doi.org/10.1017/S0305004100032655} {\bibfield
  {journal} {\bibinfo  {journal} {\emph {Mathematical Proceedings of the
  Cambridge Philosophical Society}}\ }\textbf {\bibinfo {volume} {53}}\bibfield
   {number} {\bibinfo  {number} { no.~3},\ \bibinfo {pages} {pp.~599--611}}
  (\bibinfo {year} {1957})}\BibitemShut {NoStop}%
\bibitem [{\citenamefont {Berry}(1989)}]{berry1989uniform}%
  \BibitemOpen
  \bibfield  {author} {\bibinfo {author} {\bibfnamefont {M.~V.}\ \bibnamefont
  {Berry}},\ }\bibfield  {title} {\bibinfo {title} {Uniform asymptotic
  smoothing of {{Stokes}}'s discontinuities},\ }\href
  {https://doi.org/10.1098/rspa.1989.0018} {\bibfield  {journal} {\bibinfo
  {journal} {\emph {Proceedings of the Royal Society of London. A. Mathematical
  and Physical Sciences}}\ }\textbf {\bibinfo {volume} {422}}\bibfield
  {number} {\bibinfo  {number} { no.~1862},\ \bibinfo {pages} {pp.~7--21}}
  (\bibinfo {year} {1989})}\BibitemShut {NoStop}%
\bibitem [{\citenamefont {Schomerus}\ and\ \citenamefont
  {Sieber}(1997)}]{schomerus1997bifurcations}%
  \BibitemOpen
  \bibfield  {author} {\bibinfo {author} {\bibfnamefont {H.}~\bibnamefont
  {Schomerus}}\ and\ \bibinfo {author} {\bibfnamefont {M.}~\bibnamefont
  {Sieber}},\ }\bibfield  {title} {\bibinfo {title} {Bifurcations of periodic
  orbits and uniform approximations},\ }\href
  {https://doi.org/10.1088/0305-4470/30/13/010} {\bibfield  {journal} {\bibinfo
   {journal} {\emph {J. Phys. A: Math. Theor.}}\ }\textbf {\bibinfo {volume}
  {30}}\bibfield  {number} {\bibinfo  {number} { no.~13},\ \bibinfo {pages}
  {pp.~4537--4562}} (\bibinfo {year} {1997})}\BibitemShut {NoStop}%
\bibitem [{\citenamefont {Stamnes}\ and\ \citenamefont
  {Spjelkavik}(1983)}]{stamnes1983evaluation}%
  \BibitemOpen
  \bibfield  {author} {\bibinfo {author} {\bibfnamefont {J.~J.}\ \bibnamefont
  {Stamnes}}\ and\ \bibinfo {author} {\bibfnamefont {B.}~\bibnamefont
  {Spjelkavik}},\ }\bibfield  {title} {\bibinfo {title} {Evaluation of the
  {{Field}} near a {{Cusp}} of a {{Caustic}}},\ }\href
  {https://doi.org/10.1080/713821363} {\bibfield  {journal} {\bibinfo
  {journal} {\emph {Optica Acta: International Journal of Optics}}\ }\textbf
  {\bibinfo {volume} {30}}\bibfield  {number} {\bibinfo  {number} { no.~9},\
  \bibinfo {pages} {pp.~1331--1358}} (\bibinfo {year} {1983})}\BibitemShut
  {NoStop}%
\bibitem [{\citenamefont {Chipperfield}\ \emph {et~al.}(2006)\citenamefont
  {Chipperfield}, \citenamefont {Knight}, \citenamefont {Tisch},\ and\
  \citenamefont {Marangos}}]{chipperfield2006tracking}%
  \BibitemOpen
  \bibfield  {author} {\bibinfo {author} {\bibfnamefont {L.~E.}\ \bibnamefont
  {Chipperfield}}, \bibinfo {author} {\bibfnamefont {P.~L.}\ \bibnamefont
  {Knight}}, \bibinfo {author} {\bibfnamefont {J.~W.~G.}\ \bibnamefont
  {Tisch}},\ and\ \bibinfo {author} {\bibfnamefont {J.~P.}\ \bibnamefont
  {Marangos}},\ }\bibfield  {title} {\bibinfo {title} {Tracking individual
  electron trajectories in a high harmonic spectrum},\ }\href
  {https://doi.org/10.1016/j.optcom.2006.03.078} {\bibfield  {journal}
  {\bibinfo  {journal} {\emph {Optics Communications}}\ }\textbf {\bibinfo
  {volume} {264}}\bibfield  {number} {\bibinfo  {number} { no.~2},\ \bibinfo
  {pages} {pp.~494--501}} (\bibinfo {year} {2006})}\BibitemShut {NoStop}%
\bibitem [{\citenamefont {Pedatzur}\ \emph {et~al.}(2015)\citenamefont
  {Pedatzur}, \citenamefont {Orenstein}, \citenamefont {Serbinenko},
  \citenamefont {Soifer}, \citenamefont {Bruner}, \citenamefont {Uzan},
  \citenamefont {Brambila}, \citenamefont {Harvey}, \citenamefont {Torlina},
  \citenamefont {Morales}, \citenamefont {Smirnova},\ and\ \citenamefont
  {Dudovich}}]{pedatzur2015attosecond}%
  \BibitemOpen
  \bibfield  {author} {\bibinfo {author} {\bibfnamefont {O.}~\bibnamefont
  {Pedatzur}}, \bibinfo {author} {\bibfnamefont {G.}~\bibnamefont {Orenstein}},
  \bibinfo {author} {\bibfnamefont {V.}~\bibnamefont {Serbinenko}}, \emph
  {et~al.},\ }\bibfield  {title} {\bibinfo {title} {Attosecond tunnelling
  interferometry},\ }\href {https://doi.org/10.1038/nphys3436} {\bibfield
  {journal} {\bibinfo  {journal} {\emph {Nat. Phys.}}\ }\textbf {\bibinfo
  {volume} {11}}\bibfield  {number} {\bibinfo  {number} { no.~10},\ \bibinfo
  {pages} {pp.~815--819}} (\bibinfo {year} {2015})}\BibitemShut {NoStop}%
\bibitem [{\citenamefont {Strelkov}\ \emph {et~al.}(2012)\citenamefont
  {Strelkov}, \citenamefont {Khokhlova}, \citenamefont {Gonoskov},
  \citenamefont {Gonoskov},\ and\ \citenamefont
  {Ryabikin}}]{strelkov2012highorder}%
  \BibitemOpen
  \bibfield  {author} {\bibinfo {author} {\bibfnamefont {V.~V.}\ \bibnamefont
  {Strelkov}}, \bibinfo {author} {\bibfnamefont {M.~A.}\ \bibnamefont
  {Khokhlova}}, \bibinfo {author} {\bibfnamefont {A.~A.}\ \bibnamefont
  {Gonoskov}}, \emph {et~al.},\ }\bibfield  {title} {\bibinfo {title}
  {High-order harmonic generation by atoms in an elliptically polarized laser
  field: {{Harmonic}} polarization properties and laser threshold
  ellipticity},\ }\href {https://doi.org/10.1103/PhysRevA.86.013404} {\bibfield
   {journal} {\bibinfo  {journal} {\emph {Phys. Rev. A}}\ }\textbf {\bibinfo
  {volume} {86}}\bibfield  {number} {\bibinfo  {number} { no.~1},\ \bibinfo
  {pages} {p.~013404}} (\bibinfo {year} {2012})}\BibitemShut {NoStop}%
\bibitem [{\citenamefont {Pisanty}\ and\ \citenamefont
  {Jim{\'{e}}nez-Gal{\'{a}}n}(2017)}]{Pisanty2017}%
  \BibitemOpen
  \bibfield  {author} {\bibinfo {author} {\bibfnamefont {E.}~\bibnamefont
  {Pisanty}}\ and\ \bibinfo {author} {\bibfnamefont {{\'{A}}.}~\bibnamefont
  {Jim{\'{e}}nez-Gal{\'{a}}n}},\ }\bibfield  {title} {\bibinfo {title}
  {Strong-field approximation in a rotating frame: High-order harmonic emission
  from $p$ states in bicircular fields},\ }\href
  {https://doi.org/10.1103/PhysRevA.96.063401} {\bibfield  {journal} {\bibinfo
  {journal} {\emph {Phys. Rev. A}}\ }\textbf {\bibinfo {volume} {96}}\bibfield
  {number} {\bibinfo  {number} { no.~6},\ \bibinfo {pages} {p.~063401}}
  (\bibinfo {year} {2017})}\BibitemShut {NoStop}%
\bibitem [{\citenamefont {Pisanty}\ \emph {et~al.}(2018)\citenamefont
  {Pisanty}, \citenamefont {Hickstein}, \citenamefont {Galloway}, \citenamefont
  {Durfee}, \citenamefont {Kapteyn}, \citenamefont {Murnane},\ and\
  \citenamefont {Ivanov}}]{Pisanty2018}%
  \BibitemOpen
  \bibfield  {author} {\bibinfo {author} {\bibfnamefont {E.}~\bibnamefont
  {Pisanty}}, \bibinfo {author} {\bibfnamefont {D.~D.}\ \bibnamefont
  {Hickstein}}, \bibinfo {author} {\bibfnamefont {B.~R.}\ \bibnamefont
  {Galloway}}, \emph {et~al.},\ }\bibfield  {title} {\bibinfo {title} {High
  harmonic interferometry of the lorentz force in strong mid-infrared laser
  fields},\ }\href {https://doi.org/10.1088/1367-2630/aabb4d} {\bibfield
  {journal} {\bibinfo  {journal} {\emph {New J. Phys.}}\ }\textbf {\bibinfo
  {volume} {20}}\bibfield  {number} {\bibinfo  {number} { no.~5},\ \bibinfo
  {pages} {p.~053036}} (\bibinfo {year} {2018})}\BibitemShut {NoStop}%
\bibitem [{\citenamefont {Itatani}\ \emph {et~al.}(2004)\citenamefont
  {Itatani}, \citenamefont {Levesque}, \citenamefont {Zeidler}, \citenamefont
  {Niikura}, \citenamefont {P{\'{e}}pin}, \citenamefont {Kieffer},
  \citenamefont {Corkum},\ and\ \citenamefont {Villeneuve}}]{Itatani2004}%
  \BibitemOpen
  \bibfield  {author} {\bibinfo {author} {\bibfnamefont {J.}~\bibnamefont
  {Itatani}}, \bibinfo {author} {\bibfnamefont {J.}~\bibnamefont {Levesque}},
  \bibinfo {author} {\bibfnamefont {D.}~\bibnamefont {Zeidler}}, \emph
  {et~al.},\ }\bibfield  {title} {\bibinfo {title} {Tomographic imaging of
  molecular orbitals},\ }\href {https://doi.org/10.1038/nature03183} {\bibfield
   {journal} {\bibinfo  {journal} {\emph {Nature}}\ }\textbf {\bibinfo {volume}
  {432}},\ \bibinfo {pages} {pp.~867--871} (\bibinfo {year}
  {2004})}\BibitemShut {NoStop}%
\bibitem [{\citenamefont {Za{\"i}r}\ \emph {et~al.}(2008)\citenamefont
  {Za{\"i}r}, \citenamefont {Holler}, \citenamefont {Guandalini}, \citenamefont
  {Schapper}, \citenamefont {Biegert}, \citenamefont {Gallmann}, \citenamefont
  {Keller}, \citenamefont {Wyatt}, \citenamefont {Monmayrant}, \citenamefont
  {Walmsley}, \citenamefont {Cormier}, \citenamefont {Auguste}, \citenamefont
  {Caumes},\ and\ \citenamefont {Sali{\`e}res}}]{zair2008quantum}%
  \BibitemOpen
  \bibfield  {author} {\bibinfo {author} {\bibfnamefont {A.}~\bibnamefont
  {Za{\"i}r}}, \bibinfo {author} {\bibfnamefont {M.}~\bibnamefont {Holler}},
  \bibinfo {author} {\bibfnamefont {A.}~\bibnamefont {Guandalini}}, \emph
  {et~al.},\ }\bibfield  {title} {\bibinfo {title} {Quantum {{Path
  Interferences}} in {{High-Order Harmonic Generation}}},\ }\href
  {https://doi.org/10.1103/PhysRevLett.100.143902} {\bibfield  {journal}
  {\bibinfo  {journal} {\emph {Phys. Rev. Letters}}\ }\textbf {\bibinfo
  {volume} {100}}\bibfield  {number} {\bibinfo  {number} { no.~14},\ \bibinfo
  {pages} {p.~143902}} (\bibinfo {year} {2008})}\BibitemShut {NoStop}%
\bibitem [{\citenamefont {Hoffmann}\ \emph {et~al.}(2014)\citenamefont
  {Hoffmann}, \citenamefont {Hutchison}, \citenamefont {Za{\"{\i}}r},\ and\
  \citenamefont {Marangos}}]{Hoffmann2014}%
  \BibitemOpen
  \bibfield  {author} {\bibinfo {author} {\bibfnamefont {D.~J.}\ \bibnamefont
  {Hoffmann}}, \bibinfo {author} {\bibfnamefont {C.}~\bibnamefont {Hutchison}},
  \bibinfo {author} {\bibfnamefont {A.}~\bibnamefont {Za{\"{\i}}r}},\ and\
  \bibinfo {author} {\bibfnamefont {J.~P.}\ \bibnamefont {Marangos}},\
  }\bibfield  {title} {\bibinfo {title} {Control of temporal mapping and
  harmonic intensity modulation using two-color orthogonally polarized
  fields},\ }\href {https://doi.org/10.1103/PhysRevA.89.023423} {\bibfield
  {journal} {\bibinfo  {journal} {\emph {Phys. Rev. A}}\ }\textbf {\bibinfo
  {volume} {89}}\bibfield  {number} {\bibinfo  {number} { no.~2},\ \bibinfo
  {pages} {p.~023423}} (\bibinfo {year} {2014})}\BibitemShut {NoStop}%
\bibitem [{\citenamefont {Roscam~Abbing}\ \emph
  {et~al.}(2020{\natexlab{b}})\citenamefont {Roscam~Abbing}, \citenamefont
  {Campi}, \citenamefont {Sajjadian}, \citenamefont {Lin}, \citenamefont
  {Smorenburg},\ and\ \citenamefont {Kraus}}]{RoscamAbbing2020}%
  \BibitemOpen
  \bibfield  {author} {\bibinfo {author} {\bibfnamefont {S.}~\bibnamefont
  {Roscam~Abbing}}, \bibinfo {author} {\bibfnamefont {F.}~\bibnamefont
  {Campi}}, \bibinfo {author} {\bibfnamefont {F.~S.}\ \bibnamefont
  {Sajjadian}}, \emph {et~al.},\ }\bibfield  {title} {\bibinfo {title}
  {Divergence control of high-harmonic generation},\ }\href
  {https://doi.org/10.1103/PhysRevApplied.13.054029} {\bibfield  {journal}
  {\bibinfo  {journal} {\emph {Phys. Rev. Appl.}}\ }\textbf {\bibinfo {volume}
  {13}}\bibfield  {number} {\bibinfo  {number} { no.~5},\ \bibinfo {pages}
  {p.~054029}} (\bibinfo {year} {2020}{\natexlab{b}})}\BibitemShut {NoStop}%
\bibitem [{\citenamefont {Brugnera}\ \emph {et~al.}(2011)\citenamefont
  {Brugnera}, \citenamefont {Hoffmann}, \citenamefont {Siegel}, \citenamefont
  {Frank}, \citenamefont {Za{\"{\i}}r}, \citenamefont {Tisch},\ and\
  \citenamefont {Marangos}}]{Brugnera2011}%
  \BibitemOpen
  \bibfield  {author} {\bibinfo {author} {\bibfnamefont {L.}~\bibnamefont
  {Brugnera}}, \bibinfo {author} {\bibfnamefont {D.~J.}\ \bibnamefont
  {Hoffmann}}, \bibinfo {author} {\bibfnamefont {T.}~\bibnamefont {Siegel}},
  \emph {et~al.},\ }\bibfield  {title} {\bibinfo {title} {Trajectory selection
  in high harmonic generation by controlling the phase between orthogonal
  two-color fields},\ }\href {https://doi.org/10.1103/PhysRevLett.107.153902}
  {\bibfield  {journal} {\bibinfo  {journal} {\emph {Phys. Rev. Lett.}}\
  }\textbf {\bibinfo {volume} {107}}\bibfield  {number} {\bibinfo  {number} {
  no.~15},\ \bibinfo {pages} {p.~153902}} (\bibinfo {year} {2011})}\BibitemShut
  {NoStop}%
\bibitem [{\citenamefont {Yan}\ \emph {et~al.}(2010)\citenamefont {Yan},
  \citenamefont {Popruzhenko}, \citenamefont {Vrakking},\ and\ \citenamefont
  {Bauer}}]{yan2010lowEnergy}%
  \BibitemOpen
  \bibfield  {author} {\bibinfo {author} {\bibfnamefont {T.-M.}\ \bibnamefont
  {Yan}}, \bibinfo {author} {\bibfnamefont {S.~V.}\ \bibnamefont
  {Popruzhenko}}, \bibinfo {author} {\bibfnamefont {M.~J.~J.}\ \bibnamefont
  {Vrakking}},\ and\ \bibinfo {author} {\bibfnamefont {D.}~\bibnamefont
  {Bauer}},\ }\bibfield  {title} {\bibinfo {title} {Low-{{Energy Structures}}
  in {{Strong Field Ionization Revealed}} by {{Quantum Orbits}}},\ }\href
  {https://doi.org/10.1103/PhysRevLett.105.253002} {\bibfield  {journal}
  {\bibinfo  {journal} {\emph {Phys. Rev. Letters}}\ }\textbf {\bibinfo
  {volume} {105}}\bibfield  {number} {\bibinfo  {number} { no.~25},\ \bibinfo
  {pages} {p.~253002}} (\bibinfo {year} {2010})}\BibitemShut {NoStop}%
\bibitem [{\citenamefont {Uzan}\ \emph {et~al.}(2020)\citenamefont {Uzan},
  \citenamefont {Orenstein}, \citenamefont {{Jim{\'e}nez-Gal{\'a}n}},
  \citenamefont {McDonald}, \citenamefont {Silva}, \citenamefont {Bruner},
  \citenamefont {Klimkin}, \citenamefont {Blanchet}, \citenamefont
  {{Arusi-Parpar}}, \citenamefont {Kr{\"u}ger}, \citenamefont {Rubtsov},
  \citenamefont {Smirnova}, \citenamefont {Ivanov}, \citenamefont {Yan},
  \citenamefont {Brabec},\ and\ \citenamefont {Dudovich}}]{uzan2020attosecond}%
  \BibitemOpen
  \bibfield  {author} {\bibinfo {author} {\bibfnamefont {A.~J.}\ \bibnamefont
  {Uzan}}, \bibinfo {author} {\bibfnamefont {G.}~\bibnamefont {Orenstein}},
  \bibinfo {author} {\bibfnamefont {{\'A}.}~\bibnamefont
  {{Jim{\'e}nez-Gal{\'a}n}}}, \emph {et~al.},\ }\bibfield  {title} {\bibinfo
  {title} {Attosecond spectral singularities in solid-state high-harmonic
  generation},\ }\href {https://doi.org/10.1038/s41566-019-0574-4} {\bibfield
  {journal} {\bibinfo  {journal} {\emph {Nat. Photonics}}\ }\textbf {\bibinfo
  {volume} {14}}\bibfield  {number} {\bibinfo  {number} { no.~3},\ \bibinfo
  {pages} {pp.~183--187}} (\bibinfo {year} {2020})}\BibitemShut {NoStop}%
\bibitem [{\citenamefont {Goreslavskii}\ and\ \citenamefont
  {Popruzhenko}(2000)}]{goreslavskii2000tunneling}%
  \BibitemOpen
  \bibfield  {author} {\bibinfo {author} {\bibfnamefont {S.~P.}\ \bibnamefont
  {Goreslavskii}}\ and\ \bibinfo {author} {\bibfnamefont {S.~V.}\ \bibnamefont
  {Popruzhenko}},\ }\bibfield  {title} {\bibinfo {title} {Tunneling limit in
  the theory of photoelectron rescattering by the parent ion},\ }\href
  {https://doi.org/10.1134/1.559162} {\bibfield  {journal} {\bibinfo  {journal}
  {\emph {Journal of Experimental and Theoretical Physics}}\ }\textbf {\bibinfo
  {volume} {90}}\bibfield  {number} {\bibinfo  {number} { no.~5},\ \bibinfo
  {pages} {pp.~778--787}} (\bibinfo {year} {2000})}\BibitemShut {NoStop}%
\bibitem [{\citenamefont {Rook}\ and\ \citenamefont {Figueira De
  Morisson~Faria}(2022)}]{rook2022exploring}%
  \BibitemOpen
  \bibfield  {author} {\bibinfo {author} {\bibfnamefont {T.}~\bibnamefont
  {Rook}}\ and\ \bibinfo {author} {\bibfnamefont {C.}~\bibnamefont {Figueira De
  Morisson~Faria}},\ }\bibfield  {title} {\bibinfo {title} {Exploring
  symmetries in photoelectron holography with two-color linearly polarized
  fields},\ }\href {https://doi.org/10.1088/1361-6455/ac7bbf} {\bibfield
  {journal} {\bibinfo  {journal} {\emph {J. Phys. B: At. Mol. Opt. Phys.}}\
  }\textbf {\bibinfo {volume} {55}}\bibfield  {number} {\bibinfo  {number} {
  no.~16},\ \bibinfo {pages} {p.~165601}} (\bibinfo {year} {2022})}\BibitemShut
  {NoStop}%
\bibitem [{\citenamefont {Berry}(1977)}]{berry1977focusing}%
  \BibitemOpen
  \bibfield  {author} {\bibinfo {author} {\bibfnamefont {M.~V.}\ \bibnamefont
  {Berry}},\ }\bibfield  {title} {\bibinfo {title} {Focusing and twinkling:
  Critical exponents from catastrophes in non-{{Gaussian}} random short
  waves},\ }\href {https://doi.org/10.1088/0305-4470/10/12/015} {\bibfield
  {journal} {\bibinfo  {journal} {\emph {J. Phys. A: Math. Theor.}}\ }\textbf
  {\bibinfo {volume} {10}}\bibfield  {number} {\bibinfo  {number} { no.~12},\
  \bibinfo {pages} {pp.~2061--2081}} (\bibinfo {year} {1977})}\BibitemShut
  {NoStop}%
\bibitem [{\citenamefont {{Berry}}\ and\ \citenamefont
  {{Howls}}(1991)}]{1991RSPSA.434..657B}%
  \BibitemOpen
  \bibfield  {author} {\bibinfo {author} {\bibfnamefont {M.~V.}\ \bibnamefont
  {{Berry}}}\ and\ \bibinfo {author} {\bibfnamefont {C.~J.}\ \bibnamefont
  {{Howls}}},\ }\bibfield  {title} {\bibinfo {title} {{Hyperasymptotics for
  Integrals with Saddles}},\ }\href {https://doi.org/10.1098/rspa.1991.0119}
  {\bibfield  {journal} {\bibinfo  {journal} {\emph {Proceedings of the Royal
  Society of London Series A}}\ }\textbf {\bibinfo {volume} {434}}\bibfield
  {number} {\bibinfo  {number} { no.~1892},\ \bibinfo {pages} {pp.~657--675}}
  (\bibinfo {year} {1991})}\BibitemShut {NoStop}%
\bibitem [{\citenamefont {{Dorigoni}}(2019)}]{2019AnPhy.40967914D}%
  \BibitemOpen
  \bibfield  {author} {\bibinfo {author} {\bibfnamefont {D.}~\bibnamefont
  {{Dorigoni}}},\ }\bibfield  {title} {\bibinfo {title} {{An introduction to
  resurgence, trans-series and alien calculus}},\ }\href
  {https://doi.org/10.1016/j.aop.2019.167914} {\bibfield  {journal} {\bibinfo
  {journal} {\emph {Annals of Physics}}\ }\textbf {\bibinfo {volume} {409}},\
  \bibinfo {eid} {167914} (\bibinfo {year} {2019})},\
  \href{1411.3585}{E-print}\BibitemShut {NoStop}%
\bibitem [{\citenamefont {Demekhin}\ and\ \citenamefont
  {Cederbaum}(2012)}]{Demekhin2012dynamic}%
  \BibitemOpen
  \bibfield  {author} {\bibinfo {author} {\bibfnamefont {P.~V.}\ \bibnamefont
  {Demekhin}}\ and\ \bibinfo {author} {\bibfnamefont {L.~S.}\ \bibnamefont
  {Cederbaum}},\ }\bibfield  {title} {\bibinfo {title} {Dynamic interference of
  photoelectrons produced by high-frequency laser pulses},\ }\href
  {https://doi.org/10.1103/PhysRevLett.108.253001} {\bibfield  {journal}
  {\bibinfo  {journal} {\emph {Phys. Rev. Lett.}}\ }\textbf {\bibinfo {volume}
  {108}}\bibfield  {number} {\bibinfo  {number} { no.~25},\ \bibinfo {pages}
  {p.~253001}} (\bibinfo {year} {2012})}\BibitemShut {NoStop}%
\bibitem [{\citenamefont {Vismarra}\ \emph {et~al.}(2025)\citenamefont
  {Vismarra}, \citenamefont {Bertolino}, \citenamefont {Appi}, \citenamefont
  {Plach}, \citenamefont {Guly{\'{a}}s~Oldal}, \citenamefont {Gr{\'{o}}sz},
  \citenamefont {Dolso}, \citenamefont {Poulain}, \citenamefont {Mocci},
  \citenamefont {Inzani}, \citenamefont {Biswas}, \citenamefont {De~Marco},
  \citenamefont {Zeni}, \citenamefont {Frassetto}, \citenamefont {Poletto},
  \citenamefont {Reduzzi}, \citenamefont {Borrego-Varillas}, \citenamefont
  {W{\"{o}}rner}, \citenamefont {Filus}, \citenamefont {Seres}, \citenamefont
  {J{\'{o}}j{\'{a}}rt}, \citenamefont {Major}, \citenamefont {Csizmadia},
  \citenamefont {Nisoli}, \citenamefont {Eng-Johnsson}, \citenamefont
  {Dahlstr{\"{o}}m},\ and\ \citenamefont {Lucchini}}]{Vismarra2025dynamic}%
  \BibitemOpen
  \bibfield  {author} {\bibinfo {author} {\bibfnamefont {F.}~\bibnamefont
  {Vismarra}}, \bibinfo {author} {\bibfnamefont {M.}~\bibnamefont {Bertolino}},
  \bibinfo {author} {\bibfnamefont {E.}~\bibnamefont {Appi}}, \emph {et~al.},\
  }\bibfield  {title} {\bibinfo {title} {Dynamic interference of chirped
  photoelectrons},\ }\href {https://doi.org/10.1103/73tl-w87y} {\bibfield
  {journal} {\bibinfo  {journal} {\emph {Phys. Rev. Lett.}}\ }\textbf {\bibinfo
  {volume} {135}}\bibfield  {number} {\bibinfo  {number} { no.~3},\ \bibinfo
  {pages} {p.~033202}} (\bibinfo {year} {2025})}\BibitemShut {NoStop}%
\bibitem [{\citenamefont {Itatani}\ \emph {et~al.}(2002)\citenamefont
  {Itatani}, \citenamefont {Qu{\'{e}}r{\'{e}}}, \citenamefont {Yudin},
  \citenamefont {Ivanov}, \citenamefont {Krausz},\ and\ \citenamefont
  {Corkum}}]{Itatani2002attosecond}%
  \BibitemOpen
  \bibfield  {author} {\bibinfo {author} {\bibfnamefont {J.}~\bibnamefont
  {Itatani}}, \bibinfo {author} {\bibfnamefont {F.}~\bibnamefont
  {Qu{\'{e}}r{\'{e}}}}, \bibinfo {author} {\bibfnamefont {G.~L.}\ \bibnamefont
  {Yudin}}, \emph {et~al.},\ }\bibfield  {title} {\bibinfo {title} {Attosecond
  streak camera},\ }\href {https://doi.org/10.1103/PhysRevLett.88.173903}
  {\bibfield  {journal} {\bibinfo  {journal} {\emph {Phys. Rev. Lett.}}\
  }\textbf {\bibinfo {volume} {88}}\bibfield  {number} {\bibinfo  {number} {
  no.~17},\ \bibinfo {pages} {p.~173903}} (\bibinfo {year} {2002})}\BibitemShut
  {NoStop}%
\bibitem [{\citenamefont {Moos}\ \emph {et~al.}(2020)\citenamefont {Moos},
  \citenamefont {J{\"{u}}r{\ss}},\ and\ \citenamefont
  {Bauer}}]{Moos2020intense}%
  \BibitemOpen
  \bibfield  {author} {\bibinfo {author} {\bibfnamefont {D.}~\bibnamefont
  {Moos}}, \bibinfo {author} {\bibfnamefont {H.}~\bibnamefont
  {J{\"{u}}r{\ss}}},\ and\ \bibinfo {author} {\bibfnamefont {D.}~\bibnamefont
  {Bauer}},\ }\bibfield  {title} {\bibinfo {title} {Intense-laser-driven
  electron dynamics and high-order harmonic generation in solids including
  topological effects},\ }\href {https://doi.org/10.1103/PhysRevA.102.053112}
  {\bibfield  {journal} {\bibinfo  {journal} {\emph {Phys. Rev. A}}\ }\textbf
  {\bibinfo {volume} {102}}\bibfield  {number} {\bibinfo  {number} { no.~5},\
  \bibinfo {pages} {p.~053112}} (\bibinfo {year} {2020})}\BibitemShut {NoStop}%
\end{thebibliography}%


\end{document}